\documentclass[10pt]{article}
\usepackage{bbm}
\usepackage{amsmath}
\usepackage{amsthm}
\usepackage{amssymb} 
\usepackage{mathrsfs}
\usepackage{graphicx}
\usepackage{authblk}
\usepackage[backref=none,hypertexnames=false,colorlinks=true,urlcolor=blue,linkcolor=blue,citecolor=blue]{hyperref}
\usepackage[numbers,comma,square,sort&compress]{natbib}
\usepackage[letterpaper,text={6.5in,9in},centering]{geometry}

\graphicspath{{Figs}}

\makeatletter\@addtoreset{equation}{section}\makeatother

\newtheorem{thm}{Theorem}[section]

\newtheorem{rmk}[thm]{Remark}

 {\begin{trivlist}\item[]\textbf{Acknowledgments.}}{\end{trivlist}}

\begin{document}

\title{Localized Patterns}
\author[1]{Jason J. Bramburger}
\author[2]{Dan J. Hill}
\author[3]{David J.B. Lloyd}

\affil[1]{\small Department of Mathematics and Statistics, Concordia University, Montr\'eal, QC, Canada}
\affil[2]{\small Fachrichtung Mathematik, Universit\"at des Saarlandes, 66041 Saarbr\"ucken, Germany}

\affil[3]{\small School of Mathematics and Physics, University of Surrey, Guildford, GU2 7XH, UK}

\date{}
\maketitle

\begin{abstract}
Localized patterns are coherent structures embedded in a quiescent state and occur in both discrete and continuous media across a wide range of applications. While it is well-understood how domain covering patterns (for example stripes and hexagons) emerge from a pattern-forming/Turing instability, analyzing the emergence of their localized counterparts remains a significant challenge. There has been considerable progress in studying localized patterns over the past few decades, often by employing innovative mathematical tools and techniques. In particular, the study of localized pattern formation has benefited greatly from numerical techniques; the continuing advancement in computational power has helped to both identify new types patterns and further our understanding of their behavior. We review recent advances regarding the complex behavior of localized patterns and the mathematical tools that have been developed to understand them, covering various topics from spatial dynamics, exponential asymptotics, and numerical methods. We observe that the mathematical understanding of localized patterns decreases as the spatial dimension increases, thus providing significant open problems that will form the basis for future investigations.   
\end{abstract}

\tableofcontents
\newpage
\section{Introduction}

Non-trivial behavior embedded in a quiescent state, hereby referred to as spatial localization, has long fascinated applied scientists. This stretches from trying to understand on very large spatial scales why stars form galaxies, to the formation of a Bose-Einstein condensate at tiny length-scales. Mathematical interest in localization goes back at least to the tale of John Scott Russell in 1834 and his discovery of a solitary ``wave of translation" propagating down a canal \cite{allen1998early},  which led to Joseph Boussinesq and Lord Rayleigh to develop a mathematical theory to understand the mechanisms behind Russell's observation. In the 20th century, the study of spatially periodic patterns with symmetry can be traced back to Lord Rayleigh \cite{rayleigh1916lix} and Alan Turing \cite{turing1990chemical}, including in subsequent unpublished work~\cite{Dawes2016}. It is now known that if a spatially periodic pattern and a quiescent state are both stable, the competition induced by such bistability can lead to states that admit a cooperative embedding of the pattern in the quiescent state, giving a localized pattern.  

\begin{figure}[htb]
    \centering
    \includegraphics[width=0.9\linewidth]{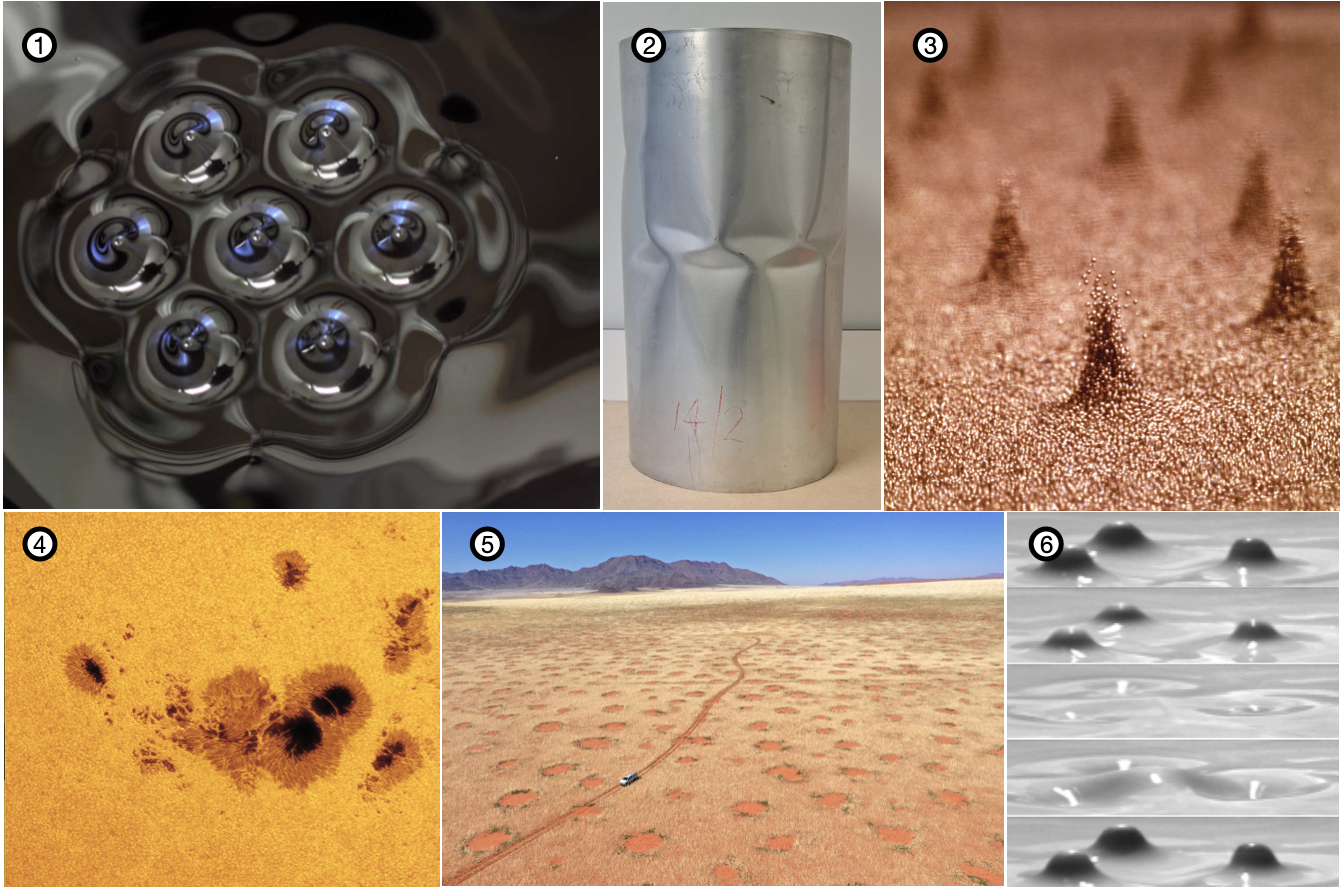}
    \caption{A collage of localized patterns found in  (1) ferrofluids~\cite{lloyd2015homoclinic}, (2) buckling of cylinders (Photo credit: M. A. Wadee),
    (3) vertically vibrated granular material (Photo credit: P.B. Umbanhowar), (4) Sun spots (By NASA Goddard Space Flight Center from Greenbelt, MD, USA - Solar Archipelago) (5) vegetation patches in arid areas (Photo credit: S. Getzin), and (6) vertically vibrated fluids~\cite{Lioubashevski1999}.}
    \label{fig:app_collage}
\end{figure}

In this review, we will focus on structures that are localized in one or more spatial dimensions embedded in a quiescent state and possess some non-trivial fine-scale regular structure, which we call a pattern. Figure~\ref{fig:app_collage} provides examples of the types of localized patterns we will concentrate on that occur in both continuum and discrete systems. Throughout this review, we will see that the field has been developed with numerical, experimental, and analytical techniques that work in tandem to better understand localization mechanisms. We highlight several reviews that came before this one \cite{dawes2010emergence,knobloch2015spatial,knobloch2008spatially,Champneys2021d,Malomed2022}, but note their focus is more rooted in the physics literature or primarily restricted to localized patterns in one spatial dimension. What differentiates this review from those that came before is that we seek to highlight the mathematical development of the subject. This includes how specific equations have been investigated for localized patterns and providing overviews of the myriad of clever mathematical and computational techniques that have been developed to study these multi-dimensional patterns.  

Localized patterns can be found in a wide range of application areas from fluids, quantum theory, biology, mechanics, sociology, to nonlinear optics. They come in the form of an oasis of vegetation in an arid landscape, a hotspot of criminal activity, or a laser pulse in plasmas. Because they arise in so many different applications, they are often referred to by different names which have historical relevance to the field. This includes solitary waves in fluid dynamics, convectons in binary fluid convection, and fairy circles in vegetation models, to name a few. To ease the reader's journey into the literature on localized patterns, we will provide references and nomenclature for each specific type of pattern as they occur in this review. This will help to connect the specific pattern to specific applications. However, we note that while the study of these patterns is motivated by application, our goal is to review all things mathematical in their investigation. This means primarily relying on phenomenological equations that describe macroscopic behavior and attempting to understand the fundamental mechanisms that drive their formation. 

{\bf The aid of computers.} 
The computational side of the study of localized pattern formation has been developed in parallel to the analytical side. This is in part due the development of numerical methods (root-finding, pseudo-arclength continuation, timestepping, etc) and software to implement it, such as AUTO \cite{doedel2007auto}. The impact of the computational side of research into localized patterns cannot be understated. Numerical exploration of spatially localized patterns has driven mathematical questions that have led to theoretical advances such as understanding the emergence of small amplitude patterns, developing center-manifold reductions, understanding flows on and near invariant manifolds, studying exponential dichotomies, predicting bifurcating patterns using symmetry, the utility of exponential asymptotic expansions, and rigorous computer-assisted proofs. The integration of newly developed software such as PDE2PATH \cite{uecker_wetzel_rademacher_2014}, VisualPDE \cite{walker2023visualpde}, and Julia's BifurcationKit.jl package \cite{veltz2020bifurcationkit}, will surely continue to drive mathematical methods for understanding novel phenomena uncovered with these programs.

\begin{figure}[t]
    \centering
    \includegraphics[width=0.9\linewidth]{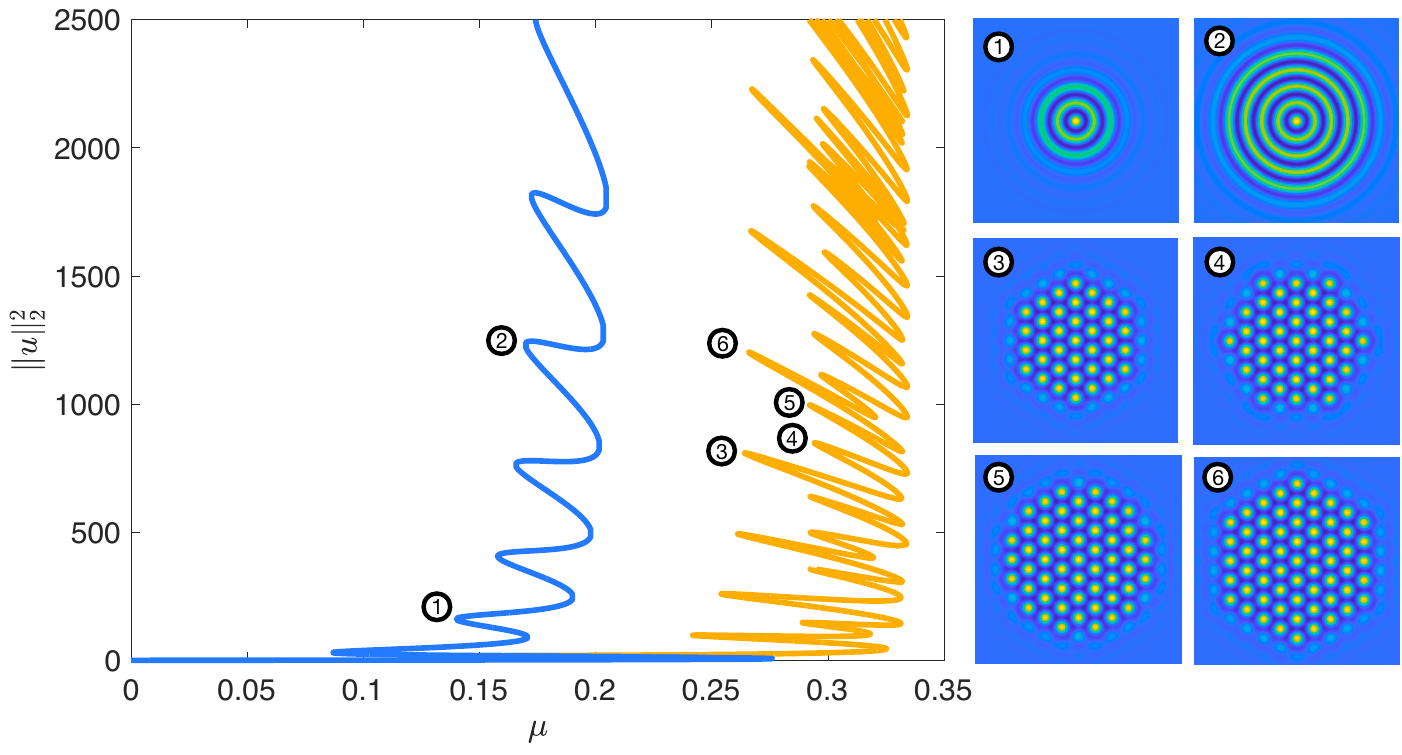}
    \caption{Numerically continued bifurcation curves of localized spot (blue) and hexagon patches (yellow) in the 2D SHE \eqref{SwiftHohenberg} with $\nu = 1.6$. Sample contour plots of the profiles are provided at folds along the curves.}
    \label{fig:hex_snake}
\end{figure}

To illustrate, Figure~\ref{fig:hex_snake} shows typical numerically continued bifurcation diagrams for localized spots and hexagon patches. In the left panel, we plot how the $L^2$-norm of the pattern (which corresponds roughly to a measure of spatial extent) changes smoothly as a control parameter is varied, while the right panels show snapshots of how the solutions change along this path. This diagram and others like it raise many mathematical questions that drive research into them. Fundamental questions arising in the discipline are:
\begin{enumerate}
    \item What are the generic conditions for localized patterns to form?
    \item What kinds of patterns will be selected in certain systems?
    \item What generic kinds of behavior do these patterns have, such as stability and bifurcations? 
    \item Can one predict the existence regions (in parameter space) and resulting bifurcation curves of these patterns?
\end{enumerate}    
It is questions like these that arise from pictures such as Figure~\ref{fig:hex_snake} that has driven so much of the mathematical investigation into localized patterns. Patterns by their very nature are enhanced by viewing them, and so computer simulation is almost necessary for their analysis.

{\bf Prototypical equations.}
What many of these applications have in common is that they are modeled by partial differential equations (PDEs), particularly reaction-diffusion 
(RD) systems, that combine spatial spreading with state-dependent nonlinearities that undergo pattern-forming instabilities and are amenable to mathematical analysis. Generally, these RD systems
take the form
\begin{equation}
    \mathbf{u}_t
    = \mathbf{D}\Delta\mathbf{u} + \mathbf{f}(\mathbf{u},\mu),
\end{equation}
where 
$\mathbf{u}=\mathbf{u}(\mathbf{x},t) \in \mathbb{R}^n$, $\mathbf{D}$ is an $n\times n$ diagonal `diffusion' matrix whose entries are all positive, 
$\Delta$ is the Laplace operator on $\mathbb{R}^d$, $\mathbf{f}$ is a smooth nonlinearity, $\mu$ is a bifurcation parameter, and the subscript $t$ denotes partial differentiation with respect to that variable. 

Beyond RD systems, much attention in the pattern formation community has been paid to the Swift--Hohenberg equation (SHE), 
whose quadratic-cubic form is
given by 
\begin{equation}\label{SwiftHohenberg}
    U_t = -(1+\Delta)^2U - \mu U + \nu U^2 - U^3,\qquad \mathbf{x}\in \mathbb{R}^{d},
\end{equation}
where $U=U(\mathbf{x},t)\in\mathbb{R}$, $\mu$ is the linear pumping or bifurcation parameter, and $\nu$ is a nonlinear damping parameter. As we will see in \S\ref{subsec:Turing}, the SHE has been introduced as a phenomenological model to better understand the pattern forming bifurcations in RD systems that Turing's pioneering work described. The SHE has the benefit of being simple enough that it can be studied analytically, while being complicated enough to generate interesting solutions. An important aspect of the SHE posed on $\mathbb{R}^d$, $1\leq d\leq3$, is that it is a gradient flow in $H^2(\mathbb{R}^d)$ since
\begin{equation}
    U_t = -\nabla\mathcal{E}(U,\mu,\nu),
\end{equation}
where the energy functional of the system is given by
\begin{equation}
    \mathcal{E}(U,\mu,\nu) = \int \bigg[\frac{(U + \Delta U)^2}{2} + \frac{\mu U^2}{2} - \frac{\nu U^3}{3} + \frac{U^4}{4}\bigg] \mathrm{d}\mathbf{x}.
\end{equation}
This gradient flow structure means that $\mathcal{E}$ decreases in time along solutions to the SHE, causing them to asymptotically settle the system into steady-states. 

\begin{figure}[t]
    \centering
    \includegraphics[width=0.9\linewidth]{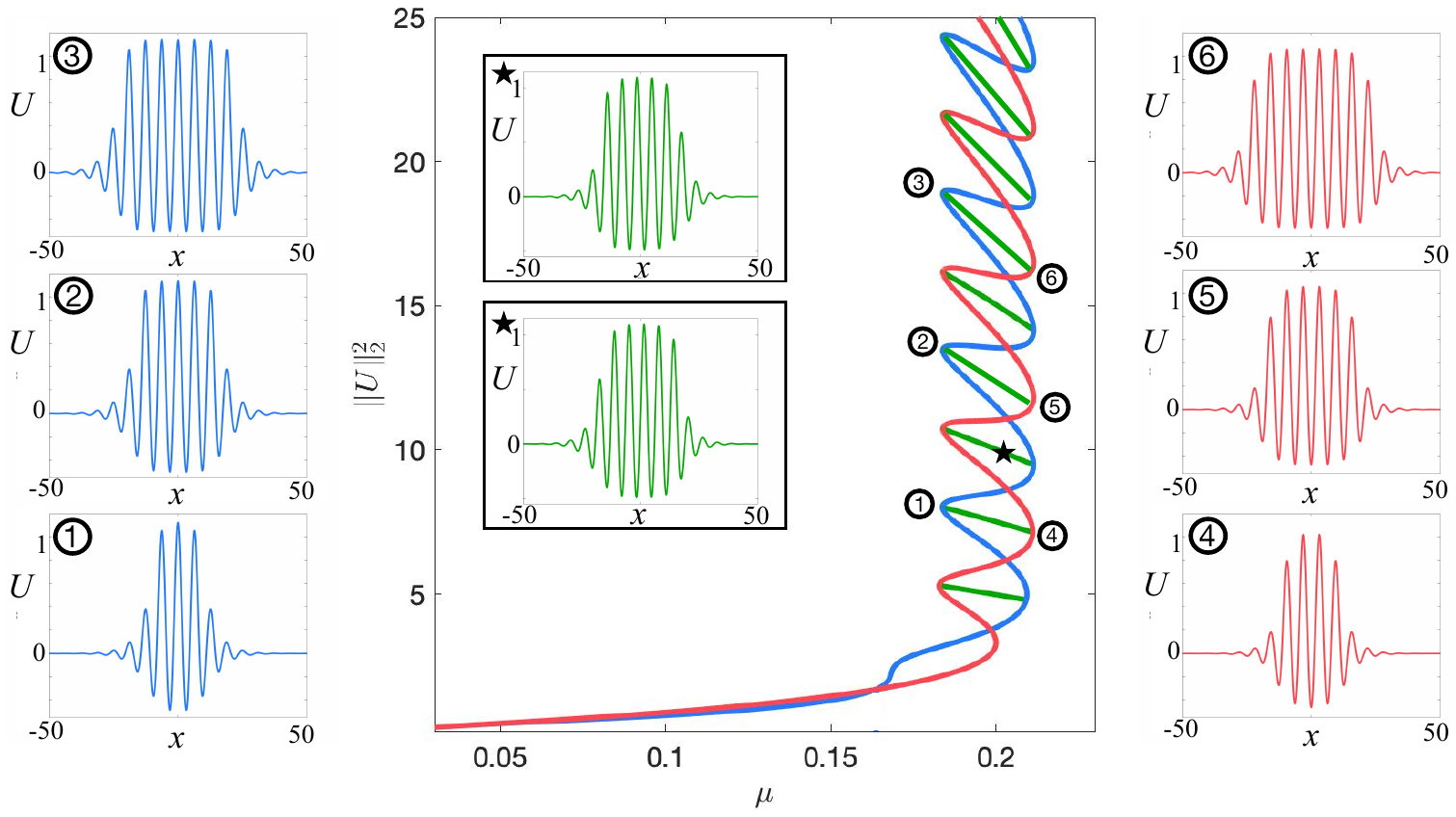}
    \caption{Snakes and ladders bifurcation diagrams are a familiar feature of localized patterns in one spatial dimension. The diagram here is for the 1D SHE \eqref{SwiftHohenberg} with $\nu = 1.6$ and features two intertwined branches of localized solutions: one with a maximum in the middle (blue) and one with a minimum (red). These branches are connected by the asymmetric `ladder' states (green). As one ascends the diagram, the localized pattern grows by adding new spatially periodic patterns (SPPs) on both sides, as exemplified with the profiles taken at successive saddle-node bifurcations on the curve.}
    \label{fig:snaking_1d}
\end{figure}

In one spatial dimension ($d=1$) the SHE supports stable spatially periodic patterns (SPPs) for $0<\mu\ll1$ and $\nu>\sqrt{27/38}$ that resemble sinusoidal functions. Since both the SPP and the trivial state $U = 0$ are stable in the same parameter regime, an interesting experiment one might undertake is to consider an initial condition in the SHE that is the SPP on a compact set and the trivial state on the rest. Now, if the energy of a SPP is larger or smaller than $0$, the energy of the trivial state $U = 0$, then one might intuit that the region of localization (the SPP) contracts or expands as time goes on. For a fixed $\nu$, the energy of a SPP will vary with $\mu$ leaving, in general, only a single point of perfect matching where it also has zero energy. This point is referred to as the {\em Maxwell point} and it occurs at $\mu = 0.2004$ when $\nu = 1.6$. This argument seems to indicate that localized SPPs should only be expected at the Maxwell point, while the numerics presented Figure~\ref{fig:snaking_1d} indicate otherwise. The reason we find such a discrepancy is that the above intuition does not account for the transition between $U = 0$ and the localized SPP components. The energy stored at these interfaces allows one to observe localized SPPs in an open parameter interval about the Maxwell point. Pomeau~\cite{Pomeau1986} was the first to postulate qualitatively that the pattern could `pin' the interfaces and create an open region in parameter space for stationary patterns to co-exist. 

One should compare the regular structure of the bifurcation curves in Figure~\ref{fig:snaking_1d} to those in Figure~\ref{fig:hex_snake}. While both present localized patterns in the SHE, the difference between them serves as a microcosm for much of the study of localized patterns. In one spatial dimension, the ``snaking'' bifurcation curves in Figure~\ref{fig:snaking_1d} are fully understood (see \S\ref{subsec:HomoclinicSnaking} below) from numerous mathematical investigations studying these localized patterns and their bifurcation curves from multiple angles. This stands in stark contrast to the mathematical understanding of the localized hexagons in Figure~\ref{fig:hex_snake}, where neither the irregular bifurcation structure nor the existence of these patterns is proven. This is a theme that runs through this review whereas one increases the dimension of the spatial setting the less is known about the localized patterns that can be found on it. 

{\bf What is not covered?}
As with all reviews, we have had to make some decisions on what to review and not. The literature on localized patterns is vast, covering decades worth of research. In terms of applications, we only give an overview of some of the main areas, instead focusing on the mathematical aspects. The focus of the review is {\it not} on solitons or solitary waves in integrable systems though a lot of the techniques described below can be applied to those systems. Instead, this review will focus on unforced non-equilibrium thermodynamical systems such as the RD systems and SHE. Additionally, we focus on the {\it existence} of localized patterns and not their stability; the literature surrounding the (in)stability of localized patterns is similarly extensive and worthy of its own review. In terms of mathematical analysis and techniques, we do not cover localized and domain-covering patterns occurring near the singular limit where one of the diffusion coefficients in the RD system is very small (though in \S\ref{subsubsec:Singular} we do describe how the localized patterns change their behavior as they move closer to the singular limit), invasion pattern-forming fronts (though we do link to this literature in \S\ref{subsec:Stripes}), numerical methods, nonlocal models, and patterns forming in the wake of an invasion front. Our focus is on localized patterns, and so we direct the reader to Cross and Hohenberg's 1993 review \cite{cross1993pattern} for a wider range of patterns than just those considered here. Furthermore, we only review mathematical concepts at the surface level, while more in-depth mathematical treatments can be found in books by Hoyle \cite{hoyle2006pattern} and Schneider \& Uecker \cite{schneider2017nonlinear}.

{\bf Outline.}
This review is organized by the dimension the localized patterns exist in. The reason for this is that different mathematical tools and techniques exist in different dimensions, and our collective understanding is greatest for one-dimensional patterns and least for three-dimensional localized patterns. In Section~\ref{sec:1D}, we review the emergence of localized patterns in one spatial dimension and the bifurcation structure known as homoclinic snaking. Section~\ref{sec:1D} is concluded by looking at variations on the standard set-up. Section~\ref{sec:fronts} then looks at two-dimensional patterns that are localized in only one direction, primarily with the non-localized direction being compact. We then move to axisymmetric localized patterns, reviewed in Section~\ref{sec:radial}, which provide a bridge from the 1D analysis to truly localized 2D patterns since one can introduce the radial independent variable and reduce the governing equations back to one spatial dimension, albeit a nonautonomous set of ODEs. This leads to Section~\ref{ssec:FullyLocal} where we review results on fully localized patterns in 2D which are not axisymmetric. Section~\ref{sec:3d} reviews localized patterns in 3D where analytic results are significantly more scattered than their 1D and 2D counterparts. Finally, we conclude in Section~\ref{sec:Conclusion} with some final thoughts and open problems.

\section{1D Localized Pattern Formation}
\label{sec:1D}

Two of the most fundamental concepts of studying pattern formation are finite-wavenumber instabilities and the Swift--Hohenberg equation. Interestingly, both of these concepts can be traced back to Alan Turing and his work on morphogenesis. In honor of Turing's contribution we typically refer to any finite-wavenumber instability in a system of RD equations as a {\em Turing 
instability
}, but as a historical note similar investigations were undertaken well before Turing in the fluids' literature by Rayleigh \cite{rayleigh1916lix} and Taylor \cite{taylor1923viii}, among others.

In this section we will review both localized pattern formation in one spatial dimension and their expected parameter-dependent existence curves. We begin with Turing's contributions in \S\ref{subsubsec:Turing} and the generalizations thereof to systems of RD equations in \S\ref{subsec:GeneralPF}. The Swift--Hohenberg equation and a variety of its solutions are discussed in \S\ref{subsec:SH1D}. We introduce the concept of {\em homoclinic snaking} in \S\ref{subsec:HomoclinicSnaking} to explain Figure~\ref{fig:snaking_1d} and conclude in \S\ref{subsec:Variations} with a brief discussion of the multitude of variations on the standard setting that have furthered our understanding of localized pattern formation in one spatial dimension.

\subsection{Turing and Finite-Wavenumber Instabilities}\label{subsec:Turing}

\subsubsection{Turing's Contribution}\label{subsubsec:Turing}

Investigations of pattern formation go back at least to Turing's seminal 1952 paper ``On the chemical basis of morphogenesis'' \cite{turing1990chemical}. To explain the emergence of complex spatial patterns, Turing considered a general two-component RD system for an activator, $u$, and inhibitor, $v$, chemical species. These two chemical species interact with reaction kinetics $f(u,v)$ and $g(u,v)$ that affect the rate at which the chemicals are produced. The chemicals are then assumed to diffuse in space, giving way to the RD model
\begin{equation}\label{e:RDsys}
    \begin{split}
        u_t =& D_u u_{xx} + f(u,v),\\
        v_t =& D_v v_{xx} + g(u,v),
    \end{split}
\end{equation}
where $D_u, D_v>0$ are the diffusion coefficients of $u$ and $v$, respectively, and subscripts on the variables $u$ and $v$ denote partial differentiation with respect to time $t \geq 0$ and space $x \in \mathbb{R}$. 

\begin{figure}[t!]
    \centering
    \includegraphics[width=\linewidth]{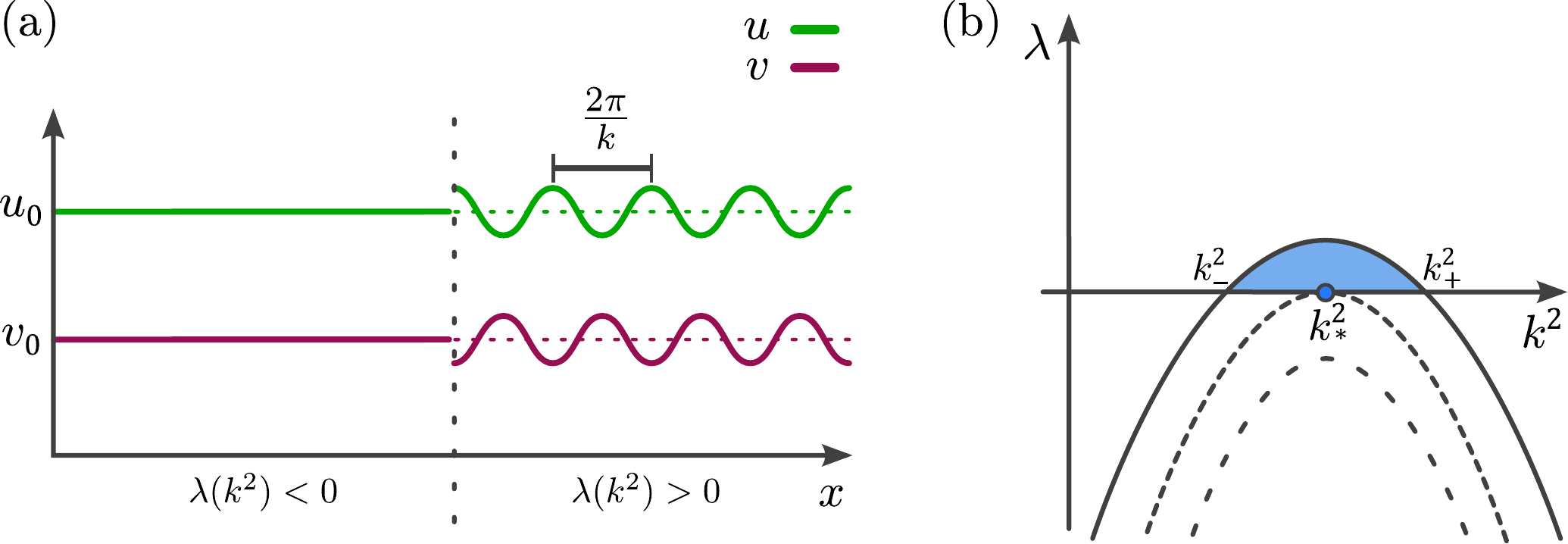}
    \caption{Turing patterns are one of the fundamental features of a system that leads to localized patterns. (a) A sketch of solutions just before (left) and after (right) bifurcation (b) A plot of the linear growth curve just before, 
    during,
    and after the Turing bifurcation.
    }
    \label{fig:Turing_bif}
\end{figure}

Turing showed that patterns can bifurcate from spatially homogeneous steady-states $u_0$ and $v_0$ of \eqref{e:RDsys}. To see this, one can linearize \eqref{e:RDsys} about these steady-states and look for exponentially growing, spatially periodic solutions of the form $(u, v)(x,t) = (u_0, v_0) + (\hat u,\hat v)e^{\lambda t}e^{ikx}+\mbox{c.c}$ where $\hat u$ and $\hat v\in\mathbb{R}$, $k\in\mathbb{R}$ is a spatial wavenumber, $\lambda\in\mathbb{C}$ is the growth rate, 
and $\mbox{c.c}$ denotes the conjugate of any complex terms.
Substituting this ansatz into \eqref{e:RDsys} and truncating at the linear terms, one is left to analyze the matrix eigenvalue problem
\begin{equation}\label{e:Turingeig}
    \lambda\begin{bmatrix}
        \hat u\\\hat v
    \end{bmatrix} = \begin{bmatrix}
        -D_uk^2 + f_u(u_0,v_0) & f_v(u_0,v_0)\\
        g_u(u_0,v_0) & -D_vk^2 + g_v(u_0,v_0)
    \end{bmatrix}\begin{bmatrix}
        \hat u\\\hat v
    \end{bmatrix}.
\end{equation}
Provided one can find a wavenumber $k \in \mathbb{R}$ such that \eqref{e:Turingeig} has a solution with positive eigenvalue $\lambda=\lambda(k^2) > 0$, then one expects to see spatially periodic patterns emerge from the homogeneous steady-states $(u_0,v_0)$, as shown in Figure~\ref{fig:Turing_bif}(a).

Turing's original paper was interested in demonstrating that diffusion could lead to pattern formation even when the spatially homogeneous steady-state is linearly stable in the absence of diffusion, i.e. $D_u = D_v = 0$. In this context, the criteria for a pattern-forming instability to occur reduces to
\[
    \begin{split}
        f_u(u_0,v_0)+g_v(u_0,v_0) &<0, \\ f_u(u_0,v_0)g_v(u_0,v_0)-f_v(u_0,v_0)g_u(u_0,v_0) &>0, \\ 
        D_u f_u(u_0,v_0) + D_u g_v(u_0,v_0)&>0,\\ (f_u(u_0,v_0)g_v(u_0,v_0)-f_v(u_0,v_0)g_u(u_0,v_0))-\frac{(D_vf_u(u_0,v_0)-D_ug_v(u_0,v_0))^2}{4D_uD_v}&<0
    \end{split}
\]
As illustrated in Figure~\ref{fig:Turing_bif}(b), by plotting $\lambda(k^2)$ against $k^2$, we see that as one passes through the Turing instability, a critical finite-wavenumber, $k=k_*$ is the first to become unstable and then a band of unstable wavenumbers occur with the most unstable (largest $\lambda > 0$) being the critical finite-wavenumber. 
Typically, the spatial patterns grow and then saturate due to the stabilizing nonlinearities in \eqref{e:RDsys} that prevent solutions from blowing up. The observed wavenumber is found to be the critical finite-wavenumber that was the first to become unstable.

\subsubsection{General Pattern-Forming Bifurcations}\label{subsec:GeneralPF}

Turing's approach can be extended to general multi-component RD equations~\cite{villar2023}, of the form
\begin{equation}\label{e:GeneralRDsystem}
    \mathbf{u}_t = \mathbf{D}\mathbf{u}_{xx} + \mathbf{f}(\mathbf{u}), \qquad \mathbf{u}\in\mathbb{R}^n.
\end{equation}
The generalization of Turing's original paper requires a finite-wavenumber instability to occur. 
We repeat Turing's arguments by looking at the linear stability of an assumed spatially homogeneous steady-state $\mathbf{u}_0$ using the ansatz $\mathbf{u}(x,t)=\mathbf{u}_0+\mathbf{\tilde u}e^{\lambda t}e^{ikx}+\mbox{c.c}$. Taking the determinant of the resulting eigenvalue problem, akin to \eqref{e:Turingeig} previously, leads to the condition
\begin{equation}
   d(\lambda,k) := \mathrm{det}(-k^2\mathbf{D}+\mathbf{f_u}(\mathbf{u}_0)-\lambda\mathbbm{1}) = 0,
\end{equation}
where $d(\lambda,k)$ is called the {\em dispersion relation}. Solving $d(\lambda,k)=0$ for $\lambda(k^2)$, the typical shape of $\lambda(k^2)$ that is required for a pattern-forming instability is shown in Figure~\ref{fig:Turing_bif}(b) and a critical wavenumber $k=k_*>0$ to yield $\lambda(k_*^2)=0$. For a pattern-forming instability, one then requires the following conditions 
\[
    d(0,k_*)=0,\qquad \frac{\partial  d(0,k_*)}{\partial \lambda} \neq 0,\qquad \Re\left(\frac{d^2\lambda(k)}{dk^2}\left|_{k=k_*}\right.\right)<0.
\]

One can reverse-engineer the above setting by looking for a single scalar equation whose linearization about the trivial state gives the same dispersion relation as Figure~\ref{fig:Turing_bif}(b). The result of this exercise is the linear SHE
\begin{equation}\label{SHlinear}
    u_t = -(q^2+\partial_x^2)^2u - \mu u.
\end{equation}
Here $k=q$ is the root of the resulting dispersion relation
\[
    d(\lambda, k) = -(q^2-k^2)^2 - \mu -\lambda
\]
and the inclusion of the parameter $\mu$ provides analogous instability criteria to the diffusion coefficients $D_{u}, D_{v}$ in \eqref{e:RDsys}. Supposing \eqref{SHlinear} represents the linearization of a nonlinear system, then a Taylor expansion of said nonlinear system about $u = 0$ would result in the (nonlinear) SHE \eqref{SwiftHohenberg}, providing justification for why the SHE is seen as a prototypical canonical form for pattern formation. 

As an interesting historical note, Turing's quest to understand finite-wavenumber instabilities and pattern formation led to his own derivation of a nonlocal variant of the SHE. Turing's version of the SHE is given by
\[
    U_t = -(1+\Delta)^2U - \mu U + U^2 - \nu\left(\frac{U^2}{1-\sigma^2\Delta} \right)U
\]
where $\sigma,\nu$ are real parameters, and arose from work in his last years developing a mathematical theory for the formation of patterns in daisies~\cite{Dawes2016}. Similarly, Swift and Hohenberg's original derivation of the equation that now bears their name had a nonlocal term that has been all but forgotten in the literature \cite{swift1977hydrodynamic}. For the rest of this section (and the paper in general), we will concentrate on variants of the SHE that do not include nonlocal terms since this is where much of the mathematical progress has been made. A brief exception will be a quick review of localized patterns in nonlocal equations found in \S\ref{subsec:Nonlocal}. Furthermore, the local versions of the SHE have a lot in common with RD systems, and so much of the discussion below also applies to equations of the form \eqref{e:RDsys}. We discuss extensions of the theory to the general RD setting in \S\ref{subsec:Variations}.

\subsection{Pattern Formation in the Swift--Hohenberg Equation}\label{subsec:SH1D}


Turing bifurcations lead to the existence of spatially-periodic steady-state solutions to RD systems. As it turns out, upon accounting for the nonlinear terms in the SHE, one can further identify the existence of spatially localized solutions emerging from the Turing bifurcation point. Formally, one carries out an asymptotic expansion of the form
\begin{equation}\label{e:TuringExpansion}
        U(x,t) = \varepsilon A(\varepsilon x,\varepsilon^2 t) e^{ikx} + \mbox{c.c}, \qquad \mu = \varepsilon^2\hat\mu,\qquad |\varepsilon|\ll1
\end{equation}
where $c.c.$ stands for complex conjugate of the previous term and $\hat\mu$ is a scaled bifurcation parameter. The function $A$ is complex and describes the envelope of the localization over the ``fast" spatial oscillations of the $e^{ikx}$ periodic pattern. A crucial part of the expansion is that the envelope function $A$ slowly evolves in space with a variable $X=\varepsilon x$ and in time with a variable $T=\varepsilon^2t$. This separation of slowly varying envelope and fast varying periodic pattern generates a multi-scales problem that can be taken advantage of when carrying out the asymptotic expansion. 

The equation governing the leading order in $\varepsilon$ evolution of $A$ is the cubic Ginzburg-Landau equation,
\begin{equation}\label{e:GZeqn}
    A_T = a A_{XX} - \hat\mu A - b |A|^2A,
\end{equation}
where $(X,T)=(\varepsilon x,\varepsilon^2 t)$ are the slowly varying independent variables. We note that for the quadratic-cubic SHE $a=4, b = 3 - 38\nu^2/9 $; see~\cite{burke2006localized}. The validity of \eqref{e:GZeqn} as a good approximation of the dynamics for $\varepsilon\ll1$ can be rigorously justified for long time scales using Gr\"onwall estimates, as was done in \cite{schneider2017nonlinear}. 

The analysis of \eqref{e:GZeqn} is typically divided into two cases: the {\em subcritical} case when $b < 0$ and the {\em supercritical} case when $b > 0$ \cite{Iooss1993}. Looking at time-independent solutions to \eqref{e:GZeqn}, we have the following special solutions for the subcritical case: 

\begin{figure}[t]
    \centering
    \includegraphics[width=0.9\linewidth]{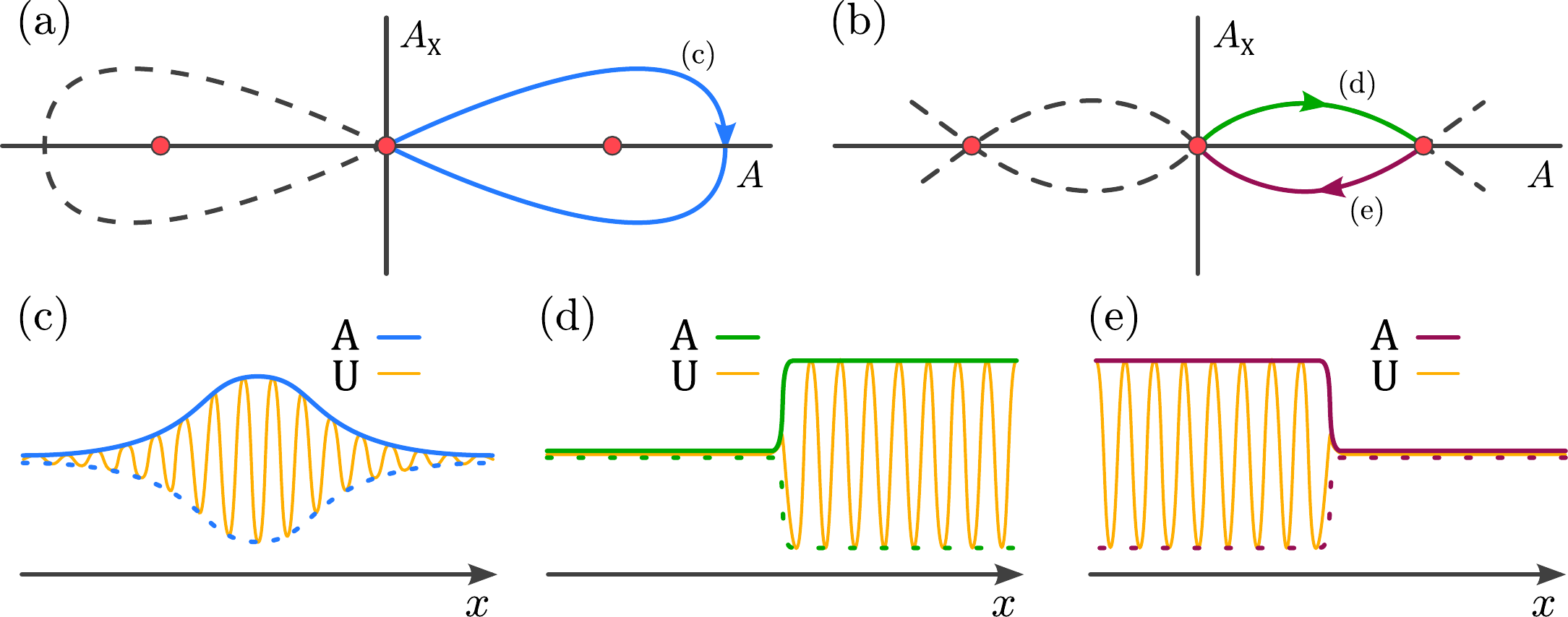}
    \caption{Phase portraits for time-independent solutions of the cubic Ginzburg--Landau equation \eqref{e:GZeqn} in the (a) subcritical case $b<0$, $\hat{\mu}>0$, and the (b) supercritical case $b>0$, $\hat{\mu}<0$. The system possesses three equilibria $A=0,\pm\sqrt{-\hat{\mu}/b}$ (red circles), and either homoclinic or heteroclinic orbits. We present profiles for (c) localized pulse (blue), (d) front (green), and (e) back (purple) solutions, where the oscillating pattern $U$ (yellow) is bounded by the amplitude $A$}
    \label{fig:GL_Sols}
\end{figure}

\begin{itemize}
    \item The trivial state $A=0$.
    \item For $\hat\mu > 0$, a nontrivial state $A=\pm\sqrt{-\hat\mu/b}$ corresponding to a spatially periodic pattern via \eqref{e:TuringExpansion}.
    \item For $\hat\mu > 0$, a localized solution $A=\pm\sqrt{-2\hat\mu/b}\,\mathrm{sech}(\sqrt{\hat\mu/a}X)$.
\end{itemize}
In the supercritical case we have the following special solutions:
\begin{itemize}
	\item The trivial state $A=0$.
    \item For $\hat\mu < 0$, a nontrivial state $A=\pm\sqrt{-\hat\mu/b}$ corresponding to a spatially periodic pattern via \eqref{e:TuringExpansion}.
    \item For $\hat\mu < 0$, front and back solutions $A=\pm\sqrt{-\hat\mu/b}\,\mathrm{tanh}(\sqrt{-\hat\mu/2a}X)$.
\end{itemize}
Hence, for localized patterns embedded in the trivial state to occur, we are required to be in the subcritical case of $b<0$. Figure~\ref{fig:GL_Sols} provides illustrations of these solutions in both the subcritical and supercritical cases.

\begin{rmk}
    The Ginzburg--Landau equation \eqref{e:GZeqn} has a phase invariance property wherein if $A$ is a solution, then so is $\mathrm{e}^{i \theta} A$ for any $\theta \in S^1$. Thus, initially one might expect that a continuum of localized patterns emerge from the trivial state at $\hat\mu = 0$ near any Turing bifurcation. However, \eqref{e:GZeqn} only constitutes a leading-order asymptotic expansion near the instability at $\mu = 0$ in the SHE and the inclusion of higher-order terms will (generically) break this phase invariance. Numerically, one observes only two localized patterns emerging from the Turing instability. In fact, due to the exponentially small corrections beyond the regular asymptotic analysis~\cite{Kozyreff2006}, at least two localized states bifurcate off the trivial state. In Figure~\ref{fig:snaking_1d}, we show a typical bifurcation diagram of time-independent states of the Swift-Hohenberg equation. 
\end{rmk}

The search for time-independent solutions to the SHE can be alternatively phrased in the language of ordinary differential equations (ODEs). Indeed, a stationary solution to the SHE satisfies $u_t = 0$, and so we are left with the fourth-order ODE in the spatial variable $x$, given by
\begin{equation}\label{e:SpatialDynamicsSHE}
    0 = -(1 + \partial_x^2)U - \mu U + \nu U^2 - U^3.
\end{equation}
Studying the SHE from this angle constitutes the {\em spatial dynamics} perspective since one can formulate the problem as a dynamical system in the single spatial variable $x$.

\begin{figure}
    \centering
    \includegraphics[width=0.7\linewidth]{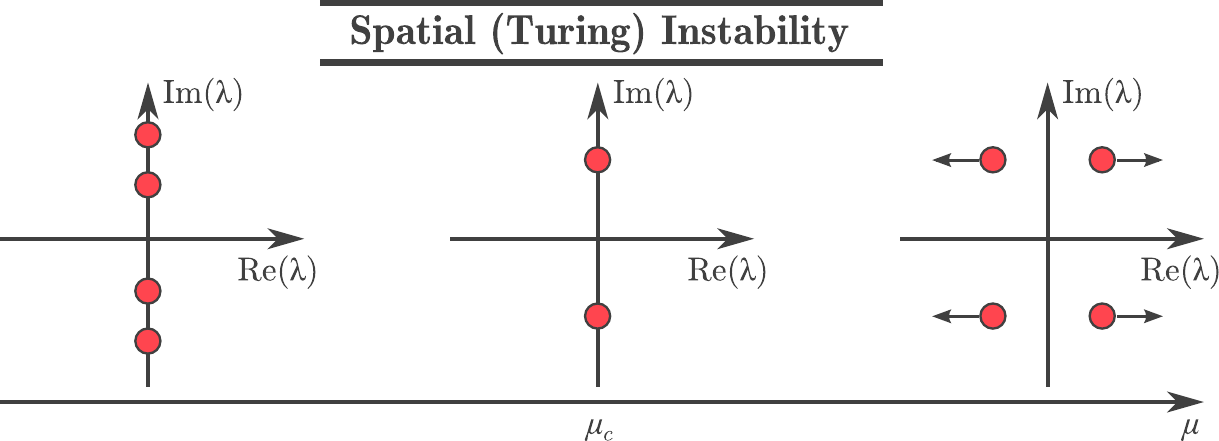}
    \caption{Spatial eigenvalues of the quiescent state for the pattern-forming/Turing instability, with critical point $\mu_c=0$.}
    \label{fig:Turing_Eig}
\end{figure}

Introducing the variables 
\begin{equation}\label{SHE_ODE_vars}
    u_1 = U, \quad u_2 = \partial_x U, \quad u_3 = (1 + \partial_{x}^2)U, \quad u_4 = \partial_x (1 + \partial_{x}^2)U,  
\end{equation}
recasts \eqref{e:SpatialDynamicsSHE}
 as a first-order system of ODEs
\begin{equation}\label{SHE_spatial_1D}
    \begin{split}
        u_1' &= u_2, \\
        u_2' &= u_3 - u_1, \\
        u_3' &= u_4, \\
        u_4' &= -u_3 - \mu u_1 + \nu u_1^2 - u_1^3,
    \end{split}
\end{equation}
where we use $'$ to denote differentiation with respect to the time-like spatial variable $x \in \mathbb{R}$.
The linearization about the trivial state $u_1 = u_2 = u_3 = u_4 = 0$ results in a matrix whose eigenvalues are given by $\sqrt{-1 \pm \sqrt{-\mu}}$ and $-\sqrt{-1 \pm \sqrt{-\mu}}$. If $\mu<0$ the eigenvalues are purely imaginary, while for $\mu > 0$ they split off the imaginary axis to have nonzero real parts; see Figure~\ref{fig:Turing_Eig}. Thus, one expects solutions of the form $e^{ix}$ to bifurcate off the trivial state. This bifurcation takes place at the Turing bifurcation point $\mu = 0$ and is known as a Hamiltonian--Hopf bifurcation \cite{gaivao2011splitting}, or sometimes a reversible 1-1 bifurcation. Such a bifurcation follows from the Hamiltonian structure of the time-independent system \eqref{e:SpatialDynamicsSHE} but can also occur in more general spatially reversible systems like the RD system \eqref{e:RDsys}; these properties of a system being reversible or Hamiltonian will be discussed in more detail in the following subsection.   

At the bifurcation point $\mu=0$ the center-eigenspace is 4 dimensional, which is the same dimension as the original ODE system and one can carry out a normal form analysis to rigorously arrive at the Hamiltonian--Hopf normal form
\begin{subequations}\label{e:HH_normal_form}
    \begin{align}
        A_x =& ik_* A + B + iAP_1 + \mathcal{R}_1,\\
        B_x =& ik_*B + \gamma_1(\mu)A + iBP_1 + AP_2 + \mathcal{R}_2
    \end{align} 
\end{subequations}
where 
\begin{equation*}
    \begin{split}
        A =& \frac{1}{2}u_1 - \frac{\mathrm{i}}{2}u_{2} - \frac{\mathrm{i}}{4}u_{4} + \mathcal{O}((|\mu| + |\mathbf{u}|)|\mathbf{u}|),\qquad\qquad
        B = - \frac{1}{4}u_{4} - \frac{\mathrm{i}}{4}u_{3} + \mathcal{O}((|\mu| + |\mathbf{u}|)|\mathbf{u}|).\\
    \end{split}
\end{equation*}
By carrying out a series of rescalings and coordinate transformations, the system above can be written in the form 
\[
    0 = aA_{xx} - A - b|A|^2A,\qquad a,b\in\mathbb{R},
\]
which brings us back to the (time-independent) Ginzburg--Landau equation \eqref{e:GZeqn}, encountered previously. With the spatial dynamics perspective, localized solutions correspond to homoclinic orbits asymptotically connecting to the trivial state $u = 0$ in \eqref{e:SpatialDynamicsSHE}. The persistence of these homoclinic orbits was proved by Iooss and P\'erou\`eme~\cite{Iooss1993} utilizing the spatial reversibility, $x\rightarrow-x$, of the system. These results provide the existence of small amplitude states that can be continued numerically to follow the localized states to larger values of $\mu$. The typical result is a {\em snaking bifurcation diagram}, as shown in Figure~\ref{fig:snaking_1d} for the SHE with $\nu = 1.6$.  

In the spatial dynamics perspective the domain covering periodic patterns correspond to limit cycle solutions to \eqref{e:SpatialDynamicsSHE}. Such solutions bifurcate subcritically from the trivial state $u = 0$ and are initially temporally unstable until they undergo a fold bifurcation, meeting with a branch of large-amplitude and temporally stable periodic solutions. For $\nu = 1.6$ this fold bifurcation point is $\mu \approx 0.25$, and from $\mu = 0$ to the fold bifurcation point there is a region of bistability between the trivial state $U = 0$ and the large-amplitude periodic solutions. As the two small amplitude localized patterns grow with increasing values of $\mu>0$, the states grow until they start to develop an interior periodic pattern embedded in the trivial state. The localized states undergo a succession of infinitely-many folds, bouncing back and forth within the bistable region, causing the periodic core to grow in spatial extent. This growth sequence of the periodic core of the localized pattern is known as {\it homoclinic snaking}, which is the focus of the following subsection and is depicted in Figure~\ref{fig:snaking_1d}.

\subsection{Homoclinic snaking}\label{subsec:HomoclinicSnaking}   

The term {\em homoclinic snaking} has arisen to describe the intricate bifurcation structure of localized steady-states in one spatial dimension. This terminology is guided by the spatial dynamics formulation, wherein localized solutions are equivalently seen as homoclinic orbits to an associated ODE \eqref{e:SpatialDynamicsSHE}. In this subsection we will build upon the spatial dynamics approach introduced in the previous subsection, while also moving away from the bifurcation regime of $\mu = 0$.

\subsubsection{Weakly Nonlinear Analysis}

\begin{figure}
    \centering
    \includegraphics[width=0.75\linewidth]{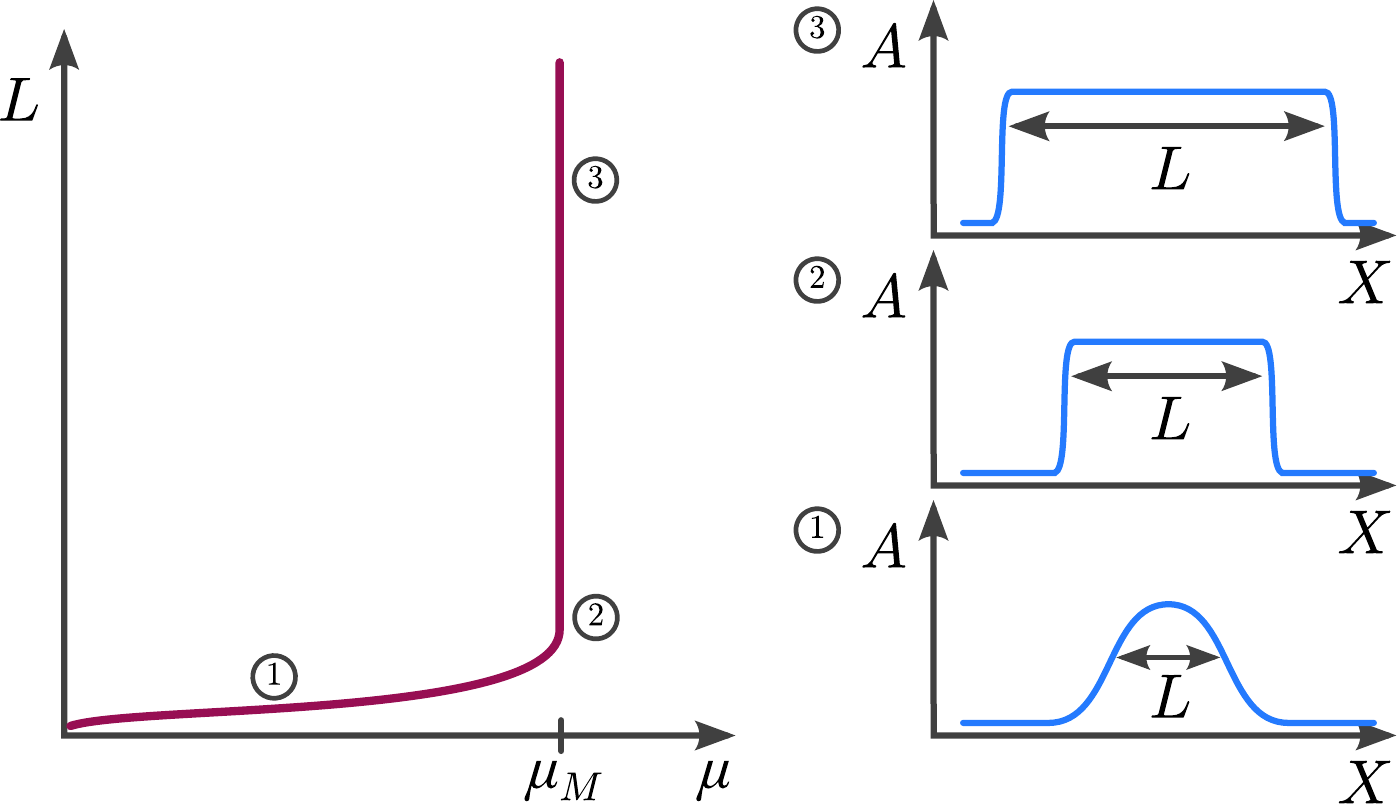}
    \caption{The cubic-quintic Ginzburg--Landau equation \eqref{e:cubic_quintic_amp_eqn} does not predict snaking bifurcation curves of localized pulse solutions.}
    \label{fig:non_snaking_GL}
\end{figure}

The weakly nonlinear analysis described in the previous subsection is unable to explain homoclinic snaking. However, it can be extended to capture the snaking behavior and its parameter dependence near the co-dimension 2 point $(\mu,\nu) = (0,\sqrt{27/38})$ where the Turing bifurcation in the SHE transitions from sub- to supercritical. The result is that the coefficient $b$ which governs the criticality of the bifurcation in \eqref{e:GZeqn} is small. This necessitates moving to the next order in the amplitude equation, resulting in the cubic-quintic Ginzburg--Landau equation~\cite{Budd2001,Budd2005,woods1999heteroclinic}
\begin{equation}\label{e:cubic_quintic_amp_eqn}
    A_T = aA_{XX} - \hat\mu A - b|A|^2A - c|A|^4A.
\end{equation}
Equation \eqref{e:cubic_quintic_amp_eqn} possesses stationary small amplitude localized solutions that bifurcate off the trivial state and grow in width as one approaches the point $\mu=\mu_M$, as illustrated in Figure~\ref{fig:non_snaking_GL}. In particular, as one approaches $\mu = \mu_M$ the bifurcating homoclinic connections grow in width, which appear to capture the growth of the periodic core of localized solutions to the SHE seen in Figure~\ref{fig:snaking_1d}. However, the monotonic approach to $\mu=\mu_M$ does not capture the snaking behavior. The point $\mu_M$ is referred to as the {\em Maxwell point}~\cite{Budd2001,Budd2005}, whose importance will be further emphasized later in this subsection.

Going to higher-order in the Ginzburg--Landau amplitude equation expansion is still unable to capture the snaking behavior. An alternative method is to use \emph{exponential asymptotics}~\cite{Kozyreff2006,Chapman2009,Dean2011,deWitt2019}. This approach focuses on the region near the co-dimension 2 point where the Turing bifurcation turns from being sub- to supercritical, at which point the width of the snaking is exponentially small and the snaking patterns are small in amplitude. In this case one attempts to explain the snaking bifurcations by expanding around the Maxwell point of the front solutions found in \eqref{e:cubic_quintic_amp_eqn}, and so the resulting analysis is highly dependent on the choice of model. Kozyreff and Chapman~\cite{Kozyreff2006,Chapman2009} carried out exponential asymptotics for the quadratic-cubic Swift-Hohenberg equation of the form
\begin{equation}
    (1+\partial_x^2)^2U + \varepsilon^4 U - 3\varepsilon\alpha U^2 + \varepsilon^2 U^3 = 0,
\end{equation}
where $0<\varepsilon\ll1$ is a small scaling parameter and $\alpha\in\mathbb{R}$. Setting $X=\varepsilon^2x$ they introduce the expansions
\begin{align}\label{e:expand_u}
    U(x) =& \sum_{n=0}^N\varepsilon^nU_n(x,X) + \delta U(x,X),\\\label{e:expand_alpha}
    \alpha =& \alpha_0 + \varepsilon^2\alpha_2 + \varepsilon^4\alpha_4 + \cdots + \delta\alpha.
\end{align}
As is typical with asymptotic series, the power series part of \eqref{e:expand_u} is divergent and hence needs to be truncated at finite $N \geq 1$ terms, so $\delta U$ captures the remainder. The power series expansion \eqref{e:expand_alpha} of $\alpha$ approximates the location of the Maxwell point, while $\delta\alpha$ describes the deviation from it. Hence, the snaking behavior of the bifurcation diagram is captured by varying $\delta\alpha$. 

Kozyreff and Chapman's work demonstrates that there is a bifurcation equation given by
\begin{equation}
    \delta\alpha = \pm \frac{c_1e^{-\pi/\varepsilon^2}}{\varepsilon^4} \cos\left(\frac{2-\varepsilon^2\beta}{4}d + \beta\ln{\varepsilon} + c_2 \right) + 2\varepsilon^2\alpha_2e^{-\varepsilon^2d},
\end{equation}
where $\beta=1/\sqrt{734}$, $c_1$ and $c_2$ are parameters that need to be  numerically computed, and $d$ is the length of the localization plateau. As the front separation distance is increased, this leads to $\delta\alpha$ varying, leading to the snaking bifurcation diagram. The $\pm$ describes two intertwining snaking curves, as in Figure~\ref{fig:snaking_1d}. 

\subsubsection{Snakes, Ladders, and Isolas of Localized Patterns}\label{sec:snakes_ladders}

The weakly nonlinear analysis in the previous section is highly model dependent and only covers small amplitude patterns. In this section we look at the general geometric arguments for homoclinic snaking that applies to large amplitude localized patterns. 
We begin by recalling the recast first-order ODE system \eqref{SHE_spatial_1D}, coming from the original fourth-order time-independent SHE ODE \eqref{e:SpatialDynamicsSHE}. Recall that \eqref{SHE_spatial_1D} relates the existence of steady-state solutions to the SHE as orbits of \eqref{SHE_spatial_1D} that remain bounded for all $x \in \mathbb{R}$, while localization requiring $U(x) \to 0$ as $x \to \pm\infty$ in the fourth-order ODE \eqref{e:SpatialDynamicsSHE} manifests itself as the condition $(u_1(x),u_2(x),u_3(x),u_4(x)) \to (0,0,0,0)$ as $x \to \pm\infty$ in \eqref{SHE_spatial_1D}. Thus, we see how the terminology homoclinic snaking arises, as we have reduced ourselves to searching for homoclinic orbits to the spatial ODE \eqref{SHE_spatial_1D}. Moreover, the SPP portion of localized states to the SHE can be interpreted as the desired homoclinic orbit undergoing an intermediary winding around a periodic orbit before mounting a return back to a neighborhood of the trivial equilibrium state. Such periodic orbits are exactly the SPPs discussed earlier in this section since spatial periodicity in \eqref{e:SpatialDynamicsSHE} is equivalent to a periodic solution of \eqref{SHE_spatial_1D}.

Due to the pattern-forming Turing bifurcation at $\mu = 0$, for certain fixed $(\mu,\nu)$ the SHE admits a one-parameter family of spatially-periodic, steady-state SPPs that can be parameterized by their period. Intuitively, localized SPPs as illustrated in Figure~\ref{fig:snaking_1d} can be obtained by truncating one of these periodic profiles over a finite patch $x \in [-L,L]$ and then gluing on the homogeneous rest state $U = 0$ for $|x| > L$. Identifying exactly which SPP should be used for this gluing procedure can be done by noticing that \eqref{e:SpatialDynamicsSHE} admits the conserved quantity
\begin{equation}\label{SHE_Hamiltonian_Space}
    \mathcal{H}(U,\mu,\nu) := U_xU_{xxx} - \frac{U_{xx}^2}{2} + U_x^2 + \frac{U^2}{2} + \frac{\mu U^2}{2} - \frac{\nu U^3}{3} + \frac{U^4}{4}.
\end{equation}
The above quantity is conserved pointwise in $x$ for solutions $U(x)$ to \eqref{e:SpatialDynamicsSHE} and so any SPP that is matched with $U = 0$ to form a localized state must lie in $\mathcal{H}(U,\mu,\nu) = \mathcal{H}(0,\mu,\nu) = 0$. As there is typically only one SPP in this zero-level set, this leads to a selection principle for localized SPPs away from their onset at $\mu = 0$ which was discussed in the previous subsection. 

The conserved quantity \eqref{SHE_Hamiltonian_Space} in $(u_1,u_2,u_3,u_4)$ coordinates takes the form
\begin{equation}\label{SHE_Hamiltonian}
    \mathcal{H}(u_1,u_2,u_3,u_4,\mu) = u_2u_4 + u_1u_3 - \frac{u_3^2}{2} + \frac{\mu u_1^2}{2} - \frac{\nu u_1^3}{3} + \frac{u_1^4}{4}.
\end{equation}
when using the variables \eqref{SHE_ODE_vars}. For a fixed $(\mu,\nu)$ this conserved quantity restricts the flow of \eqref{SHE_spatial_1D} to the (generically) three-dimensional level sets of $\mathcal{H}$. In particular, since we seek orbits that are homoclinic to $(u_1,u_2,u_3,u_4) = (0,0,0,0)$, we restrict to $\mathcal{H}^{-1}(0)$. Moreover, the $\mathbb{Z}_2$-symmetry of taking $x \to -x$ in \eqref{e:SpatialDynamicsSHE} manifests itself as a reversible symmetry in \eqref{SHE_spatial_1D}, with reverser given by
\begin{equation}
    \mathcal{R}(u_1,u_2,u_3,u_4) = (u_1,-u_2,u_3,-u_4).
\end{equation}
This means that if $\mathbf{u}(x) = (u_1(x), u_2(x), u_3(x), u_4(x))^T$ is a solution of \eqref{SHE_spatial_1D}, then so is $\mathcal{R}\mathbf{u}(-x)$, while a solution is said to be symmetric if $\mathbf{u}(x) = \mathcal{R}\mathbf{u}(-x)$ for all $x \in \mathbb{R}$. Symmetric solutions are equivalently defined by having 
\[
    \mathbf{u}(0) \in \mathrm{Fix}(\mathcal{R}) := \{(a,0,b,0):\ a,b\in\mathbb{R}\},
\]
the set of points in $\mathbb{R}^4$ left fixed after applying $\mathcal{R}$ and correspond to even solutions of \eqref{e:SpatialDynamicsSHE}, i.e. $\mathbf{u}(x) = \mathbf{u}(-x)$. It is exactly the symmetric localized solutions that lie along the snaking branches in Figure~\ref{fig:snaking_1d}, while the asymmetric states, i.e. non-symmetric, are the ladder states.  

\begin{figure}
    \centering
    \includegraphics[width=\linewidth]{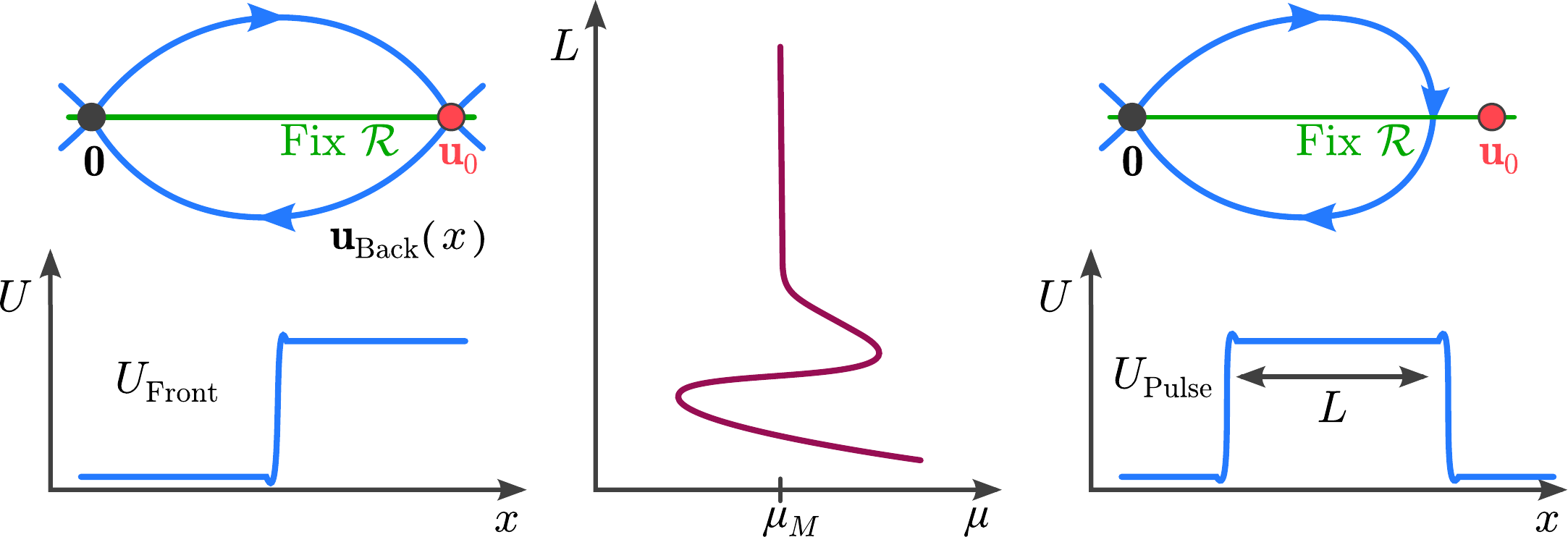}
    \caption{Localized front solutions to the SHE \eqref{e:SpatialDynamicsSHE} are interpreted as heteroclinic and homoclinic orbits in the ODE \eqref{SHE_spatial_1D}. Left: A front solution (bottom) connects the background state $U=0$ and a non-trivial uniform state $U=\tilde{U}$, interpreted as a heteroclinic orbit in four-dimensional phase-space (top). Middle: A bifurcation diagram for localized pulses, with the width $L$ of the bifurcating pulses plotted against the bifurcation parameter $\mu$. Right: A localized pulse solution (bottom) of width $L$ corresponds to a bifurcating homoclinic orbit in four-dimensional phase-space (top).}
    \label{fig:non_snaking}
\end{figure} 

Before approaching the homoclinic orbits that result in snaking, we begin by describing a simpler case in which snaking does not occur. We consider a front solution $U_{\mathrm{Front}}(x)$ for \eqref{e:SpatialDynamicsSHE} that connects the trivial state $U=0$ to a non-trivial equilibrium $\tilde{U}$. Such a solution corresponds to a heteroclinic orbit $\mathbf{u}_{\mathrm{Front}}(x)$ in the four-dimensional system \eqref{SHE_spatial_1D}, connecting the trivial state $\mathbf{u}=\mathbf{0}$ and a non-trivial equilibrium \linebreak $\mathbf{u}=\mathbf{u}_0:=(\tilde{U},0,\tilde{U},0)$. Due to the reversible symmetry from \eqref{SHE_spatial_1D}, there also exists a corresponding heteroclinic orbit connecting $\mathbf{u}_0$ to the trivial state, thus forming a heteroclinic cycle. It can be shown using techniques from \cite{Knobloch2005Homoclinic} that a family of symmetric homoclinic orbits $\mathbf{u}_{\mathrm{Pulse}}(x)$ to the trivial state $\mathbf{u}=\mathbf{0}$ bifurcate from the above heteroclinic orbit. These orbits correspond to localized pulse solutions $U_{\mathrm{Pulse}}(x)$ to \eqref{e:SpatialDynamicsSHE}, characterized by their width $L$, which exist for a parameter value $\mu=\mu(L)$, where $\mu(L)\to\mu_{F}$ as $L\to\infty$. The existence and bifurcation structure of these front and pulse solutions are summarized in Figure~\ref{fig:non_snaking}. The crucial difference here is that the bifurcation curve oscillates around the Maxwell point before converging to the Maxwell point (unlike in Figure~\ref{fig:non_snaking_GL}) since the linearization about the equilibria has spatially complex eigenvalues.

\begin{figure}
    \centering
    \includegraphics[width=0.9\linewidth]{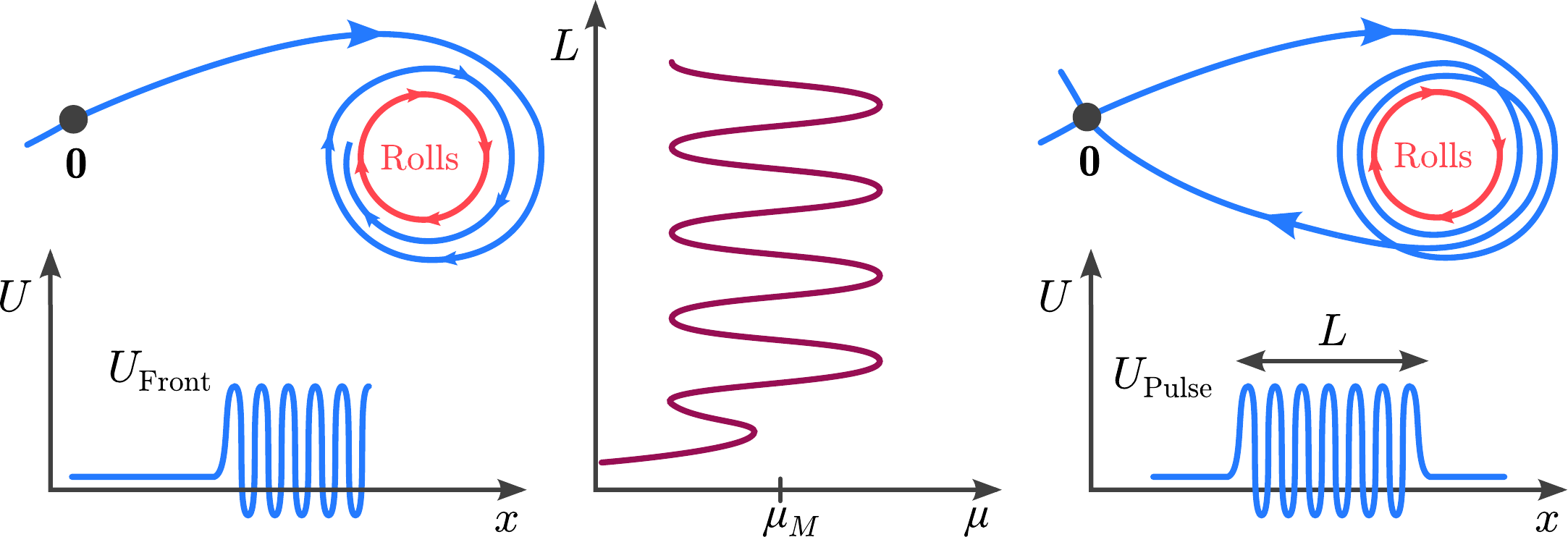}
    \caption{The work of Beck et al. \cite{beck2009snakes} uses spatial dynamics to frame localized spatially periodic patterns (SPPs) as homoclinic orbits in the ODE \eqref{SHE_spatial_1D}. Left: A front pattern (bottom) is used to transit between the background state $U = 0$ and the SPP in phase-space (top). Middle: Ascending the snaking bifurcation diagram is equivalent to extending the length of the localization plateau. Right: Localized patterns are homoclinic to $U = 0$ and wrap around the periodic orbit, with more wraps leading to a longer plateau.}
    \label{fig:snakingDiag}
\end{figure}

We now turn our attention back to the case when localized SPPs undergo homoclinic snaking. 
This spatial dynamics approach to obtain the existence of the desired homoclinic orbits to \eqref{SHE_spatial_1D} requires assuming the existence of a heteroclinic connection between a periodic orbit and $\mathbf{u} = 0$ \cite{beck2009snakes,aougab2019isolas,knobloch2011isolas,sandstede2012snakes}. In practice this assumption can be verified numerically. Such heteroclinic connections are pinned front solutions to \eqref{e:SpatialDynamicsSHE} and reversible symmetry implies that a connection from a SPP to $U = 0$ exists if and only if a connection from $U = 0$ to the SPP exists. In the geometry of the 4-dimensional phase-space of \eqref{SHE_spatial_1D}, these heteroclinic orbits provide a method of transiting in phase-space from $\mathbf{u} = 0$ to the periodic orbit and back, as illustrated in Figure~\ref{fig:snakingDiag}. Beck et al. \cite{beck2009snakes} make the hypothesis that the parameter values $\mu$ at which a heteroclinic connection exists in \eqref{SHE_spatial_1D} can be parameterized by the phase along the periodic orbit, $\varphi$, where the connection enters a tubular neighborhood of the periodic orbit through $\mu = z(\varphi)$. The result is that symmetric homoclinic orbits that spend time $2L$ in the neighborhood of the periodic orbit exist at a parameter value $\mu$ if and only if
\begin{equation}
    \mu = z(L + \varphi_0) + \mathcal{O}(\mathrm{e}^{-\eta L}),
\end{equation}
for some $\eta > 0$ and $\varphi_0 \in \{0,\pi\}$. Each choice of $\varphi_0$ gives way to one of the distinct snaking branches in Figure~\ref{fig:snaking_1d}. Most importantly, this result gives that the bifurcation structure of symmetric localized states to \eqref{e:SpatialDynamicsSHE} is entirely predicted by the response to parameters of the front/back solutions. Moreover, the condition $\mu = z(\varphi)$ was loosened in \cite{aougab2019isolas} to allow for an implicit relationship between $\mu$ and $\varphi$ and returned identical findings.    

\begin{figure}[t!]
    \centering
    \includegraphics[width=\linewidth]{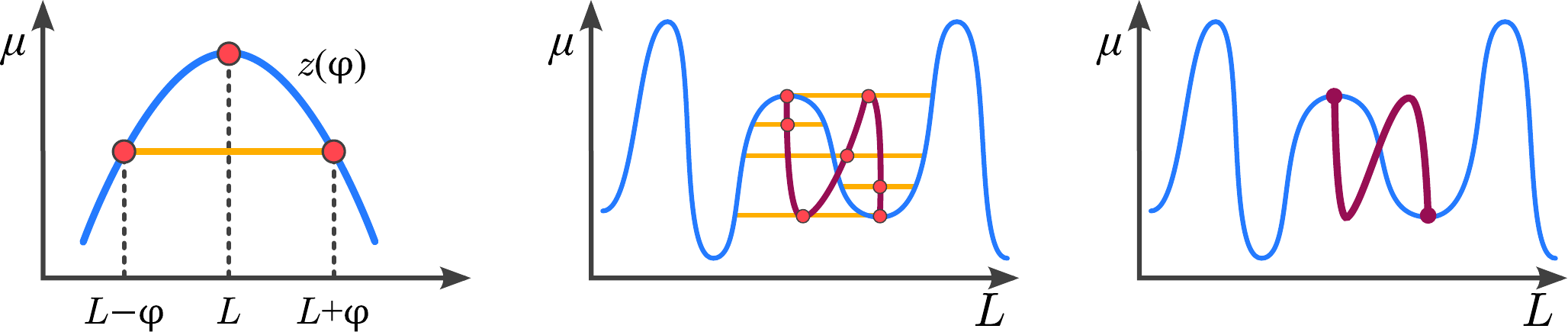}
    \caption{Solutions in $(L,\varphi)$-space satisfying \eqref{e:ladderz} can be found graphically by identifying two intersection points with $z$ along a fixed horizontal line whose abscissas give $L \pm \varphi$. Moving the horizontal line up and down generates the entire bifurcation branch, as depicted in the center panel, with the resulting bifurcation curve on the right. If $\varphi$ reaches $\pi$ the branch terminates on a second snaking curve, as in Figure~\ref{fig:snaking_1d}.}
    \label{fig:AsymPredict}
\end{figure}

The function $\mu = z(\varphi)$ further provides a complete description of the asymmetric ladder states as well. Asymmetric homoclinic orbits of \eqref{SHE_spatial_1D} that spend $2L$ units of $x$ in the neighborhood of the periodic orbit exist for $(L,\mu)$ if and only if $(L,\mu) = (L,z(L + \varphi))$ with some $\varphi \in S^1$ satisfying 
\begin{equation}\label{e:ladderz}
    z(L + \varphi) = z(L - \varphi)
\end{equation}
Satisfying the condition on $\varphi$ can be done graphically, as demonstrated in Figure~\ref{fig:AsymPredict}, allowing one to predict the shape of ladder states without explicitly computing them. 

\begin{figure}[t]
    \centering
    \includegraphics[width=0.75\linewidth]{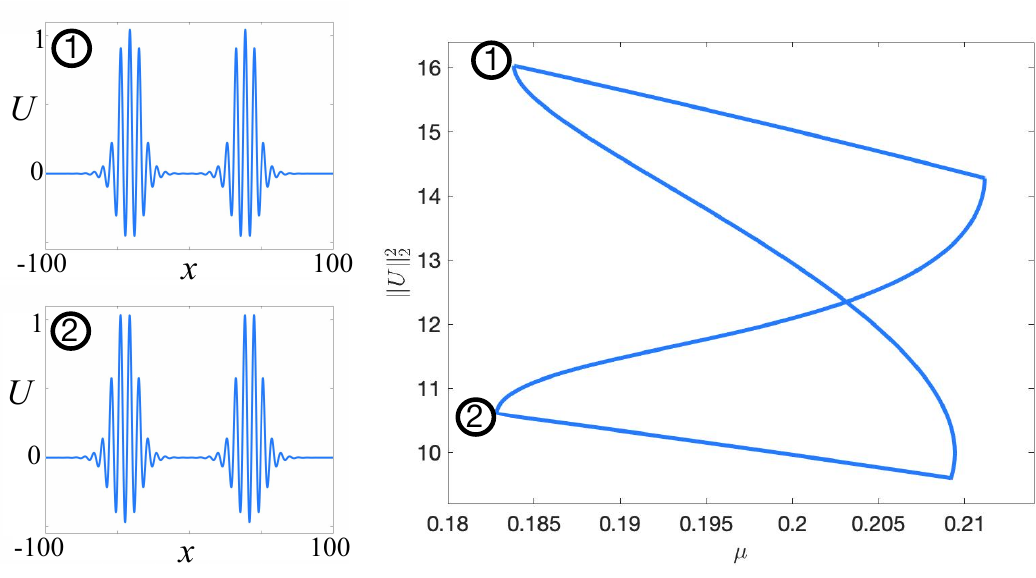}
    \caption{An isola (closed bifurcation curve) of multipulse solutions to the SHE with $\nu = 1.6$ with two representative solutions from the leftmost saddle-node bifurcation points along the isola.}
    \label{fig:multipulse}
\end{figure}

General results for Hamiltonian ODEs indicate that the presence of homoclinic orbits is accompanied by infinitely many more $N$-homoclinic orbits for every $N \geq 2$ \cite{harterich1998cascades,devaney1976homoclinic}. These $N$-homoclinic orbits manifest themselves as localized solutions with $N$ distinct regions of localization, each separated by the profile approximately at the trivial background state $U = 0$. These {\em multipulse} solutions are well-documented in the literature and numerical investigations have indicated that they all lie along isolas \cite{knobloch2011isolas,van2002spatially,wadee2002solitary,burke2009multipulse}, regardless of the bifurcation structure of the front/back solutions used to prove their existence. Typical multipulses with $n = 2$ to the SHE are presented in Figure~\ref{fig:multipulse} along with the isola they lie along. Knobloch et al. proved this for $N = 2$ \cite{knobloch2011isolas}, while Bramburger has proved this for all $N \geq 2$ for spatially discrete lattice equations \cite{bramburger2021isolas}. However, a Hamiltonian structure in the spatial dynamical system is not necessary to observe snaking bifurcations and isolas of multipulses, as has been documented, for example, in the Lugiato--Lefevre equation \cite{parra2016dark}.

Extensions of the work in \cite{beck2009snakes} abound. For example, \cite{aougab2019isolas} shows that if the Floquet exponents of the periodic orbits are negative then all localized states must lie along isolas, regardless of the front/back bifurcation structure. Burke and Dawes examined localized states in a non-variational SHE model and compared them to the SHE studied herein \cite{burke2012localized}.
Various symmetry-breaking results have been documented as well \cite{knobloch2012non,makrides2014predicting,sandstede2012snakes}, while PDE stability of localized solutions was examined numerically in \cite{burke2006localized} and analytically in \cite{makrides2019existence}. Knobloch, Uecker, and Wetzel further examined a variant of the SHE with a degree 7 polynomial nonlinearity to find the existence of localized solutions that replace the background state $U = 0$ with another smaller amplitude SPP \cite{Knobloch_2019}. Using the above spatial dynamical framework one can interpret these defect patterns as trajectories that are homoclinic to a periodic orbit but spend an intermediary portion of the orbit wrapping around a different one. All of the above spatial dynamics analysis is therefore applicable to understand these defect patterns as well. 

\subsubsection{Lattice Dynamical Systems}\label{subsubsec:Lattice1D}

\begin{figure}[t]
    \centering
    \includegraphics[width=\linewidth]{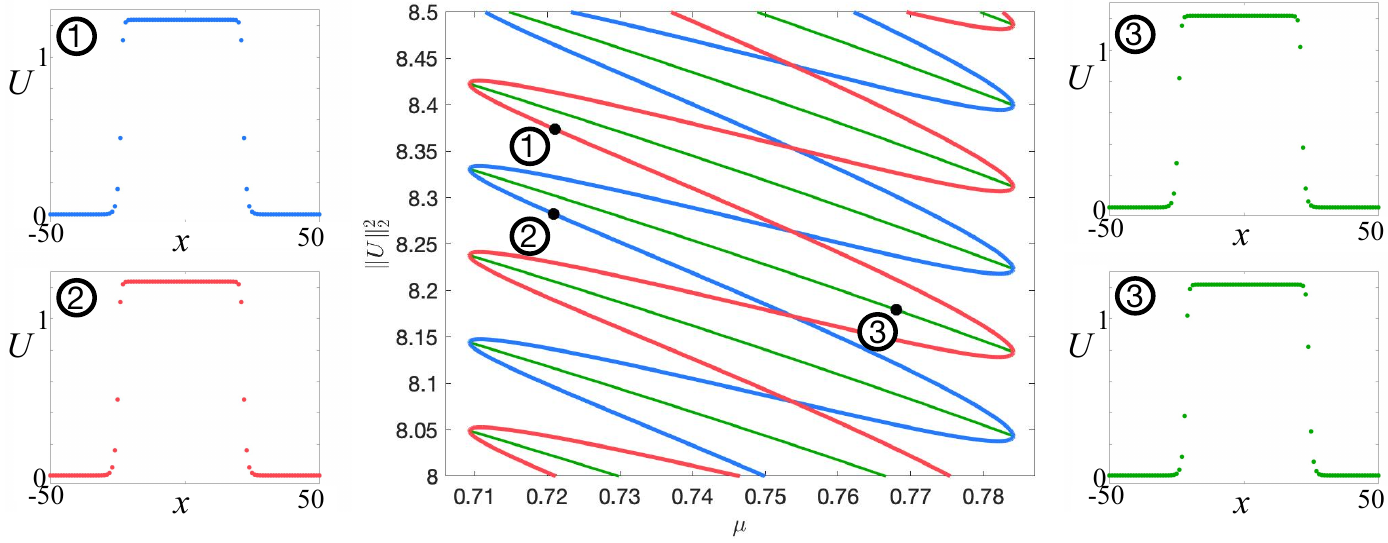}
    \caption{Snaking localized `top-hat' steady-states of the lattice system \eqref{Lattice1D} with $d = 0.5$. As in the continuous spatial setting, there are two branches of snaking symmetric states (red and blue) with asymmetric ladder states that connect them near the folds (green).}
    \label{fig:1DLattice}
\end{figure}

Much of the aforementioned analysis on localized pattern formation summarized in this section requires one to hypothesize the existence of the necessary heteroclinic connections. Although this can be established numerically, there are cases where it can be proven explicitly. A particular example is that of lattice dynamical systems, where localized patterns have been known to exist for some time \cite{chong2009multistable,kusdiantara2017homoclinic,papangelo2017snaking,taylor2010snaking,woods1999heteroclinic,AvitabileBump2023,avitable2017macroscopic, whiteley2022modelling,schmidt2020bumps}. To illustrate, consider the lattice dynamical system
\begin{equation}\label{Lattice1D}
    \dot U_n = d(U_{n+1} + U_{n-1} - 2U_n) - \mu U_n + 2U_n^3 - U_n^5, \qquad n \in \mathbb{Z},
\end{equation}
where $U_n \in \mathbb{R}$ is the state at each lattice point $n \in \mathbb{Z}$, $d > 0$ is the strength of interaction between the nearest-neighbors on the lattice, and $\mu$ is again our bifurcation parameter. Figure~\ref{fig:1DLattice} demonstrates how spatially localized `top-hat' steady-states of \eqref{Lattice1D} arrange themselves in the familiar snakes and ladders bifurcation diagram. 

Taking $\dot U_n = 0$ for all $n$ and setting $(u_n,v_n) = (U_{n-1},U_n)$ results in the spatial dynamical formulation of \eqref{Lattice1D}, given by the map
\begin{equation}\label{Lattice_map}
    \begin{split}
        u_{n + 1} &= v_n, \\
        v_{n+1} &= 2v_n - u_n + \frac{1}{d}(\mu v_n - 2v_n^3 + v_n^5).
    \end{split}
\end{equation}
As in \eqref{SHE_spatial_1D}, localized solutions of \eqref{Lattice1D} come in the form of bounded orbits of \eqref{Lattice_map} homoclinic to the origin. The correspondence between the continuous and discrete settings can be furthered since the map \eqref{Lattice_map} can be used as an example of a Poincar\'e map near the periodic orbits of \eqref{SHE_spatial_1D} that exhibits a homoclinic tangle and leads to the plethora of homoclinic orbits that correspond to localized solutions. Bramburger and Sandstede \cite{bramburger2020spatially} used the (discrete) spatial dynamics formulation \eqref{Lattice_map} to provide a complete explanation of the snaking structure of localized solutions, with \cite{bramburger2021isolas} extending these results to prove that multipulses must always lie along isolas. Most importantly though, these studies employed perturbation theory and Lyapunov--Schmidt reductions to completely identify the bifurcation of front/back solutions of \eqref{Lattice1D} when $0 < d \ll 1$. These results therefore provide a method of explicitly checking all hypotheses required to obtain the results of \cite{bramburger2020spatially,bramburger2021isolas}, thus bridging the gap from the theory to application in specific systems.

\subsection{Variations on the Standard Set-Up}\label{subsec:Variations}

While the SHE has provided an excellent test bed to explore and develop the mathematical theory of localized patterns, there have been several extensions to more general classes of equations and other types of 1D spatially localized patterns. In this section, we briefly review extensions of the theory to general RD systems, nonlocal systems, systems with large-scale modes, time-periodic spatially localized patterns, connections to the singular perturbation limit analysis and invasion fronts outside the snaking region. 

\subsubsection{General Reaction-Diffusion Systems}

Localized patterns have been observed in a wide variety of RD models~\cite{dawes2016localised,Koga1980,vanag2007,Ward_2018,Gandhi2018,saadi2022,parra-rivas2021,Saadi2023,hill2021localised} where homoclinic snaking is found to be ubiquitous. The weakly nonlinear analysis of \S\ref{subsec:SH1D} can be extended for general RD systems \eqref{e:GeneralRDsystem} by letting
\[
    \mathbf{u} = \varepsilon A(\varepsilon x)\mathbf{\hat U}e^{ikx} + c.c.,\qquad A\in\mathbb{C}
\]
where $\mathbf{\hat U}$ is a left eigenvector of $\mathbf{D}^{-1}\mathbf{f}_{\mathbf{u}}(\mathbf{u}_{0})$. Carrying out the asymptotic analysis, one finds at $\mathcal{O}(\varepsilon^3)$ the amplitude equation to be the Ginzburg--Landau equation \eqref{e:GZeqn}. Similarly, a center-manifold reduction~\cite{Vanderbauwhede1989} can be carried out to yield the same normal form as the Hamiltonian--Hopf normal form \eqref{e:HH_normal_form}. All subsequent analysis of the amplitude equation or normal form proceed as before. 

The spatial dynamical analysis of Beck et al.~\cite[\S6.2]{beck2009snakes} and Knobloch and Wagenknecht~\cite{Knobloch2005Homoclinic} can be carried over to higher dimensional ODEs with and without Hamiltonian structure. One major exception is that the loss of a Hamiltonian or conservative structure in the resulting spatial dynamical system leads the asymmetric ladder states to become dynamic. That is, they move with nonzero speed either to the left or right in the spatial domain while maintaining their localized profile~\cite{Houghton2011}. 

\subsubsection{Nonlocal Systems}\label{subsec:Nonlocal}

Beyond PDEs, localized patterns have been extensively investigated in nonlocal models. The most studied nonlocal model exhibiting localized structures is the Wilson--Cowan--Amari model~\cite{Amari1977,Wilson1973}, given by
\begin{equation}\label{e:WCA}
    U_t = -U + W\ast S,
\end{equation}
where $U = U(\mathbf{x},t)$ describes the evolution of the average membrane voltage potential of a neuronal population at a position $\mathbf{x}$ on the cortex, $W = W(\mathbf{x},\mathbf{x}')$ is the connectivity function of the neuronal population, the symbol $\ast$ represents a convolution integral, and $S = S(U)$ describes the neuronal firing rate. When $W$ has a rational Fourier transform, one can transform \eqref{e:WCA} via Fourier transform to a PDE~\cite{Laing2003} and apply all the previous analyses. Numerical continuation routines have shown that the bifurcation diagrams are just as rich as the PDE set up~\cite{Laing2003,Coombes2003,Faye2013,rankin2014continuation}. In the special case where $S$ is a Heaviside function, Avitabile \& Schmidt~\cite{avitabile2015snakes} were able to explicitly construct bifurcations equations for the snaking diagram.

\subsubsection{Slanted Snaking and Finite Domains}\label{ss:slant}

\begin{figure}[ht!]
    \centering
    \includegraphics[width=0.9\linewidth]{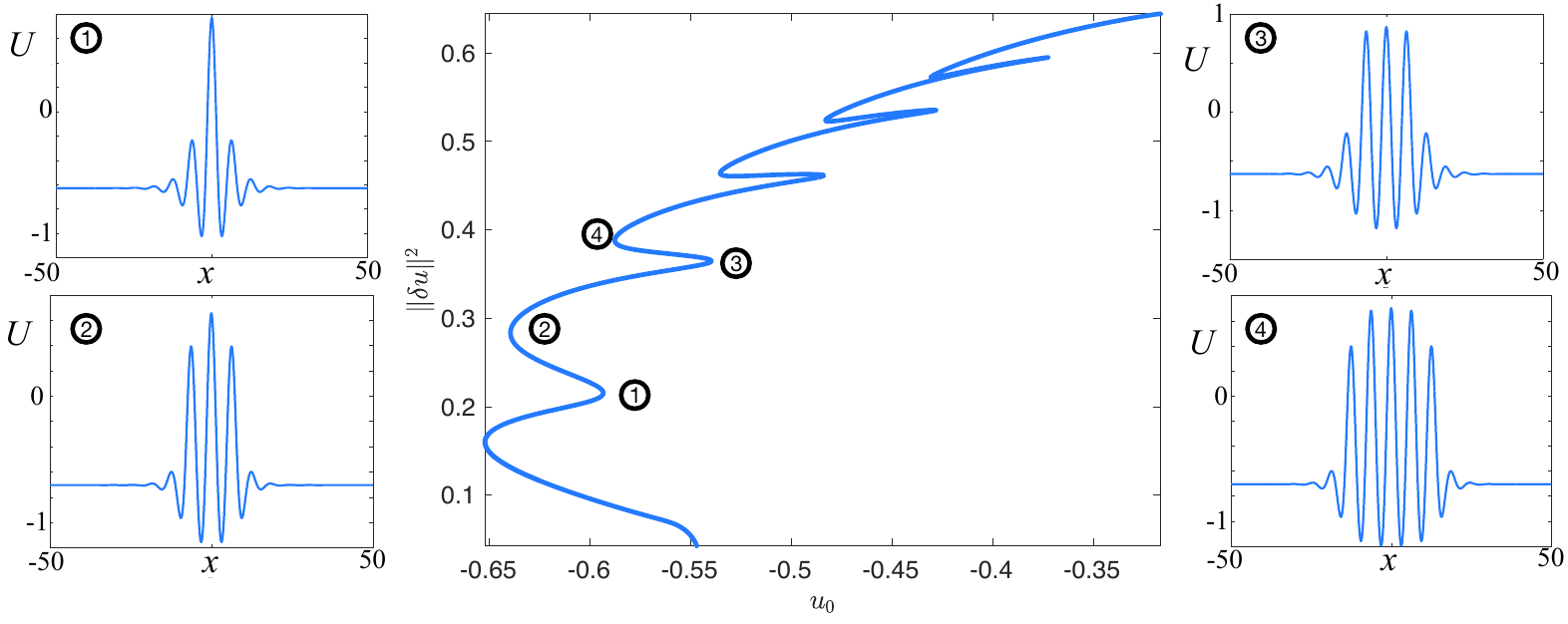}
    \caption{Slanted snaking of stationary localized solutions for the conserved SHE equation \eqref{e:conSHE} with $\mu=-0.9$ and domain length $L=100$. The solution measure is $\|\delta u\|_2 = \|u - u_0\|_2$ where $u_0=\frac{1}{L}\int_0^Lu\mathrm{d}x$.}
    \label{fig:slanted_snake}
\end{figure}

The standard bifurcation diagram where the snaking occurs between two fixed limits in parameter space can become slanted when the model has an additional conserved quantity and the computation is done over a finite spatial domain~\cite{Dawes2008,Morgan2014,dawes2010localized,Knobloch2016_con,Knobloch2017,beaume_bergeon_kao_knobloch_2013,Bortolozzo2008,Thiele2013}. The simplest extension to the SHE that has this feature is the phase-field crystal model (or conserved SHE)
\begin{equation}\label{e:conSHE}
    U_t = -\Delta\left[-(q^2+\Delta)^2U - \mu U + \nu U^2 - U^3 \right],
\end{equation}
which conserves the total mass $\int_0^LU \mathrm{d}x$ on the bounded domain $x\in[0,L]$. At $\mu=0$, the dispersion relation $d(\lambda,k) = -k^2(q^2 - k^2) - \lambda$ has a neutral mode at $k=0$ and at $k=|q|$, with the former due to the additional conserved quantity. In Figure~\ref{fig:slanted_snake}, we show a typical bifurcation diagram that is observed wherein the snaking `slants' towards the right while undergoing successive folds. The requirement for the domain to be finite is to guarantee that the conserved quantity is always finite for bounded solutions of the conserved SHE, while on an infinite domain one would have regular snaking. 

Various weakly nonlinear analyses, such as \cite{Matthews_2000}, have been carried out to understand \eqref{e:conSHE}. One can expand $U$ as 
\[
    U(x,t) = \varepsilon A(X,T)e^{iqx} + \varepsilon^2B(X,T) + \varepsilon^2 C(X,T)e^{2iqx}+c.c. + \mathcal{O}(\varepsilon^3),
\]
where $X=\varepsilon x,T=\varepsilon^2t,\mu=\varepsilon^2\hat\mu$. At $\mathcal{O}(\varepsilon^2)$ one identifies the presence of $e^{2iqx}$ terms, which allows one to deduce that $C=-\nu A^2/9$. At the next order, $\mathcal{O}(\varepsilon^3)$, one then arrives at the amplitude equations
\begin{subequations}
    \begin{align}
        A_T =& 4A_{XX} - \hat\mu A - (3-2\nu^2/9)|A|^2A - 2\nu AB,\\
        B_T =& B_{XX} + 2\nu(|A|^2)_{XX}.
    \end{align}
\end{subequations}
Unfortunately, these equations do not possess explicitly known localized solutions and still have to be solved numerically; see for instance~\cite{Dawes2008}. Developing a rigorous theory for the existence and description of slanted snaking still remains an open problem. 

We also briefly mention mass conserving systems that also display localized patterns (see for instance~\cite{Brauns2020,Mori2008} where there has been a considerable amount of analysis carried out.

\subsubsection{Oscillons and Breathers}\label{subsec:1DOscillons}

While all the structures we have looked at so far are stationary, time-periodic spatially localized structures have been extensively studied in a wide-range of models. These structures are called either oscillons or breathers, depending on the context they arise in. Oscillons can arise in nonlinear systems from a uniform background state under parametric forcing whose frequency is often close to the frequency of the pattern oscillation in space (the wavenumber of the SPP) \cite{alnahdi2014localized,Alnahdi2018,Ma2016}. Such localized time-periodic patterns have been observed in Newtonian fluids \cite{arbell2000temporally,germain2009global,venkataramani2001pattern,xia2012oscillon}, chemical reactions \cite{petrov1997resonant,vanag2001oscillatory,vanag2004stationary}, granular layers \cite{umbanhowar1996localized}, and colloidal suspension \cite{Lioubashevski1999}.  

In RD systems, the main instabilities of a spatially homogeneous steady-state giving rise to temporally oscillating spatially localized patterns are the subcritical Hopf bifurcation and Turing--Hopf instability; see~\cite{Iooss1991} for an analysis of the Navier--Stokes equations in these limits. Weakly nonlinear analysis and center-manifold reduction techniques have been used to investigate the emerging spatially localized structures. Despite the wealth of experimental evidence, the study of these structures in the RD set up is very sparse. 

In periodically forced systems, such as the Faraday wave problem, simplified models have been proposed to better understand oscillons. An example is the forced complex Ginzburg--Landau equation (CGL)
\begin{equation}\label{e:FCGL}
  U_t = (1+i\alpha)U_{xx} + (-\mu+i\omega)U - (1+i\beta)|U|^2U + \gamma \bar U,\qquad U\in\mathbb{C}  
\end{equation}
where $\alpha,\beta,\gamma,\mu,\omega\in\mathbb{R}$ are system parameters. Similar to the contexts that we have already seen the Ginzburg--Landau equation arise throughout this section, the equation \eqref{e:FCGL} is argued in \cite{elphick1987normal,coullet1992strong} to be a model for a temporally forced spatially homogeneous Hopf bifurcation. 
Here $\omega$ is related to the offset of temporal forcing frequency, 
while $\gamma$ is related to the forcing amplitude. The parameter $\gamma$ can always be assumed positive due to the gauge symmetry $(\gamma,U) \mapsto (\gamma\mathrm{e}^{\mathrm{i}\phi},U\mathrm{e}^{\mathrm{i}\phi/2})$ for any $\phi \in \mathbb{R}$. 

\begin{figure}
    \centering
    \includegraphics[width=\linewidth]{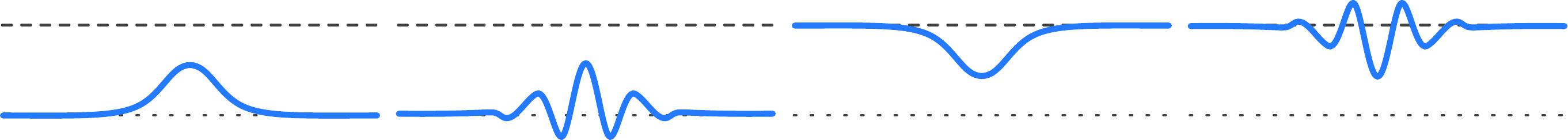}
    \caption{Profiles of standard and reciprocal oscillons with oscillatory and monotonic tails. The lower dotted line represents $u=0$, the upper dashed line $u=u_{\mathrm{unif}}\neq0$ }
    \label{fig:oscillons}
\end{figure}

In the framework of \eqref{e:FCGL}, oscillons correspond to localized solutions that are time-independent. The work \cite{burke2008classification} demonstrated that there are two different types of localized steady-states in \eqref{e:FCGL}, as depicted in Figure~\ref{fig:oscillons}. The first is the usual localized solution that limits to $U =0$ as $x \to \pm \infty$. The second limits to a nontrivial background state $U_\mathrm{unif} \neq 0$ as $x \to \pm \infty$ and were referred to as {\em reciprocal oscillons}, also {\em superoscillons} in earlier work \cite{blair2000patterns}. Both of these types of oscillons can have either decay monotonically to $U = 0$ as $x\to \pm \infty$, leading to spatial profiles resembling Gaussians/solitons, or damped oscillations into $U = 0$, leading to spatial profiles resembling the localized SPPs of the SHE.  

Parallel investigations of oscillons were carried out by Rucklidge and Silber~\cite{rucklidge2009} who postulated a phenomenological PDE to understand Faraday waves. Their model is given by
\begin{equation}\label{e:ruckPDE}
    U_t = (\mu+i\omega)U + (\alpha+i\beta)\nabla^2U + (\gamma+i\delta)\nabla^4 U + a_1U^2 + a_2|U|^2 + a_3|U|^2U + i\Re(U)f(t),
\end{equation}
where $a_i\in\mathbb{C}$ and $f(t+2\pi)=f(t)$ is a periodic function. Model \eqref{e:ruckPDE} was investigated for localized patterns in a series of papers~\cite{alnahdi2014localized,Alnahdi2018}. In the first paper~\cite{alnahdi2014localized}, Alnahdi studied the special case of \eqref{e:ruckPDE} where patterns occur when the critical spatial wavenumber is zero by setting $\gamma=\delta=0$ and setting the quadratic terms to zero i.e. $a_1=a_2=0$. Linearising about $U=0$ and looking at solutions of the form $U(x,t)=e^{\sigma t + ikx}$ yields the dispersion relation for the growth rate, $\sigma$, given by
\[
    \sigma = \mu - \alpha k^2 + i(\omega - \beta k^2).
\]
Setting $\alpha>0$ and $\omega$ close to 1 so that a forcing $f(t)=F\cos(2t)$ drives a subharmonic response with frequency 1, one can find a Hopf bifurcation at $\mu=0$ with the critical wavenumber at zero. Setting
\[
    U(x,t) = \varepsilon A(X,T)e^{it} + c.c.+\mathcal{O}(\varepsilon^2),\qquad X=\varepsilon x,T=\varepsilon^2 t, 
\]
and carrying out an asymptotic analysis in the limit of weak forcing and weak damping (i.e. setting $\omega = 1+\varepsilon^2\nu,F = \varepsilon\tilde F,\mu = \varepsilon^2\tilde\mu$), at $\mathcal{O}(\varepsilon^3)$ the forced CGL \eqref{e:FCGL} is found (where the $U$ is replaced with $A$) and existing results for localized patterns in the forced CGL could be leveraged. Furthermore, it is possible to reduce the forced CGL near onset to 
\[
    B_T = aB_{XX} + bB + cB^3,
\]
where $a,b,c$ are real parameters whose values are determined by the parameters in the CGL. In the limit of strong damping corresponding to setting $f(t) = F_c\cos(2t)(1+\varepsilon^2\lambda)$, where $F_c$ is the critical forcing which must be determined numerically, Alnahdi et al. were able to carry out a direct reduction to the Allen--Cahn equation skipping the forced CGL reduction step altogether. 

The second paper~\cite{Alnahdi2018} investigating \eqref{e:ruckPDE} studies the case when the spatial wavenumber is non-zero at onset by reintroducing the fourth-order spatial derivatives. A formal weakly nonlinear analysis is carried out in the limit of weak damping, weak detuning, weak forcing, small group velocity, and small amplitude taking $\mu = \varepsilon$ as the bifurcation parameter, by setting
\[
    U(x,t) = \varepsilon (A(X,T)e^{i(t+x)} + B(X,T)e^{i(t-x)}) + \mathcal{O}(\varepsilon^2),\qquad X=\varepsilon x, T= \varepsilon^2 t,
\]
where $A$ and $B$ are slowly varying complex amplitudes. Substituting this ansatz into \eqref{e:ruckPDE} with the specified parameter choices and carrying out the weakly nonlinear analysis, one finds at $\mathcal{O}(\varepsilon^3)$ a coupled forced CGL system of the form
\begin{equation}\label{e:coupledFCGL}
    \begin{split}
        A_T =& (\rho + i\nu)A - 2(\alpha+i\beta)A_{XX} + v_gA_X + (\tau+\mathrm{i}\delta)(|A|^2+2|B|^2)A + i\Gamma \overline{B},\\
        B_T =& (\rho+i\nu)B - 2(\alpha+i\beta)B_{XX}-v_gB_X + (\tau+\mathrm{i}\delta)(2|A|^2+|B|^2)B + i\Gamma \overline{A},
    \end{split}
\end{equation}
where all parameters are real and $v_g=\frac{d\sigma}{dk}|_{k=k^*}$ is the scaled group velocity evaluated at the critical spatial wavenumber $k^*$. The system \eqref{e:coupledFCGL} can then be subsequently reduced to a real coupled Ginzburg-Landau equation, where explicit localized solutions could be found. Localized solutions of \eqref{e:coupledFCGL} remain largely unexplored, particularly from the spatial dynamics perspective. 

Lastly, time periodic forcing has also been studied by~\cite{gandhi_knobloch_beaume_2015,gandhi_beaume_knobloch_2016} where they investigated the SHE with $\mu=\mu(t)$ time-periodic and oscillating around the snaking region (and just outside the fold limits). They were able to find breathing localized patterns that start to invade the trivial state and then retreat.

\subsubsection{Transition to the Singular Limit}\label{subsubsec:Singular}

The localized patterns that are the focus of this review primarily emerge from pattern-forming Turing instabilities. However, another class of localized patterns are those that arise in singularly perturbed models, where a high-order spatial derivative is multiplied by a small parameter. These patterns are well-studied and have an extensive literature, including the works~\cite{doelman2003,doelman1997,doelman2015,ward2002}.  The study of such patterns deserves its own review article, and so our focus here will be exclusively on how the localized patterns that emerge from a Turing instability connect with those near the singular limit, as studied in \cite{champneys2021,champneys2021b,champneys2021c,saadi2022,Lloyd2013}. 

\begin{figure}[htb]
    \centering
    \includegraphics[width=\linewidth]{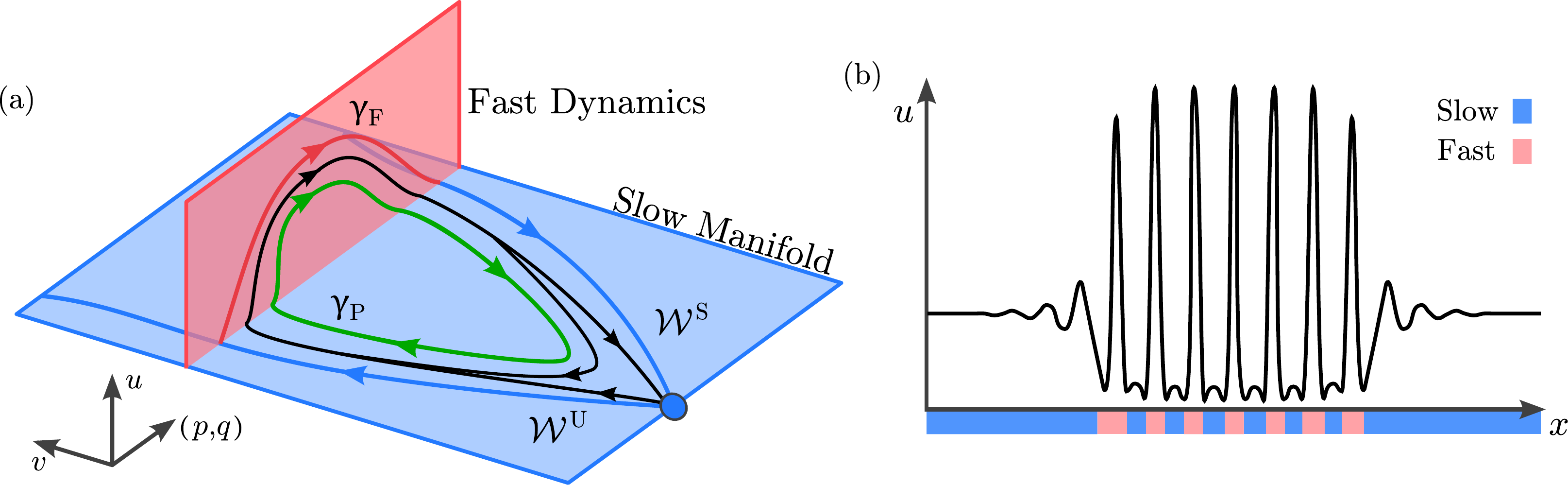}
    \caption{Localized patterns in fast-slow systems. Left: the phase-space dynamics. Here $\gamma_{\mathrm{F}}$ is a homoclinic orbit in the fast regime, $\gamma_P$ is a periodic orbit, and $\mathcal{W}^{S/U}$ are the stable and unstable manifolds of the saddle on the slow manifold. Right: The localized PDE solution.}
    \label{fig:singular_Limit}
\end{figure}
The general class of two-component RD system near a singular limit can be written in the form
\begin{equation}\label{e:singular_sys}
    \begin{split}
        u_t =& \delta^2\Delta u +  a - cu + dv + g(u,v),\\
        v_t =& \Delta v + b + eu - fv + h(u,v),
    \end{split}
\end{equation}
where $a,b,c,d,e,f\in\mathbb{R}$, $0<\delta\ll 1$ is the singular perturbation parameter, and $g,h$ are nonlinear functions of $u,v$. A typical choice for $g$ and $h$ are $g(u,v) = -h(u,v) = u^2v$, which includes the classic Schnakenberg, Brusselator, Glycolysis, Selkov--Schnakenberg, and root-hair models~\cite{avitabile2018spot,champneys2021b}. Other models that have been studied in this context include the cell-polarity model~\cite{champneys2021c} where one has 
\[
    g(u,v) = -h(u,v) = \gamma\frac{u^2v}{1+u^2} - \eta u + v,\qquad a\equiv d\equiv e\equiv f \equiv 0,\qquad \gamma,\eta\in\mathbb{R},
\]
and the urban-crime model of Short et al.~\cite{Lloyd2013}
\begin{equation}\label{e:crime_sys}
    \begin{split}
        u_t =& \delta^2\Delta u - u + a + uv,\\
        v_t =& \nabla\cdot\left( \nabla v - \frac{2v}{u}\nabla u\right) - uv + b-a.
    \end{split}
\end{equation}
Setting $\delta = 0$ in any of the above models removes the highest spatial derivative in $u$, thus making the parameter regime $0 < \delta \ll 1$ a singularly perturbed RD system. In each of these singularly-perturbed RD models one can encounter homoclinic snaking that bears a significant resemblance to that which we encounter in the SHE. As one observes in Figure~\ref{fig:singular_Limit}, the oscillating pattern in the middle of the localized structure has a characteristic sharp spike with a little bump at the bottom, which is the result of the singular perturbation in the small $\delta$ parameter regime. 

The spiking of the localized structures in \eqref{e:singular_sys} can best be explained through the `fast-slow' structure of the resulting spatial dynamical system. In the case of one spatial dimension, one sets the time derivative on the left-hand side of \eqref{e:singular_sys} to zero and rewrites the system as the first-order system of ODEs. In the limit of $\delta$ small, there exists a slow manifold that possesses a saddle equilibrium. Rescaling $x=\delta y$ and taking $\delta=0$, one finds a homoclinic orbit which describes the fast spikes in the oscillating profile of the snaking localized pattern. The geometry of this fast-slow orbit is shown in Figure~\ref{fig:singular_Limit}. 

\begin{figure}
    \centering
    \includegraphics[width=0.7\linewidth]{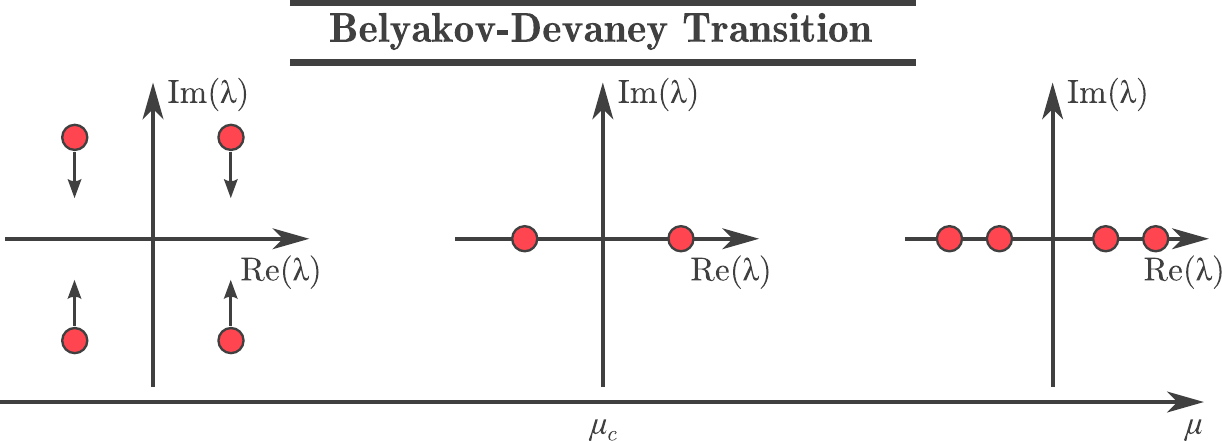}
    \caption{Spatial eigenvalues of the quiescent state for the Belyakov--Devaney transition showing the double-real-eigenvalue transition. We note that this is not a local bifurcation, as no eigenvalues cross the imaginary axis.}
    \label{fig:BD_transition}
\end{figure}

This fast-slow analysis provides a partial explanation for the snaking patterns that are observed, but it is only able to predict the existence of one small and one large amplitude solution~\cite{saadi2022} and not the snaking curves. What is typically observed in these models is that, as one approaches the singular limit from the Turing instability, the uniform state's spatial eigenvalues go through a transition from complex conjugates to purely real eigenvalues. This is known as a Belyakov--Devaney transition;
see Figure~\ref{fig:BD_transition} for an illustration.

To better understand the effect the Belyakov--Devaney transition has on localized patterns in singularly-perturbed reaction-diffusion systems, Verschueren and Champneys~\cite{champneys2021c} carried out a return map analysis near the transition point. They were able to show that there can only be two primary homoclinic orbits when the spatial eigenvalues of the uniform state are real. However, when the eigenvalues are complex there are infinitely-many homoclinic orbits, giving way to the snaking solutions. Furthermore, they were able to describe how the solutions in the snaking region are destroyed as one crosses the Belyakov--Devaney transition.

\subsubsection{Invasion Fronts Outside the Snaking Region}\label{subsubsec:1DInvasion}

\begin{figure}[t]
    \centering
    \includegraphics[width=0.9\linewidth]{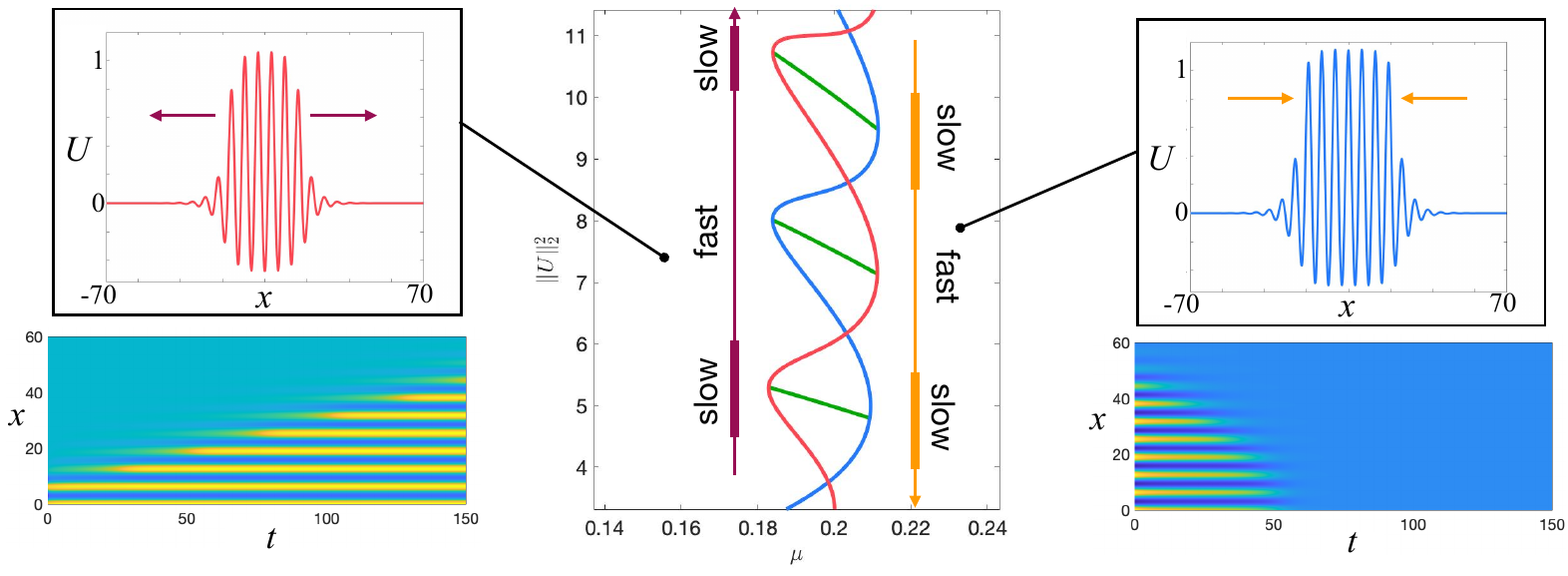}
    \caption{Outside of the snaking parameter regime localized SPPs are dynamic, growing for $\mu$ to the left (red) of the snaking branches and collapsing in on itself to the right (blue). This growth is non-uniform, alternating between fast and slow expansions, depending on if the pattern is close to a steady-state at a nearby saddle-node. }
    \label{fig:invasion_1d}
\end{figure}

Throughout this review we primarily focus on steady-state patterns. However, some work has been done to understand dynamic localized patterns. For example, outside of the snaking parameter regime in the SHE one can find that either the SPPs expand into the background state or collapse upon themselves as time progresses, as illustrated in Figure~\ref{fig:invasion_1d}. Despite the snaking region of parameter space typically being much smaller than the invasion regions, significantly less is known about these patterns and processes. 

The spreading of SPPs into the trivial state is fundamentally different in nature to the spreading of the trivial state. That is, in the case where the trivial state invades the SPP patterns, the speed of the invasion is dictated by the wavelength of the SPPs themselves. However, for the SPP invasion process, since we are in the bistable region, the wavelength of the deposited pattern and the time between stripes being deposited are selected~\cite{lloyd2019,Aranson2000}. The SPP invasion fronts do not invade at a constant speed and instead have a ``helical" invasion process, coming as a more complicated version of the ``bottlenecks'' one encounters in parameter regimes near a saddle-node bifurcation \cite[Section~4.3]{strogatz2018nonlinear}. Precisely, the non-constant traveling speed of the front is understood by looking at the fronts near the snaking region, where one would expect the invasion process to slow down as one passes near a fold bifurcation and then speed up between the folds, as illustrated in Figure~\ref{fig:invasion_1d}. There has been a lot of recent progress on the analysis of pattern-forming invasion fronts; see \cite{Goh_2023} for a review. 

Near the edge of the snaking region, Burke and Knobloch~\cite{burke2006localized} carried out a semi-analytical asymptotic analysis to predict the transition time, $T$, between two successive folds. They did so by considering a solution to the SHE \eqref{SwiftHohenberg} of the form
\begin{equation}\label{e:1D_semi_ana_ansatz}
    U(x,t) = U_0(x) + |\varepsilon|^{1/2}U_1(x,t),\qquad \mu = \mu_{\mathrm{fold}}+\varepsilon,\qquad |\varepsilon|\ll 1,
\end{equation}
where $U_0(x)$ is the stationary, localized front at the fold on the snaking curve. Substituting \eqref{e:1D_semi_ana_ansatz} into the SHE \eqref{SwiftHohenberg}, one finds that $U_1$ can be decomposed at leading order into
\[
    U_1(x,t) = a(t)U_e(x) + b(t)U_o(x) + c(t)U_n(x) + \mathcal{O}(|\varepsilon|^{1/2}),
\]
where $U_e(x)$ is the even eigenfunction of the linearization of the stationary SHE about the fold front whose eigenvalue vanishes at the fold, $U_o(x)$ is the odd eigenfunction that tracks the even mode ever more closely as one moves up the snake, and $U_n(x)$ is the neutrally stable odd eigenfunction $U_o'(x)$. The calculation can be simplified by taking $b\equiv c\equiv 0$ to yield a nonlinear ODE for $a(t)$. This ODE can be solved by calculating the time it takes $a(t)$ to pass from $-\infty$ to $+\infty$. While this approximation provides a strong prediction for the invasion speed of the front, there are some problems with it. Primarily, it says nothing about the selected wavelengths of the stripe pattern, while the ansatz only has leading order terms and no $\mathcal{O}(\varepsilon)$ correction. 

\begin{figure}[htb]
    \centering
    \includegraphics[width=0.7\linewidth]{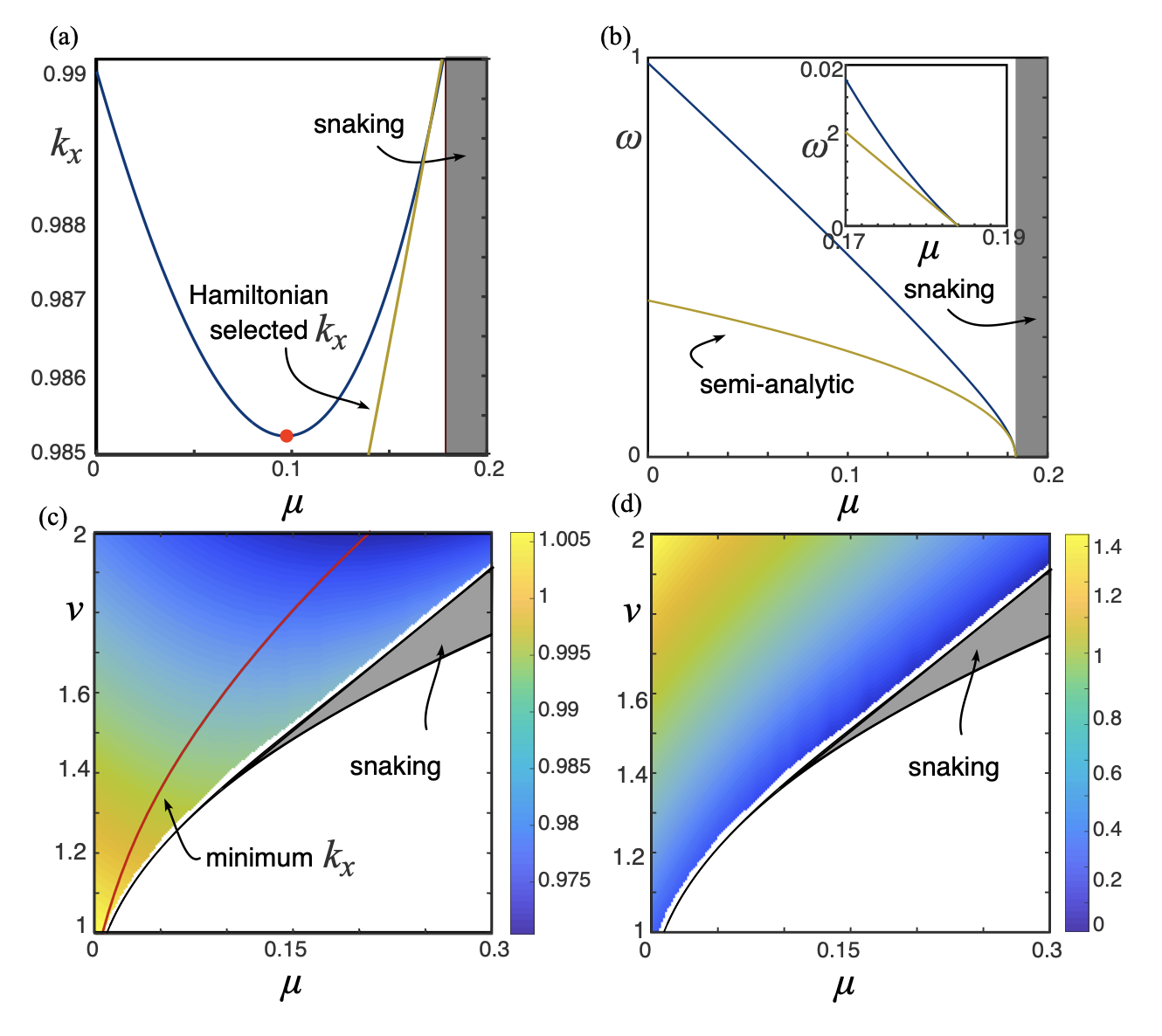}
    \caption{Invasion front outside the homoclinic snaking region in the bistable region for the quadratic-cubic Swift-Hohenberg equation. (a) \& (b) show the selected wavenumber and invasion frequency for $\nu=1.6$ in \eqref{SwiftHohenberg}. Panels (c) and (d) show the two-parameter diagrams of the selected wavenumbers and temporal frequency for~\eqref{SwiftHohenberg}. Reproduced from~\cite{lloyd2019}. 
    }
    \label{fig:invasionfront_bif}
\end{figure}
Numerical continuation methods to compute these fronts rely on a coordinate transformation to put the helical invasion process into a fixed frame of reference. This is achieved by introducing the coordinates $(\rho,\tau)=(k_x(x-ct),\omega t)$ where $\omega=ck_x$. Letting $U(x,t)=U(\rho,\tau)$ transforms the SHE to  
\[
    \omega(U_\rho-U_\tau)-(1+k_x^2\partial_\rho^2)^2U -\mu U + f(U) = 0,\qquad (\rho,\tau)\in[-\infty,\infty]\times(0,2\pi],
\]
subject to the boundary conditions
\begin{equation}\label{e:1D_invasion}
    U(\rho,\tau)=U(\rho,\tau+2\pi),\qquad \lim_{\rho\rightarrow\infty}U(\rho,\tau) =0,\qquad \lim_{\rho\rightarrow-\infty}(U(\rho,\tau)-U_p(\rho+\tau;k_x)) = 0, 
\end{equation}
with $U_p$ being the stripes in the co-moving frame with a wavenumber $k_x$ that is selected by the invasion front. To have an amenable numerical scheme, Lloyd~\cite{lloyd2019} employed a far-field core decomposition approach where we decompose $U$ as follows
\[
    U(\rho,\tau) = U_p(\rho+\tau;k_x)\chi(x) + w(\rho,\tau),
\]
with $\chi(\rho) = (1 + \mathrm{tanh}(m(\rho - d)))/2$. The unknown function $w(\rho,\tau)$ is known as the core function, which can be found by solving the PDE, yielded from substitution of the above ansatz into the SHE and subtracting out the equation for $U_p$, given by 
\begin{equation}
   \mathbb{L}[U_p(\rho+\tau;k_x)\chi(\rho)+w(\rho,\tau;\omega)]+f(U_p(\rho+\tau;k_x)\chi(\rho)+w(\rho,\tau;\omega))-\chi(\mathbb{L}U_p +f(U_p))=0. 
\end{equation}
where $\mathbb{L} := \omega(\partial_\rho + \partial_\tau)-(1+k_x^2\partial_\rho^2)^2 - \mu$. For computation the $\rho$ domain is truncated to $[-L,L]$. Two other conditions are also imposed to select $\omega$ and $k_x$ 
\begin{equation}
    \int_0^{2\pi}\int_{-L}^L([\partial_\rho-\partial_\tau]U^{\mathrm{old}})(U-U^{\mathrm{old}})\,\mathrm{d}\rho\,\mathrm{d}\tau=0,\qquad \int_0^{2\pi}\int_{-L}^{-L+2\pi/k_x}(\partial_\rho U_p)w \,\mathrm{d}\rho\,\mathrm{d}\tau = 0.
\end{equation}
The first condition is a standard phase condition used to select front speeds as described in~\cite{doedel2007auto}, while the second condition is a 0th order phase condition to make the core-function $w(\rho,\tau)$ orthogonal to the translation eigenfunction of the stripes.

Numerically, one finds the bifurcation diagram for the invasion fronts as shown in Figure~\ref{fig:invasionfront_bif} for the quadratic-cubic and cubic-quintic SHE in the bistable region. It is found that as one leaves the edge of the snaking region and the selected wavenumber that is the same as the selected wavenumber of the periodic orbit of the localized stationary pattern at the edge of the snaking region, the wavenumber decreases as $\mu$ decreases to a minimum value and then rises again. The invasion fronts can be continued for $\mu<0$ until they reach the convective instability threshold where there is a transition to a pulled-front, which was analyzed in~\cite{Avery2023}. The invasion speed monotonically increases as $\mu$ is decreased from the edge of the snaking region. No other bifurcations are found for the invasion fronts, and this picture is found to be the same in the cubic-quintic SHE. Invasion fronts have yet to be explored using numerical bifurcation methods in more complicated models.

\section{Localization in Only One Direction}\label{sec:fronts}

Localized solutions to mathematical equations with localization in only one unbounded direction have been well-documented in the literature \cite{avitabile2010snake,beck2009snakes,burke2007homoclinic,Hunt2000,Escaff2009,Malomed1990,Hagberg2006,Malomed2021}. Much like the axisymmetric solutions we will see in Section~\ref{sec:radial}, the analysis of patterns localized in only one direction can often be approached through similar methods to the 1D spatial setting. In this subsection we review much of the recent progress on understanding the existence, bifurcation structure, and dynamics of planar patterns that are localized in only one direction. 

We will primarily concentrate on localized patterns in the cubic-quintic SHE on an infinite cylinder
\begin{equation}\label{e:Swift-Hohenberg_2D_cubic_quintic}
    U_t = -(1+\partial_x^2+\partial_y^2)^2U - \mu U + \nu U^3 - U^5,\qquad (x,y)\in \mathbb{R} \times S^1,
\end{equation}
where $S^1=\mathbb{R}/2L\mathbb{Z} \simeq [-L,L]$ is the circle of length  $2L > 0$. If not stated otherwise, we will use $L = \pi$ in for numerical demonstrations. The reason for using the cubic-quintic SHE here is that numerical investigations have revealed that it is easier to find stable patterns localized in the $x$-direction in this model. 

We begin this section in \S\ref{subsec:Stripes} with localized stripe patterns wherein the localized pattern takes the form of a 1D SPP extended trivially in either the direction of $x$, $y$, or a combination thereof. \S\ref{subsec:HexFront} reviews the work on patterns with hexagonal patches arranged along the region of localization, while \S\ref{subsec:Spatial2D} provides the spatial dynamics interpretation of these stripe and front patterns. Finally, in \S\ref{subsec:LatticePinning} we review similar work on planar lattices.

\subsection{Localized Stripe Fronts}\label{subsec:Stripes}

\begin{figure}[t]
    \centering
\includegraphics[width=0.8\linewidth]{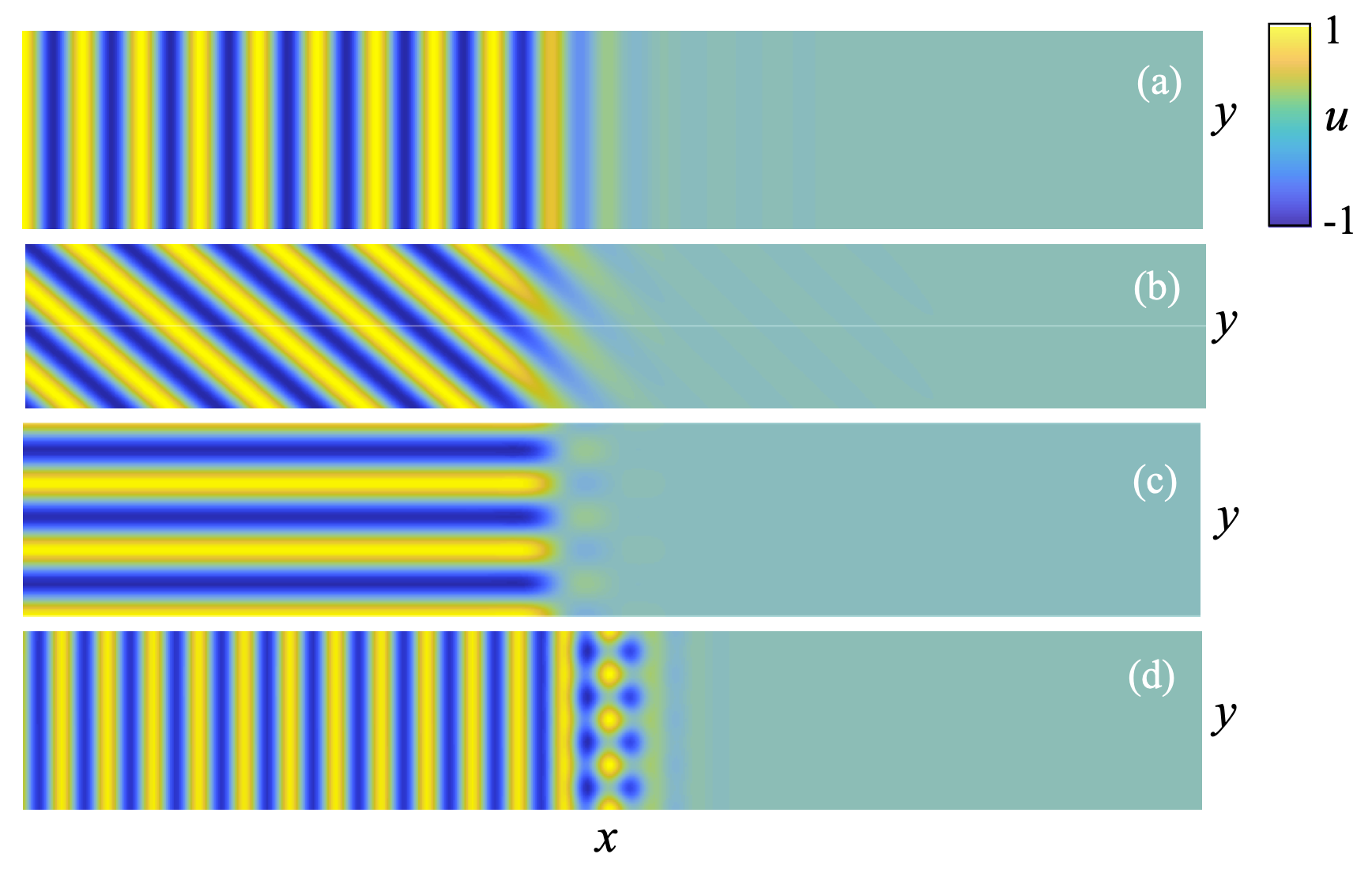}
    \caption{Four different types of localized planar stripe patterns (a) parallel to the front interface, (b) oblique to the front interface, (c) perpendicular to the front interface, and (d) ``almost planar" fronts. Patterns are symmetric over the left figure edge with only half the pattern plotted to emphasize localization. Image originally appears in \cite{lloyd2019}}
    \label{fig:stripe_types}
\end{figure}

\begin{figure}[t]
    \centering
    \includegraphics[width=0.9\linewidth]{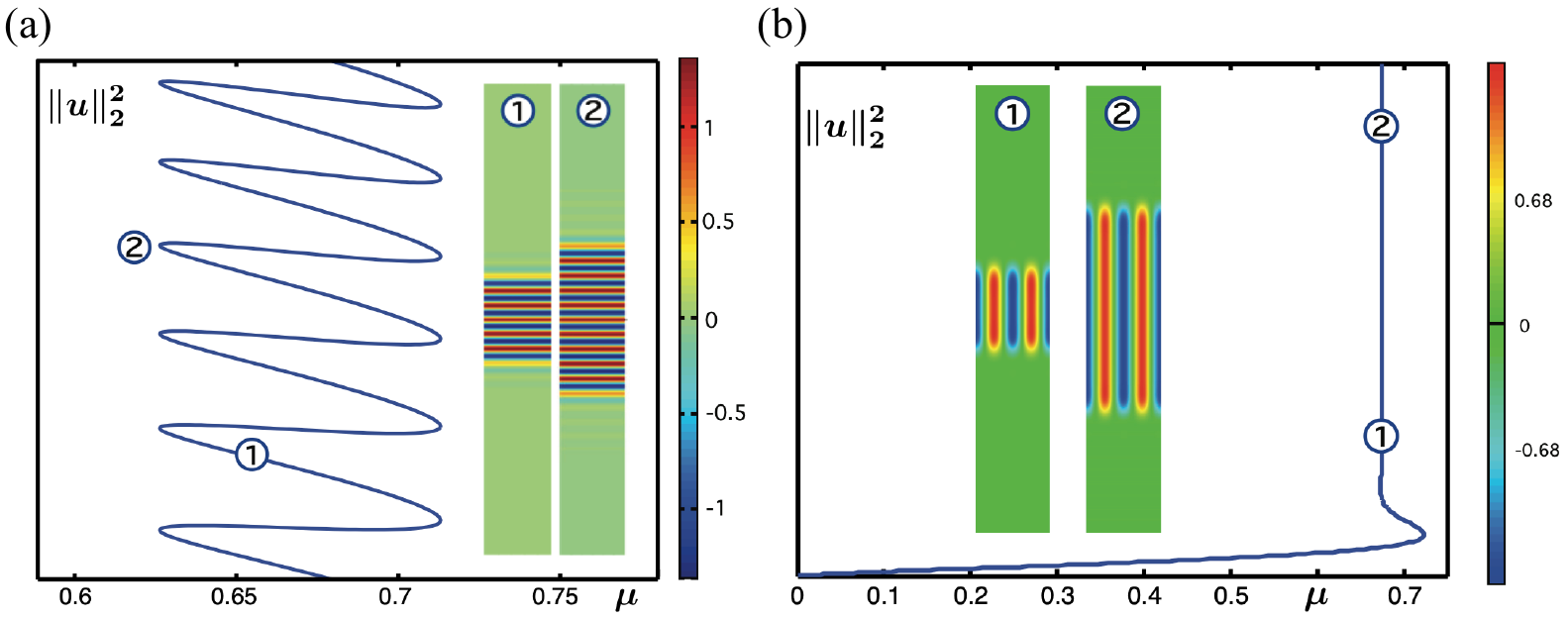}
    \caption{(a) Snaking and (b) Non-snaking bifurcation diagrams of parallel and perpendicular localized stripes in~\eqref{e:Swift-Hohenberg_2D_cubic_quintic} with $\nu=2$. Image originally appears in \cite{avitabile2010snake}. }
    \label{fig:localised_stripes_1}
\end{figure}

The simplest form of 2D planar localized patterns in \eqref{e:Swift-Hohenberg_2D_cubic_quintic} are those that have no $y$-dependence. Such a restriction trivially reduces one to the 1D setting, with localization only taking place in the $x$-direction. As shown in Figures~\ref{fig:stripe_types}(a) and \ref{fig:localised_stripes_1}(a), the result is a localized stripe pattern whose stripes are parallel to the front interface. Since these structures are equivalent to the 1D patterns of Section~\ref{sec:1D}, they exhibit the same snaking bifurcation curves as the 1D patterns.   

The 2D cylindrical SHE \eqref{e:Swift-Hohenberg_2D_cubic_quintic} also exhibits localized patterns that are genuinely two-dimensional. For example, numerical investigations have revealed the existence of localized stripe patterns that are perpendicular to the front interface, as shown in Figure~\ref{fig:stripe_types}(c). Continuing these patterns reveals that they do not snake, but instead smoothly grow the stripes onto the quiescent state, as show in Figure~\ref{fig:localised_stripes_1}(b). In an effort to understand such patterns, one may carry out a formal weakly nonlinear analysis, similar to that in \S\ref{subsec:SH1D}, by expanding
\[
    U(x, y, t) =\varepsilon A(\varepsilon x,\varepsilon^4 t)e^{i \pi y/L} + c.c. + \mathcal{O}(\varepsilon^2),\qquad \mu = \varepsilon^4\tilde\mu,
\]
and putting this into \eqref{e:Swift-Hohenberg_2D_cubic_quintic} with $\nu=\varepsilon^2\tilde\nu$. At order $\mathcal{O}(\varepsilon^5)$ one again arrives at a Ginzburg--Landau amplitude equation of the form
\begin{equation}\label{e:perp_amp}
    A_T = -A_{XXXX} - \tilde\mu A + 3\tilde\nu|A|^2A - 10|A|^4A,
\end{equation}
where $(X,T) = (\varepsilon x,\varepsilon^4 t)$. Stationary solutions of \eqref{e:perp_amp} are not known analytically, but one can show that the equilibrium states have complex spatial eigenvalues and we observe the non-snaking explained in \S\ref{sec:snakes_ladders}.

Further numerical investigations indicate that one can find localized patterns involving stripes with any orientation between the parallel and perpendicular configurations, known as oblique stripes~\cite{lloyd2019}. Small amplitude localized oblique stripes can be captured similarly to the perpendicular patterns above with the expansion
\[
    U(x, y, t) = \varepsilon A(\varepsilon^2x,\varepsilon^4 t) e^{ikr} + c.c. + \mathcal{O}(\varepsilon^2),
\]
where $k$ is the critical linear wavenumber at bifurcation, $r = \cos(\alpha)x - \sin(\alpha)y$, and $\alpha$ is the orientation of the interface. With this ansatz $\alpha = 0$ corresponds to the parallel ($y$-independent) stripes and $\alpha = \frac{\pi}{2}$ corresponds to the perpendicular stripes. The amplitude equation is again found at $\mathcal{O}(\varepsilon^5)$ to be
\begin{equation}\label{e:amp_oblique_stripes}
  A_T = 4(\cos(\alpha))^2A_{XX} - \tilde\mu A + 3\tilde\nu |A|^2A - 10|A|^4A.  
\end{equation}
For $\alpha=0$, the amplitude equation \eqref{e:amp_oblique_stripes} reduces to the 1D amplitude equation encountered in \eqref{e:GZeqn}. As $\alpha$ varies from $0$ to $\pi/2$ the coefficient in front of the second derivative term in \eqref{e:amp_oblique_stripes} tends to zero, in turn leading to sharper localized fronts. It can be shown that oblique stripes do not snake using either energy arguments involving \eqref{e:2D_Ham} and \eqref{e:2D_momentum}~\cite{lloyd2019} or by dimensional analysis since we expect the intersection between the center-unstable and stable manifolds for such fronts in a spatial dynamics setting is not transverse due to the 2D kernel from the $x$ and $y$ derivatives.

\begin{figure}[t]
    \centering
\includegraphics[width=0.9\linewidth]{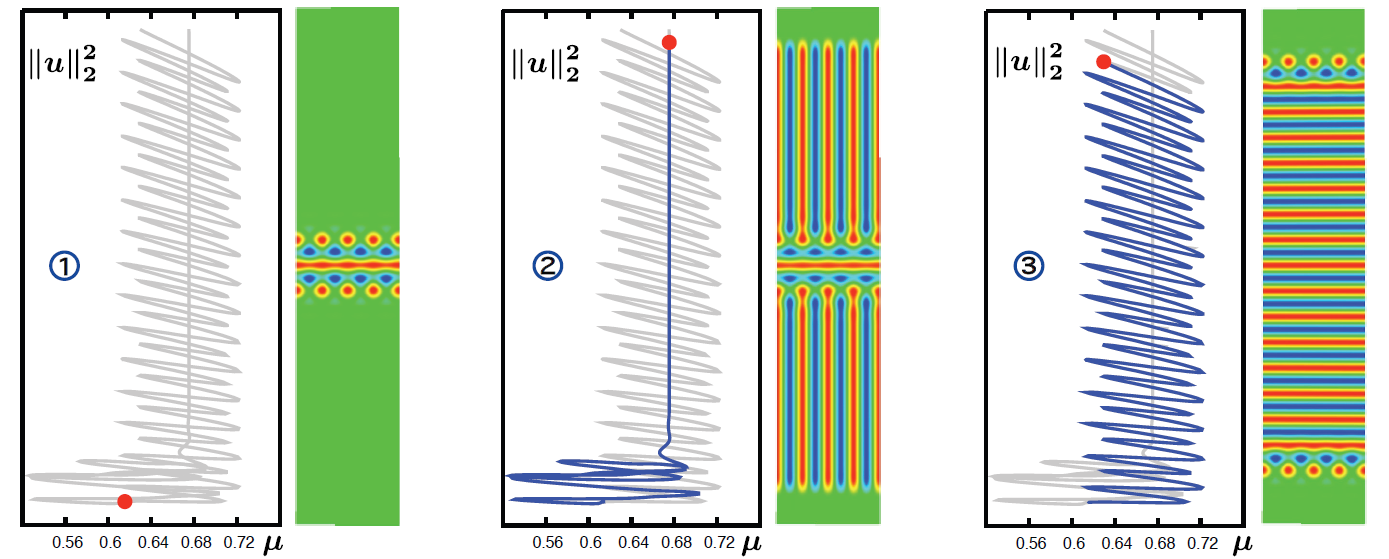}
    \caption{Bifurcation diagram of almost planar fronts in \eqref{e:Swift-Hohenberg_2D_cubic_quintic} with $\nu=2$. Localized patterns can either snake or not snake depending on the interface with the trivial state. Image originally appears in \cite{avitabile2010snake}.}
    \label{fig:bif_almost_planar}
\end{figure}

\begin{figure}[t]
    \centering
    \includegraphics[width=0.7\linewidth]{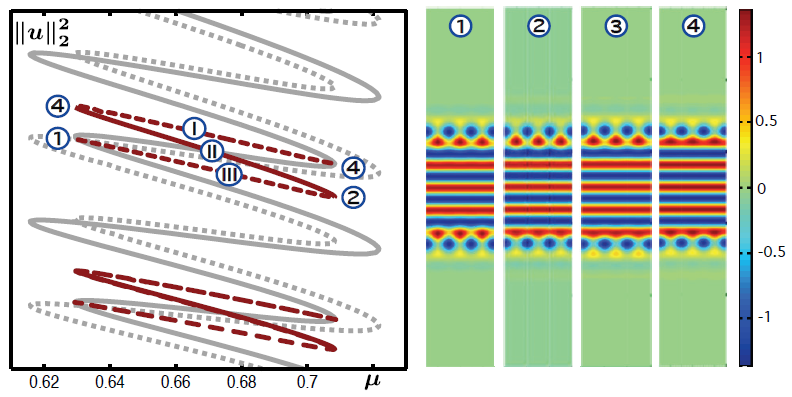}
    \caption{Almost planar stripes ladder structure for \eqref{e:Swift-Hohenberg_2D_cubic_quintic} with $\nu=2$. The ladder structure of the asymmetric localized state can be found from the symmetric snaking diagram using the formula~\eqref{e:ladderz}. Image originally appears in \cite{avitabile2010snake}.}
    \label{fig:almost_planar_stripe_ladder}
\end{figure}

Beyond parallel, oblique, and perpendicular localized stripe patterns, there is another entirely distinct type of pattern to \eqref{e:Swift-Hohenberg_2D_cubic_quintic}. These localized patterns are referred to as {\em almost planar} fronts, comprised of a core of parallel stripes to the front interface with square cells occurring along the length of the interface. Examples are shown in Figure~\ref{fig:bif_almost_planar} and were initially found by looking at bifurcations off the parallel stripe fronts. It was found that far away from bifurcation these structures can lie on a single bifurcation curve that can either snake or not \cite{avitabile2010snake}. Interestingly, we start to see a significant difference from the typical 1D snaking in that instead of there being a regular ascension of the bifurcation diagram by bouncing between two different fold values, there now appears to be four fold limits. This leads to a more interesting bifurcation structure for the ladder/asymmetric states that can be predicted from the theory described in \S\ref{sec:snakes_ladders}, as shown in Figure~\ref{fig:almost_planar_stripe_ladder}.

\begin{figure}[t]
    \centering
    \includegraphics[width=\linewidth]{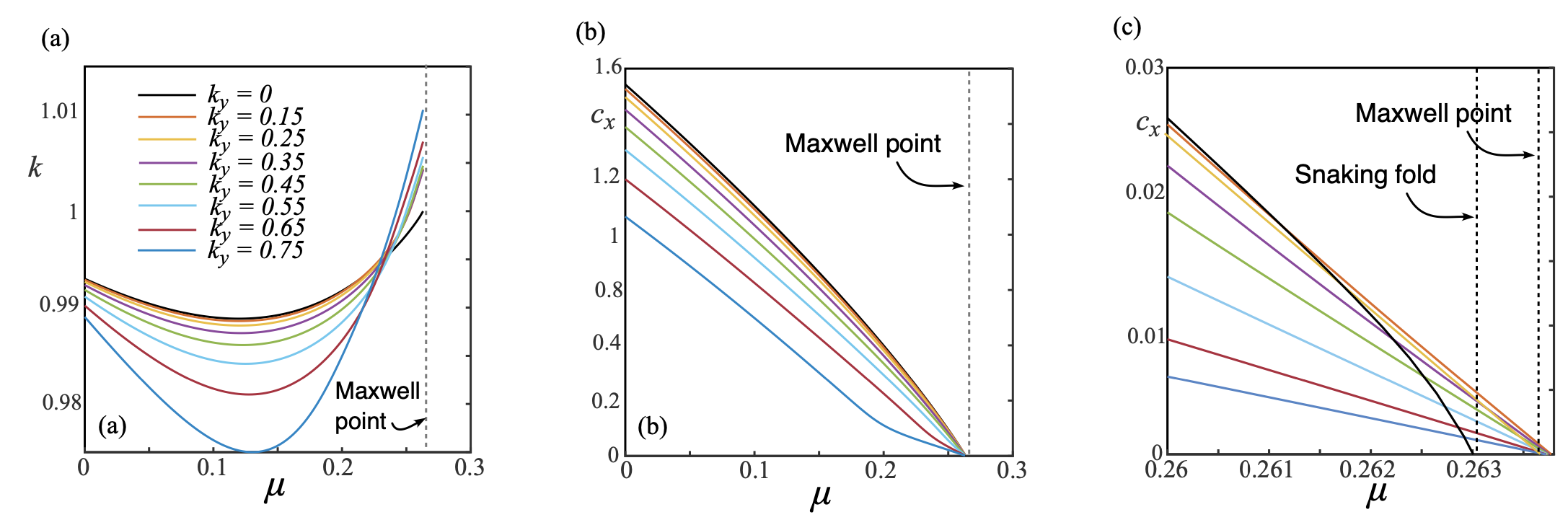}
    \caption{Bifurcation diagram for oblique invasion stripes for \eqref{e:Swift-Hohenberg_2D_cubic_quintic} with $\nu=1.25$ and $k_y>0$. The parallel stripe invasion front is shown for $k_y=0$. Panel (a) shows the selected wavenumber $k=\sqrt{k_x^2+k_y^2}$, panel (b) shows the wavespeed $c_x$ selection and panel (c) shows a zoom-in near the Maxwell point showing that the parallel front ceases at the edge of the snaking region while the oblique fronts continue to the 1D stripe Maxwell point. Image originally appears in \cite{lloyd2019}.}
    \label{fig:oblique_non_snake}
\end{figure}

Outside of the snaking regime these patterns become dynamic, much like the 1D case of \S\ref{subsubsec:1DInvasion}. Figure~\ref{fig:oblique_non_snake} presents results from numerically solving the Swift--Hohenberg equation for invading oblique stripe fronts. Observe that all of these fronts terminate very close to the Maxwell point (they are slightly off due to numerical approximation error) and the parallel front terminates at the edge of the snaking region. 
For invading/retreating perpendicular stripes, it is observed that as the wavelength of the stripes is shortened, the invading fronts can exist further for more negative $\mu$ values. The almost planar stripe fronts are also found to invade and bifurcate subcritically off the parallel invading fronts and are found to restabilizes in a fold and be the stable invading fronts close to the snaking region.

\subsection{Localized Hexagon Fronts}\label{subsec:HexFront}

\begin{figure}[t]
    \centering
    \includegraphics[width=0.75\linewidth]{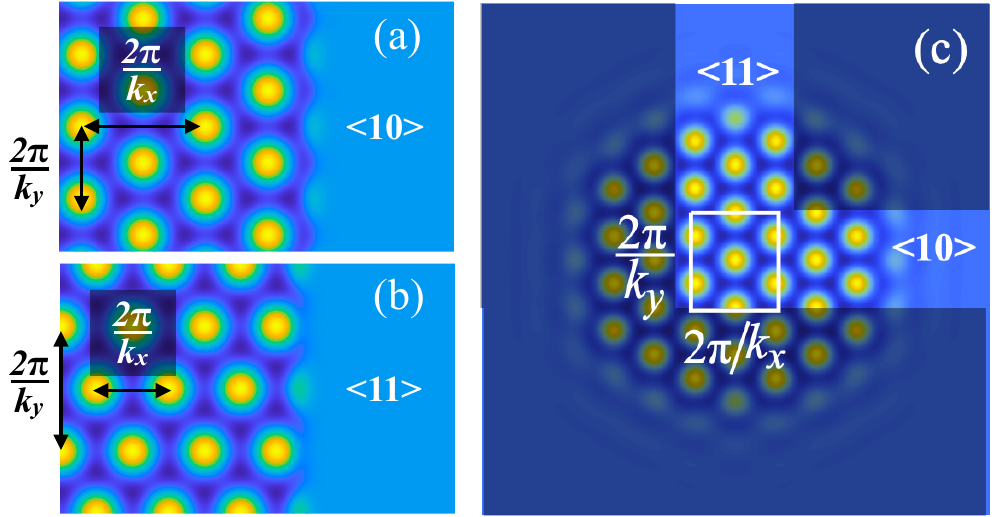}
    \caption{(a) A $\langle10\rangle$ hexagon front, (b) a $\langle11\rangle$ hexagon front, and (c) a localized hexagon patch made up of combinations of a $\langle10\rangle$- and $\langle11\rangle$-fronts. Image originally appears in \cite{lloyd2021}.}
    \label{fig:hex_front}
\end{figure}

So far in this subsection we have been introduced to localized stripe patterns. These localized patterned states are nearly one-dimensional and do not necessarily showcase the variety of localized patterns that can arise in two spatial dimensions. Here we will detail another type of localized pattern to \eqref{e:Swift-Hohenberg_2D_cubic_quintic}, this time with the localized patterned state being one that is fundamentally two-dimensional. Precisely, we review the emergence of localized dihedral hexagon patterns, localized in a single direction and known as planar fronts. Examples of these states can be found in Figure~\ref{fig:hex_front}. One key observation is that the orientation of the hexagonal lattice with respect to the front leads to different types of front interfaces. Two key hexagon fronts are called the $\langle10\rangle$ and $\langle11\rangle$-fronts, as defined by the Bravais--Miller index~\cite{lloyd2008localized}. Other localized hexagon front orientations are also possible, but not plotted here for brevity. These fronts also play a fundamental role in localized hexagon patches described in \S\ref{ssec:FullyLocal} since one can view a hexagon patch as being made up of different combinations of these hexagon fronts. 

We begin with the weakly nonlinear analysis of these planar hexagon fronts. This can be used to understand the variety of different fronts possible between domain covering hexagon patterns and the trivial background state. There are infinitely many (discrete) orientations of the front interface to the hexagon lattice that are possible and these fronts are not all the same. The weakly nonlinear analysis starts by expanding a solution to \eqref{e:Swift-Hohenberg_2D_cubic_quintic} as
\begin{equation}\label{e:hex_amp_ansatz}
    U(x,y) = \varepsilon \left(\sum_{i=1}^3A_i(X,T)e^{ik_i\cdot \mathbf{r}} + c.c.\right) + \mathcal{O}(\varepsilon^2),\qquad X=\varepsilon x,\quad T = \varepsilon^2t,\quad 0 < \varepsilon\ll 1
\end{equation}
where $\mathbf{r}=x\hat x + y\hat y$, $\hat x = (\cos\alpha,-\sin\alpha)$,$\hat y = (-\sin\alpha,\cos\alpha)$, $k_1 = (-1,0),k_2=(\frac12,\frac{\sqrt{3}}{2})$, $k_3=(\frac12,-\frac{\sqrt{3}}{2})$, and $\alpha$ describes the orientation of the hexagonal pattern to the direction $X$. Working in the parameter regime $(\mu,\nu)=(\varepsilon^2\tilde\mu,\varepsilon\tilde\nu)$ allows one to derive a system of amplitude equations for the $A_1,A_2,$ and $A_3$ profiles at $\mathcal{O}(\varepsilon^3)$. The equation for $A_1$ is given by
\begin{equation}\label{e:hex_amp}
    \partial_T A_1 = M_1\partial^2_X A_1 - \left(\tilde\mu+3(|A_1|^2 +2(|A_2|^2 + |A_3|^2)) \right)A_1 + 2 \tilde{\nu} \bar{A}_2 \bar{A}_3,
\end{equation}
while the other equations are given by cycling the indices of $A_i$ and $M_i$. In the above equation we have defined $M_i=4(k_i\cdot\hat x)$ and we use a bar to denote complex conjugation. The hexagon pattern state is given by the solution $A_1=A_2=A_3=A_h\in\mathbb{R}$, meaning that $A_h$ solves
\[
    - \left(\tilde\mu+15 A_h^2 \right)A_h+2\tilde\nu A_h^2 = 0.
\]
Unlike the one-dimensional equivalents we have previously encountered for these Ginzburg--Landau-type amplitude equations, there are no explicitly known stationary homoclinic or heteroclinic solutions to \eqref{e:hex_amp} that can be used to prove the existence of localized hexagon front patterns taking the form \eqref{e:hex_amp_ansatz}. However, the amplitude equations \eqref{e:hex_amp} do possess a gradient structure, meaning that one can compute the Maxwell point for the hexagons and trivial state to find that a stationary heteroclinic connection is likely to occur around $\mu_M = (8/135)\,\nu^2 + \mathcal{O}(\nu^3)$.

\begin{figure}[t]
    \centering
    \includegraphics[width=0.9\linewidth]{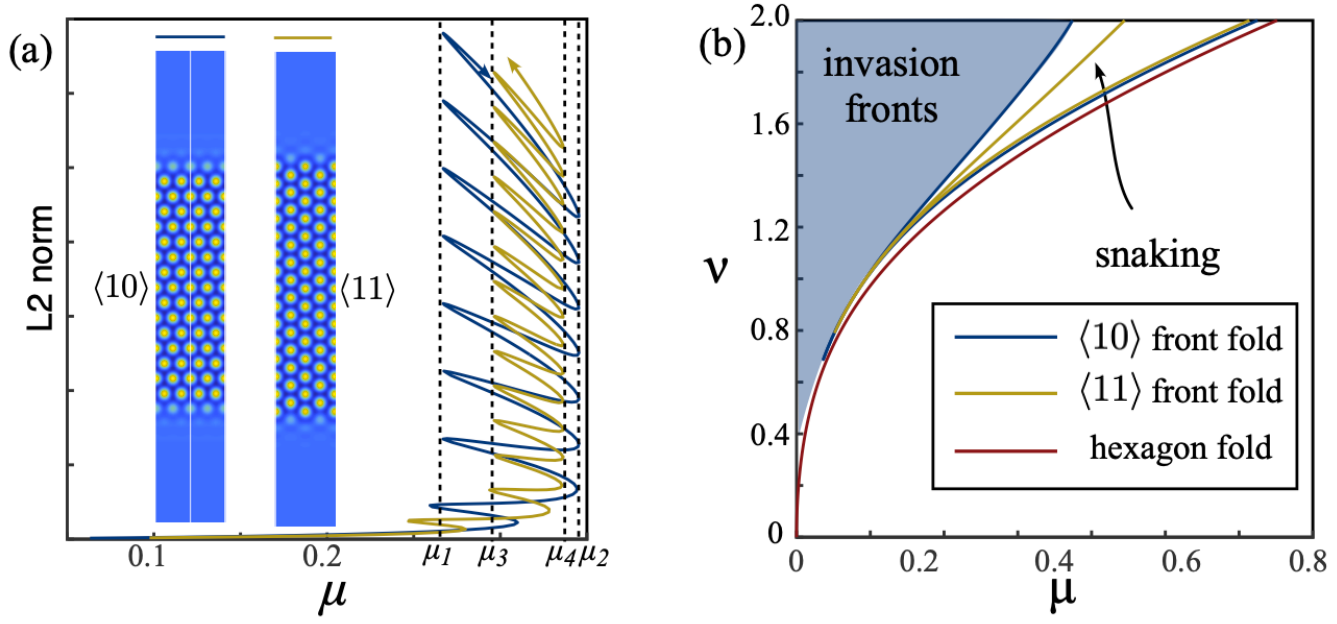}
    \caption{(a) Hexagon front snaking bifurcation diagram for the $\langle10\rangle$- and $\langle11\rangle$-fronts of \eqref{e:Swift-Hohenberg_2D_cubic_quintic} with $\nu=1.6$ for $k_y = \frac12$ and $\sqrt{3/2}$. (b) A bifurcation diagram in the parameters $(\mu,\nu)$ for the stationary hexagon fronts showing how the widths of the snaking regions vary and the invasion front region. Image originally appears in \cite{lloyd2021}.}
    \label{fig:hexagon_front_snaking_1}
\end{figure}

Based on the intuition built from the one-dimensional spatial setting, one expects that the Maxwell point lies firmly inside the region of existence for the planar hexagon fronts. Numerical continuation of these localized hexagon fronts up from the small-amplitude parameter regime confirms this and reveals that these structures exhibit a similar snaking structure to the 1D localized patterns encountered in Section~\ref{sec:1D}, as can be seen in Figure~\ref{fig:hexagon_front_snaking_1}. We see that entire rows of cells are added at the interface. Numerically, both the orientation of the front interface with the hexagon lattice as well as how squeezed the hexagons are 
significantly affect the widths of the snaking regions. 

\begin{figure}[t]
    \centering
    \includegraphics[width=0.9\linewidth]{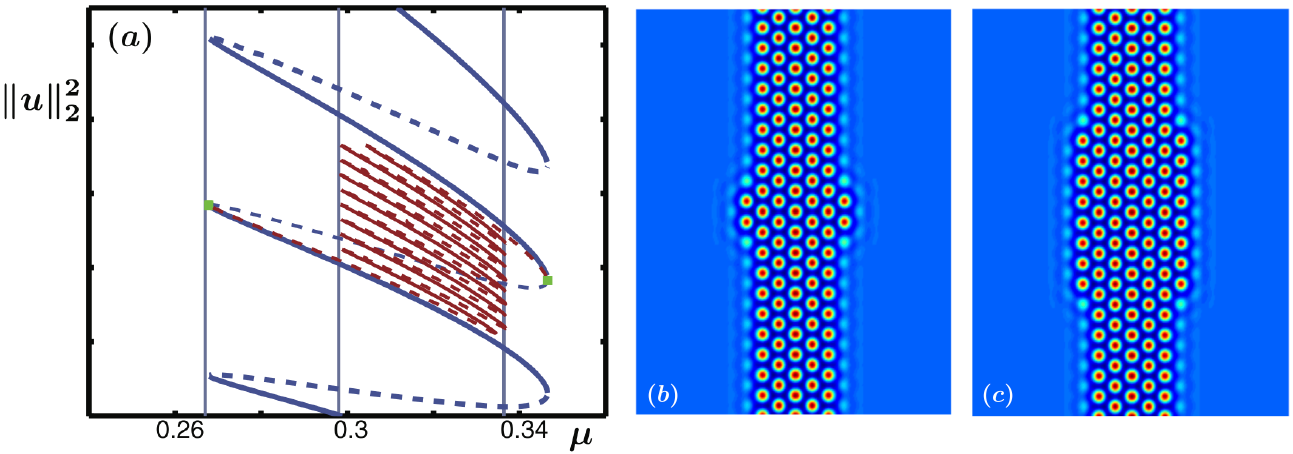}
    \caption{A bifurcation diagram of almost-planar hexagon pulses shown in red with the blue curve depicting the regular $\langle10\rangle$-front. Panels (b) and (c) show two of the almost-planar hexagon pulses for $\mu=0.3$ on the red bifurcation curve. Image originally appears in \cite{lloyd2008localized}. }
    \label{fig:hexagon_front_spot_snaking}
\end{figure}

Along with these snaking bifurcation curves, pairs of cells can be added to the front to grow the pattern between the extremal folds of the snaking curves. These branches appear similar to the ladder states in the 1D setting, as shown in Figure~\ref{fig:hexagon_front_spot_snaking}, but cannot be captured by the spatial dynamics prediction that is featured next in the following subsection. The result is that this process of adding individual pairs along the front creates a mini snaking diagram within the full snaking diagram. This process appears to be crucial in understanding the snaking of fully localized patches of hexagon pattern seen in \S\ref{ssec:FullyLocal}.

Kozyreff and Chapman~\cite{Kozyreff2013} extended their exponential asymptotic analysis from 1D (see \S\ref{subsec:SH1D}) to better understand these localized hexagon fronts. They focused their analysis on the co-dimension 2 point where the domain-filling hexagons transition from bifurcating super- to sub-critically from the trivial state. Their analysis is presented for more general systems, but in the case of the SHE \eqref{e:Swift-Hohenberg_2D_cubic_quintic} this is done by setting $\mu=\varepsilon^2$ and $\nu=\sqrt{3/4}\varepsilon\tilde\nu$. These results show that the width of the parameter regime in $\mu$ for which these localized hexagon fronts exist, denoted $\delta\nu(\varepsilon) \geq 0$, is approximately given by
\[
    \delta\nu(\varepsilon) \propto e^{-1.5\pi|\Delta k|/\varepsilon},
\]
where $(q_x+\Delta k)^2 + q_y^2 = 1$ for appropriate orientations $(q_x,q_y) = m_1k_1+m_2k_2$ and $m_i\in\mathbb{Z}$. This result states that the orientation of the front interface to the hexagonal lattice with the largest pinning region occur for the smallest $\Delta k$, which turns out to be the $\langle10\rangle$-front. The next smallest is the $\langle11\rangle$-front with other orientations having larger values. This work was then corroborated via numerical experiments in \cite{Kozyreff2013} and holds in general near any Turing or finite-wavenumber instability occurring in a RD system. 

Nearly a decade after Kozyreff and Chapman's work, Boissoni\'ere, Choksi, and Lessard were able to rigorously prove the existence of localized hexagon fronts in a phase-field crystal model~\cite{Boissoniere_2022}. This result was achieved using rigorous numerics via a Newton--Kantorovich argument applied to a phase-field crystal model whose Euler--Lagrange equation is similar to the SHE. Moreover, Boissoni\'ere et al. proved the existence of a variety of different localized patches, as well as the existence of hexagon-to-square fronts.

\begin{figure}[htb]
    \centering
    \includegraphics[width=0.9\linewidth]{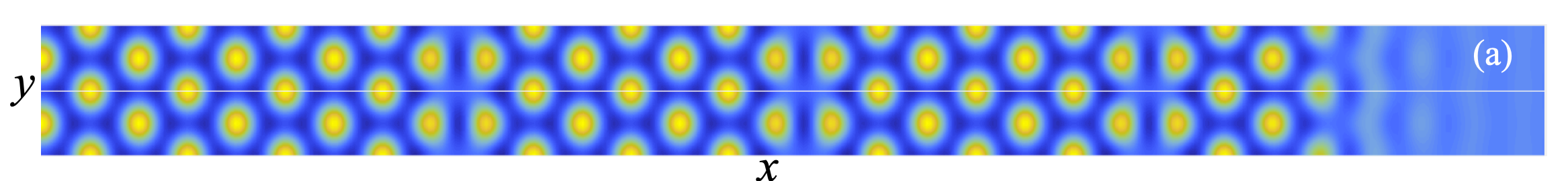}
    \caption{A snapshot from temporal simulations of a $\langle10\rangle$ hexagon front with defects in the SHE with $(\mu,\nu)=(0.05,1.6)$ and $k_y = 0.72$. Image originally appears in \cite{lloyd2021}.}
    \label{fig:hexagon_invasion_defect}
\end{figure}

Much like we saw in \S\ref{subsubsec:1DInvasion} for 1D localized patterns, the localized hexagon patterns can be found to invade or retreat outside their snaking region. When invading, the fronts select an average front speed and a wavenumber $k_x$ for a fixed size of the periodic strip, denoted $L = \pi/k_y$, in \eqref{e:Swift-Hohenberg_2D_cubic_quintic}; see Figure~\ref{fig:hex_front}. In particular, these fronts select a `squeezed' hexagon cellular pattern in the far-field. 
It is found that the $\langle10\rangle$-invasion fronts can develop periodic defects due to a saddle-node bifurcation of a periodic orbit. These defects can be observed in Figure~\ref{fig:hexagon_invasion_defect} where we provide a snapshot of an invading hexagon front. 

\subsection{Spatial Dynamics}\label{subsec:Spatial2D}

The spatial dynamics approach to localized structures can also be applied to prove the existence and bifurcation structure of the stripe and hexagon patterns encountered previously in this subsection. For exposition, we follow \cite{beck2009snakes} and again consider the 2D SHE on an infinite cylinder \eqref{e:Swift-Hohenberg_2D_cubic_quintic}. Time-independent solutions of \eqref{e:Swift-Hohenberg_2D_cubic_quintic} can be interpreted as bounded solutions in $x \in \mathbb{R}$ evolving in the phase-space $H^3(S^1)$. Precisely, the spatial dynamics formulation for \eqref{e:Swift-Hohenberg_2D_cubic_quintic} is given by the first order system 
\begin{equation}\label{e:spatial_dyn_stripes}
    \begin{split}
        u_1' &= u_2, \\
        u_2' &= u_3, \\
        u_3' &= u_4, \\
        u_4' &= -(1 + \partial_y^2)^2 u_1 - 2(1 + \partial_y^2)u_3 - \mu u_1 + \nu u_1^3 - u_1^5,
    \end{split}
\end{equation}
where we use $'$ to denote differentiation with respect to the unbounded spatial variable $x$ to emphasize the dynamical systems formulation. Any solution $\mathbf{u}(x,y) = (u_1(x,y),u_2(x,y),u_3(x,y),u_4(x,y))$ is, for each fixed $x$, a function of $y$ that lies in the infinite-dimensional phase-space $\mathcal{X} := H^3(S^1) \times H^2(S^1) \times H^1(S^1) \times L^2(S^1)$. 

Despite the fact that \eqref{e:spatial_dyn_stripes} possesses an infinite-dimensional phase-space, it bears many of the same features as its finite-dimensional counterpart \eqref{SHE_spatial_1D}. First, \eqref{e:spatial_dyn_stripes} can be written as a Hamiltonian system of the form
\[
    q_1' = \frac{\partial \mathcal{H}}{\partial p_1}, \qquad  q_2' = \frac{\partial \mathcal{H}}{\partial p_2}, \qquad p_1' = -\frac{\partial \mathcal{H}}{\partial q_1}, \qquad p_2' = -\frac{\partial \mathcal{H}}{\partial q_2},
\]
where 
\[
    (q_1, q_2, p_1, p_2) = (u_1, u_2, -(u_4 + 2(1 + \partial_y^2)u_2), u_3) \in H^3(S^1) \times H^2(S^1) \times L^2(S^1) \times H^1(S^1),
\]
and 
\begin{equation}\label{e:2D_Ham}
  \mathcal{H}(\mathbf{q}) = \int_{-L}^{L} \left[q_2 (p_1 + (1+\partial_y^2)q_2) + \frac{p_2^2}{2} - \frac{[(1+\partial_y^2)q_1]^2}{2} - \frac{\mu q_1^2}{2} + \frac{\nu q_1^4}{4} - \frac{q_1^6}{6}  \right] \mathrm{d}y.
\end{equation}
As well as the Hamiltonian structure, system \eqref{e:spatial_dyn_stripes} has another conserved quantity, given by
\begin{equation}\label{e:2D_momentum}
    S(\mathbf{u}) = \int_{-L}^{L}\left[[(1 + \partial_{y}^2) u_{1} + u_{3}]\partial_{y}u_{2} + u_{1}\partial_{y}[(1 + \partial_{y}^2) u_{2} + u_{4}]\right]\mathrm{d}y.
\end{equation}
Moreover, the presence of only even derivatives in $x$ in \eqref{e:Swift-Hohenberg_2D_cubic_quintic} endows \eqref{e:spatial_dyn_stripes} with a reversible structure with reverser given by $\mathcal{R}(u_1,u_2,u_3,u_4) = (u_1,-u_2,u_3,-u_4)$, while the presence of the additional bounded spatial variable $y$ now gives an additional $\mathbb{Z}_2$ symmetry whose action is generated by $\tau: \mathbf{u}(y) \mapsto \mathbf{u}(L - y)$.

Thus, we see that many of the same ingredients that were exploited in \cite{beck2009snakes} (and summarized in \S\ref{subsec:HomoclinicSnaking}) are present to prove the existence of localized SPPs in 1D. Much like the 1D setting, localized structures of \eqref{e:Swift-Hohenberg_2D_cubic_quintic} correspond to orbits homoclinic to $\mathbf{0} \in \mathcal{X}$, i.e. $\mathbf{u}(x,\cdot) \to \mathbf{0}$ as $x \to \pm \infty$. One needs to be careful here though as \eqref{e:spatial_dyn_stripes} is ill-posed as an initial-value problem. However, the theory developed in \cite{peterhof1997exponential,sandstede2004defects} guarantees that stable and unstable manifolds exist for the quiescent state $\mathbf{0} \in \mathcal{X}$, providing the necessary phase-space structure for homoclinic orbits to exist. Beyond this, first recall from \S\ref{subsec:HomoclinicSnaking} that spatial dynamics interprets localized solutions as homoclinic orbits wrapping a periodic orbit to produce the localized SPPs. The same interpretation now holds here where the localized pattern comes from the homoclinic orbit of \eqref{e:spatial_dyn_stripes} wrapping around a periodic in $x$ solution $\gamma(x;y)$. Again, the theory in \cite{sandstede2004defects} guarantees the existence of stable and unstable manifolds to $\gamma(x;y)$ in the phase-space $\mathcal{X}$ that can be used to prove the existence of a homoclinic orbit that wraps around $\gamma$ (in $x$), producing the desired localized solution to \eqref{e:Swift-Hohenberg_2D_cubic_quintic}.  

With all these pieces, Beck et al. \cite{beck2009snakes} were able to extend their 1D analysis to prove the existence of localized snaking solutions to \eqref{e:Swift-Hohenberg_2D_cubic_quintic}. This work comes as a complete extension of their work in one spatial dimension, while providing the necessary hypotheses and verification thereof to achieve these results in some generality. Important to this review is that much, if not all, of the 1D theory can be carried over to the multidimensional spatial setting when only one variable is unbounded. Unfortunately, if $y$ in \eqref{e:Swift-Hohenberg_2D_cubic_quintic} is taken not to be periodic but infinitely extended in one or both directions then the necessary theory put forth in \cite{peterhof1997exponential,sandstede2004defects} does not apply. Thus, the spatial dynamics interpretation ultimately fails for understanding planar pattern formation outside of some limited scenarios.

\subsection{Orientation-Dependent Pinning}\label{subsec:LatticePinning}

Another investigation of patterns localized in a single direction on a planar landscape can be found in the work of Dean et al. \cite{dean2015} for lattice dynamical systems. Precisely, this work was concerned with spatially-discrete systems of the form
\begin{equation}\label{2Dlattice}
    \dot{U}_{n,m} = d[U_{n+1,m} + U_{n-1,m} + U_{n,m+1} + U_{n,m-1} - 4U_{n,m}] + F(U_{n,m},\mu), \quad (n,m)\in\mathbb{Z}^2,
\end{equation}
where $F(U,\mu)$ is any bistable nonlinearity. This includes the two bistable nonlinearities we have encountered with the SHE: $F(U,\mu) = - \mu U + \nu U^2 - U^3$ and $F(U,\mu) - \mu U + \nu U^3 - U^5$. System \eqref{2Dlattice} represents a standard 5-point discretization of the Laplace operator and is the 2D version of the lattice system encountered in \S\ref{subsubsec:Lattice1D}.

Patterns of the planar lattice system \eqref{2Dlattice} that are localized in one direction manifest themselves as planar monotone fronts and backs on the lattice glued together. That is, they are solutions of \eqref{2Dlattice} taking the form $U_{n,m} = A(n\cos(\psi) + m\sin(\psi),t)$, for an angle $\psi \in [0,2\pi)$ and profile $A(z,t) \to 0$ as $|z| \to \infty$. Under this ansatz, the system \eqref{2Dlattice} becomes
\begin{equation}\label{LatticePhi}
    \dot{A}(z,t) = d[A(z+\cos(\psi),t) + A(z-\cos(\psi),t) + A(z+\sin(\psi),t) + A(z-\sin(\psi),t) - 4A(z,t)] + F(A(z,t),\mu), 
\end{equation}
where we set $z = n\cos(\psi) + m\sin(\psi)$. Notice that if $\tan(\psi) = b/a$ for $(a,b)\in\mathbb{Z}^2\setminus\{0,0\}$ with $\gcd(|a|,|b|) = 1$ then we can interpret \eqref{LatticePhi} as a one-dimensional lattice equation indexed by the countable set
\begin{equation}
    \Phi = \bigg\{\frac{na + mb}{\sqrt{a^2 + b^2}}\bigg|\ (n,m)\in\mathbb{Z}^2\bigg\}.
\end{equation}
In contrast, if $\tan(\psi)\notin\mathbb{Q}$ then the set $\{n\cos(\psi) + m\sin(\psi)|\ (n,m) \in \mathbb{Z}^2\}$ is a dense subset of $\mathbb{R}$. Thus, unlike $\Phi$ above, there is no well-defined lattice spacing when $\tan(\psi)$ is irrational. Already one can see that orientation of the pattern plays a major role as rational directions, i.e. $\tan(\psi) \in \mathbb{Q} \cup \{\pm\infty\}$, result in a one-dimensional lattice equation, but irrational directions do not. In the context of front propagation, the regular lattice spacing in rational directions leads to wave pinning while irrational directions do not \cite{hoffman2010universality,hupkes2011propagation,chow1996dynamics,chow1998traveling}.

Using matched exponential asymptotics, \cite{dean2015} investigates the existence of solutions to bistable lattice equations that are localized in one direction. These results demonstrate that with $d \gg 1$ the snaking parameter regime (in $\mu$) of localized steady-state solutions to \eqref{LatticePhi} has nonzero width so long as $\tan(\psi) = b/a \in \mathbb{Q}\cup\{\pm\infty\}$. This region of parameter space depends explicitly on the orientation through the term $\sqrt{a^2 + b^2}$, wherein smaller effective lattice spacing $(a^2 + b^2)^{-1/2}$ leads to smaller existence regions in parameter space. This echoes similar results for propagating monotone fronts \cite{van1998traveling}. This work can similarly be applied to hexagonal lattice discretizations of the Laplace operator, where again one finds that orientations leading to well-defined one-dimensional lattice equations on the hexagonal lattice have exponentially small pinning regions with similar dependence on the effective lattice spacing.

\section{Axisymmetric Patterns}\label{sec:radial}

A natural way of extending the study of 1D patterns to higher spatial dimensions is to consider \emph{axisymmetric} patterns. As solutions to mathematical equations an axisymmetric pattern depends only on the distance from the origin in a multi-dimensional space, essentially making them one dimensional in a radial variable. Axisymmetric patterns abound in both experiment and theory, having been documented in dryland vegetation models \cite{Getzin2016FairyCircles,Meron2007,Kletter2012}, nonlinear optics \cite{mcsloy2002computationally,KIVSHAR1998Dark,menesguen2006optical,parra-rivas2021,Bortolozzo2009}, vibrating media \cite{Alnahdi2018,umbanhowar1996localized,Ma2016,Lioubashevski1999}, neural field equations \cite{Faye2013,Faye2013Poincare,rankin2014continuation}, phase-field crystal models \cite{Ophaus2021localizedPFC,Holl2021}, and several other examples from the physical and social sciences \cite{vanag2007,Barkman2020,Lloyd2013}. 

Mathematically, an axisymmetric pattern over the spatial domain $x \in \mathbb{R}^n$ takes the form $U(x) = U(r)$ where $r:=|x|\geq0$. For exposition, axisymmetric steady-state solutions of the $n$-dimensional SHE satisfy the ODE (now in the radial variable $r$) given by 
\begin{equation}\label{SHE_nDim}
    0 = - \bigg(1 + \frac{n - 1}{r}\partial_r + \partial^{2}_{r}\bigg)^2U - \mu U + \nu U^2 - U^3,
\end{equation}
where $U = U(r)$. Following as we did previously, we may introduce the variables 
\begin{equation}\label{SHE_Rad_ODE_vars}
    u_1 = U, \quad u_2 = \partial_r U, \quad u_3 = \bigg(1 + \frac{n - 1}{r}\partial_r + \partial_{r}^2\bigg)U, \quad u_4 = \partial_r \bigg(1 + \frac{n - 1}{r}\partial_r + \partial_{r}^2\bigg)U 
\end{equation} 
to result in the four-dimensional radial ODE 
\begin{equation}\label{SHE_spatial_Rad}
    \begin{split}
        u_1' &= u_2, \\
        u_2' &= u_3 - u_1 - \frac{n - 1}{r}u_2, \\
        u_3' &= u_4, \\
        u_4' &= -u_3 - \mu u_1 + \nu u_1^2 - u_1^3 - \frac{n - 1}{r}u_4,
    \end{split}
\end{equation}
where $' = \frac{\mathrm{d}}{\mathrm{d} r}$. Prior to moving on one should note the difference between \eqref{SHE_spatial_Rad} and \eqref{SHE_spatial_1D}. In spatial dimensions $n > 1$ one introduces an inhomogeneity and a singularity at $r = 0$ (equivalently $x = 0$) into the spatial ODE of the SHE. As we will review in this section, the subtle difference between \eqref{SHE_spatial_Rad} and its one-dimensional counterpart \eqref{SHE_spatial_1D} has drastic consequences on the approach and results for the existence and bifurcation structure of radially-localized patterns.

\subsection{Emergence of Axisymmetric Patterns}\label{sec:AxiEmerg}

A natural focal point for the analysis of solutions to PDEs is near bifurcation points, and this is exactly where we begin the review of axisymmetric patterns in the SHE. As we saw previously with localized patterns in 1D, the natural organizing center for the SHE is the Turing bifurcation point $\mu = 0$. The goal of works such as \cite{lloyd2009localized,mccalla2013spots} was to rigorously establish the existence of localized axisymmetric patterns to the SHE by demonstrating their emergence in a bifurcation at $\mu = 0$ from the trivial steady-state $U = 0$. Such states are interpreted as bounded solutions to the spatial ODE \eqref{SHE_spatial_Rad} whose maximum value over all of $r \geq 0$ scales with $|\mu| \ll 1$, and who converge exponentially fast to $(u_1,u_2,u_3,u_4) = (0,0,0,0)$ as $r \to \infty$.  

A major advancement toward applying spatial dynamics to examine the existence and bifurcation structure of axisymmetric patterns was the work of Scheel~\cite{scheel2003radially}. Scheel introduced a novel extension of standard dynamical systems techniques to systems such as \eqref{SHE_spatial_Rad}, which we now call {\em radial spatial dynamics}. There is a qualitative difference between the spatial eigenvalues of \eqref{SHE_spatial_Rad} when $r\ll1$ and $r\gg1$; small-amplitude solutions should exhibit algebraic behavior for small-to-moderate values of $r$ (the `core' region) and exponential behavior for large values of $r$ (the `far-field' region). Then, the set of localized axisymmetric solutions to \eqref{SHE_spatial_Rad} can be thought of as the intersection of the set of solutions to \eqref{SHE_spatial_Rad} that remain bounded in the core region and the set of solutions that decay exponentially to zero as $r\to\infty$ in the far-field region.

Using Scheel's radial spatial dynamics, Lloyd and Sandstede \cite{lloyd2009localized} proved the existence of axisymmetric, localized solutions to the planar ($n=2$) SHE \eqref{SHE_nDim}. One such solution is described in the following theorem. This solution is referred to as Spot A for reasons that will become clear shortly.

\begin{thm}[Spot A, {\cite[Theorem~2]{lloyd2009localized}}]\label{thm:SpotARad} 
	Fix $\nu\neq0$. Then, there exist constants $\mu_0,r_0,r_1 > 0$ such that \eqref{SHE_nDim} with $n = 2$ has a steady-state solution of the form $U(r) = u_A(r)$, where
\begin{equation}\label{RadialProfile:SpotA}
    u_{A}(r) = \frac{\sqrt{6}}{\nu \sqrt{\pi}}
    \begin{cases}
         \mu^{\frac{1}{2}}\sqrt{\frac{\pi}{2}}\,J_{0}(r) + \mathcal{O}(\mu) & 0\leq r\leq r_{0},  \\
         \mu^{\frac{1}{2}} r^{-\frac{1}{2}}\,\cos(r - \frac{\pi}{4}) + \mathcal{O}(\mu), & r_{0} \leq r \leq r_1\mu^{-\frac{1}{2}},\\
         \mu^{\frac{1}{2}} r^{-\frac{1}{2}}\,\mathrm{e}^{(r_1- \mu^{\frac{1}{2}}r)/2}\,\cos(r - \frac{\pi}{4}) + \mathcal{O}(\mu), & r_1\mu^{-\frac{1}{2}} \leq r,
    \end{cases}
\end{equation} 
for each $\mu\in(0,\mu_{0})$, and $J_{0}$ is the $0^\text{th}$ order Bessel function of the first kind.
\end{thm} 

\begin{figure}
    \centering
   \includegraphics[width=\linewidth]{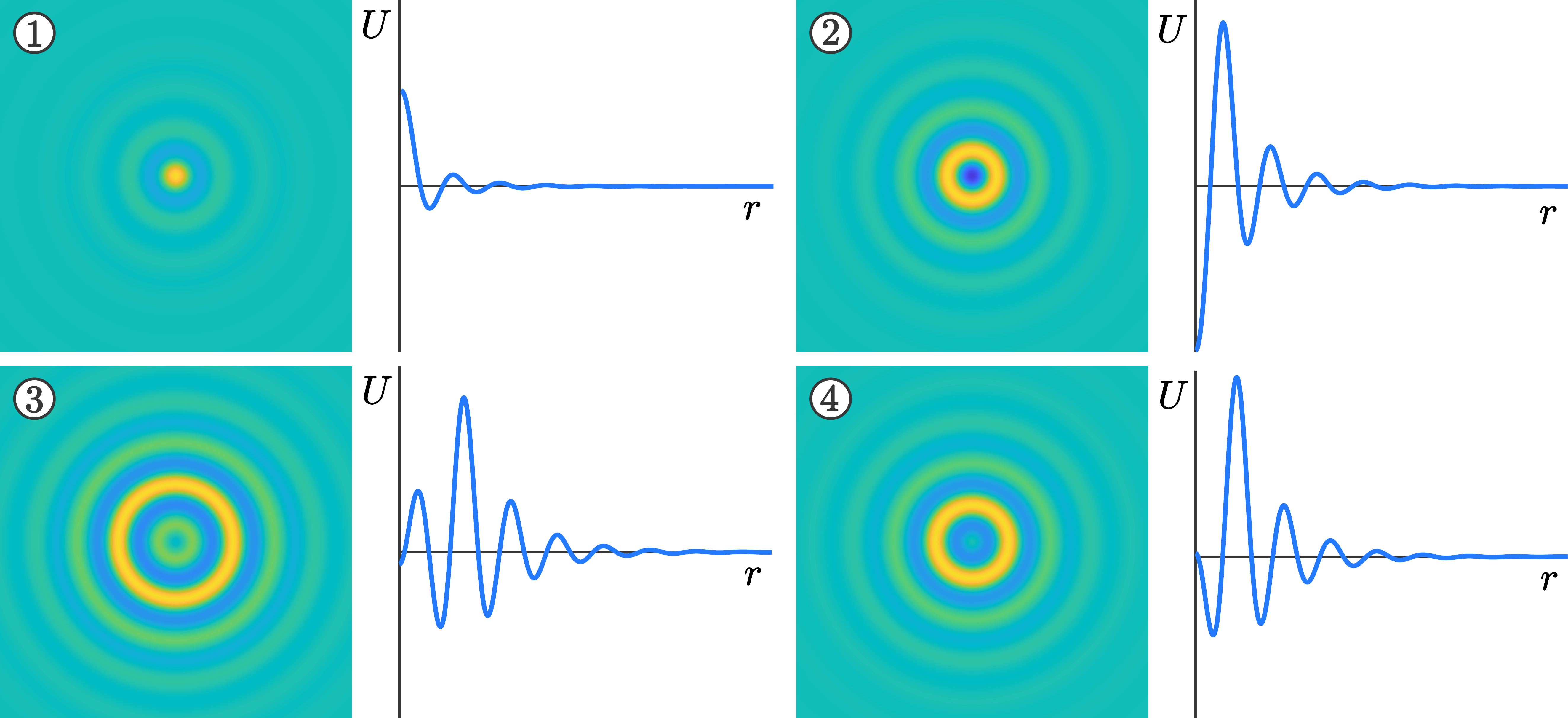}
    \caption{Contour plots and radial profiles of localized axisymmetric patterns. including (1) Spot A, (2) Spot B, (3) Ring A, and (4) Ring B. Each pattern is numerically obtained as a solution of \eqref{SHE_nDim} with $n=2$, $\gamma=1.6$ and $\mu\approx 0.08$. }
    \label{fig:Rad_SpotA}
\end{figure}

One should immediately note the $\mathcal{O}(|\mu|^\frac{1}{2})$ scaling of the core of the Spot A solution in Theorem~\ref{thm:SpotARad}. Such a scaling is typical of codimension-one saddle-node, pitchfork, and Turing bifurcations. However, Lloyd and Sandstede identified another type of solution, {\em Rings}, that exhibit a slightly unusual $\mathcal{O}(|\mu|^\frac{3}{4})$ scaling in the core instead. Spot A and Ring type solutions are compared in Figure~\ref{fig:Rad_SpotA}, where one notices that the Spot A solutions have a global maximum at $r = 0$. The ring vanishes at $r = 0$ and has global extrema away from the origin, leading to the ring structure that is exhibited in contour plots of the solution. 

The existence of rings relies on performing a normal form transformation in the far-field. The resulting leading-order approximation for the radial dynamics in the far-field are governed by the nonautonomous Ginzburg--Landau equation 
\begin{equation}\label{Ginzburg-Landau_Rad}
    0=a\left(\frac{\mathrm{d}}{\mathrm{d}s} + \frac{1}{2s}\right)^2 q(s) - q(s) - b |q(s)|^2 q(s) + \mathcal{O}(\mu^{\frac{1}{2}})
\end{equation}
where $s:=\mu^{\frac{1}{2}}r$ is the rescaled radial coordinate in the far-field for $\mu \geq 0$. A key difficulty in studying localized axisymmetric patterns in \eqref{SHE_nDim} lies in proving the existence of exponentially localized solutions to this Ginzburg--Landau equation. Precisely, the combination of nonautonomous and nonlinear terms results in a differential equation that is intractable to standard existence theorems. The existence of a localized solution to \eqref{Ginzburg-Landau_Rad} was consequently left as an unproven hypothesis in early works on localized axisymmetric patterns \cite{lloyd2009localized,mccalla2013spots}, until van den Berg et al. verified this hypothesis with a computer-assisted proof \cite{vandenberg2015Rigorous}. They showed that so long as 
\begin{equation}\label{nuSubcritical}
    \frac{3}{4} -\frac{19}{18}\nu^2 < 0 \implies |\nu| > \sqrt{27/38}   
\end{equation} 
there exists a nontrivial solution on $0 \leq r < \infty$ with the following asymptotic expansions
\begin{equation}\label{GinzbLand:asymp}
    q(r) = \begin{cases}
        q_0 r^{\frac{1}{2}} + \mathcal{O}(r^{\frac{3}{2}}), &r\to0,\\
        \left(q_+ + \mathcal{O}(\mathrm{e}^{-r/2})\right) r^{-\frac{1}{2}}\mathrm{e}^{-r/2}, &r\to\infty,
    \end{cases}
\end{equation}
for some constants $q_0>0$, $q_{+}\neq0$. With this solution $q(r)$ we now provide the existence result for rings. 

\begin{thm}[Rings, {\cite[Theorem~1]{lloyd2009localized}}]\label{thm:RingRad} 
	Fix $|\nu|>\sqrt{27/38}$ and let $q$ be the solution of \eqref{Ginzburg-Landau_Rad} with asymptotics \eqref{GinzbLand:asymp}. Then, there exist constants $\mu_0,r_0,r_1,q_0 > 0$ such that \eqref{SHE_nDim} with $n = 2$ has two steady-state solutions of the form $U(r) = \pm u_R(r)$, where
\begin{equation}\label{RadialProfile:Ring}
    u_{R}(r) = \begin{cases}
         2\mu^{\frac{3}{4}}\sqrt{\frac{\pi}{2}} q_0r J_{1}(r) + \mathcal{O}(\mu), & 0\leq r\leq r_{0},  \\
         2\mu^{\frac{3}{4}} q_0 r^{\frac{1}{2}} \sin(r - \frac{\pi}{4}) + \mathcal{O}(\mu), & r_{0} \leq r \leq r_1\mu^{-\frac{1}{2}},\\
         2\mu^{\frac{1}{2}}q(\mu^{\frac{1}{2}}r)\sin(r - \frac{\pi}{4}) + \mathcal{O}(\mu), & r_1\mu^{-\frac{1}{2}} \leq r,
    \end{cases}
\end{equation} 
for each $\mu\in(0,\mu_{0})$, and $J_{1}$ is the $1^\text{st}$ order Bessel function of the first kind.
\end{thm} 

Notice that these solutions bifurcate from $\mu = 0$ in pairs, giving that rings arise via a pitchfork bifurcation from the Turing instability in the SHE. Moreover, the condition $|\nu| > \sqrt{27/38}$ arises from \eqref{nuSubcritical} and translates to the requirement that the SPPs (Turing patterns) in the SHE bifurcate subcritically. In what follows we refer to solutions along the bifurcating $u_R(r)$ branch as Ring A and the solutions along the $-u_R(r)$ branch as Ring B. 

Shortly after Theorems~\ref{thm:SpotARad} and \ref{thm:RingRad} were published, numerical continuations of these patterns in the SHE revealed the existence of another type of localized solution: Spot B profiles \cite{mccalla2010snaking}, as shown in panel (2) of Figure~\ref{fig:Rad_SpotA}. The results of these continuations will be the subject of the next subsection, but what is most important for now is that the Spot B solutions are entirely distinct from the Spot A and ring solutions. Like the other localized axisymmetric solutions in this section, they emerge from the Turing bifurcation point $\mu = 0$, but their distinction comes from the fact that they exhibit a $\mathcal{O}(|\mu|^{\frac{3}{8}})$ scaling for small $0 < \mu \ll 1$. The existence of these patterns was proven in \cite{mccalla2013spots} and again relies on the existence of the solution $q(r)$ to \eqref{Ginzburg-Landau_Rad}. The result is summarized in the following theorem.

\begin{thm}[Spot B, {\cite[Theorem~1]{mccalla2013spots}}]\label{thm:SpotBRad}
    Fix $|\nu|>\sqrt{27/38}$. Then, there exist constants $\mu_0,r_0 > 0$ such that \eqref{SHE_nDim} with $n = 2$ has a steady-state localized solution $U(r) = u_B(r)$ for all $\mu \in (0,\mu_0)$ with   
    \begin{equation}
        u_B(r) \sim -\mu^{\frac{3}{8}} J_{0}(r) + \mathcal{O}(\sqrt{\mu})
    \end{equation}
    uniformly on bounded intervals $[0,r_0]$ as $\mu \to 0$. 
\end{thm}

The above results have focused only on the planar $(n = 2)$ SHE. However, many of the techniques developed for the planar system carry over to the SHE in higher spatial dimensions. This was demonstrated by McCalla and Sandstede who further proved the existence of Spot A and Spot B solutions that bifurcate from $\mu = 0$ in the SHE with $n = 3$ \cite{mccalla2013spots}. Both have profiles that are given by the spherical Bessel function $j_0(r) = \sin(r)/r$, as opposed to the Bessel function $J_0(r)$ when $n = 2$ and again Spot A solutions exhibit the $\mathcal{O}(|\mu|^\frac{1}{2})$ scaling. With $n = 3$ the Spot B solutions now exhibit a $\mathcal{O}(|\mu|^\frac{1}{4})$ scaling. In fact, these results indicate that Spot B patterns in $\mathbb{R}^n$ should scale like $\mathcal{O}(|\mu|^\frac{5 - n}{8})$. Along with the provided results for $n = 2,3$, this agrees with the fact that for $n = 1$ there are two solutions that bifurcate from $\mu = 0$ with $\mathcal{O}(|\mu|^\frac{1}{2})$ scaling. One barrier to using radial spatial dynamics to prove the existence of axisymmetric solutions in the SHE with $n \geq 4$ is determining the profile of the solutions, potentially as a function of $n$. These profiles are determined by the expansion of the core manifold near $r = 0$, while finding and matching with bounded nontrivial solutions of the far-field equations may again require bounded solutions of \eqref{Ginzburg-Landau_Rad} or a related equation. Finally, rings have not been proven to bifurcate from $\mu = 0$ in the SHE for any $n > 2$ and so their existence in higher space dimensions remains an open question.

We briefly mention that away from the Turing instability, axisymmetric localized spots have been extensively investigated in spatially singularly perturbed PDEs; see for example~\cite{Chen2011,Heijster2014,avitabile2018spot,Ward_2018,Wong2020,Wong2022,Byrnes2023,Wei2003,Wei1999} both in terms of existence and (in)stability analysis. These spots can be stationary or translating, and rely on a core (known as either inner/fast solutions) and far-field (outer/slow solutions) matching procedure using either matched asymptotics or geometric singular perturbation theory. Localized rings have also been studied in near the singular limit~\cite{Kolokolnikov2006,Moyles2017,Kolokolnikov2022} using similar techniques. Localized spots have also been investigated on the sphere~\cite{Rozada2014,Jamieson2016}.

\subsection{Axisymmetric Snaking Branches}\label{subsec:RadialSnaking}

\begin{figure}[t]
    \centering
    \includegraphics[width=0.7\linewidth]{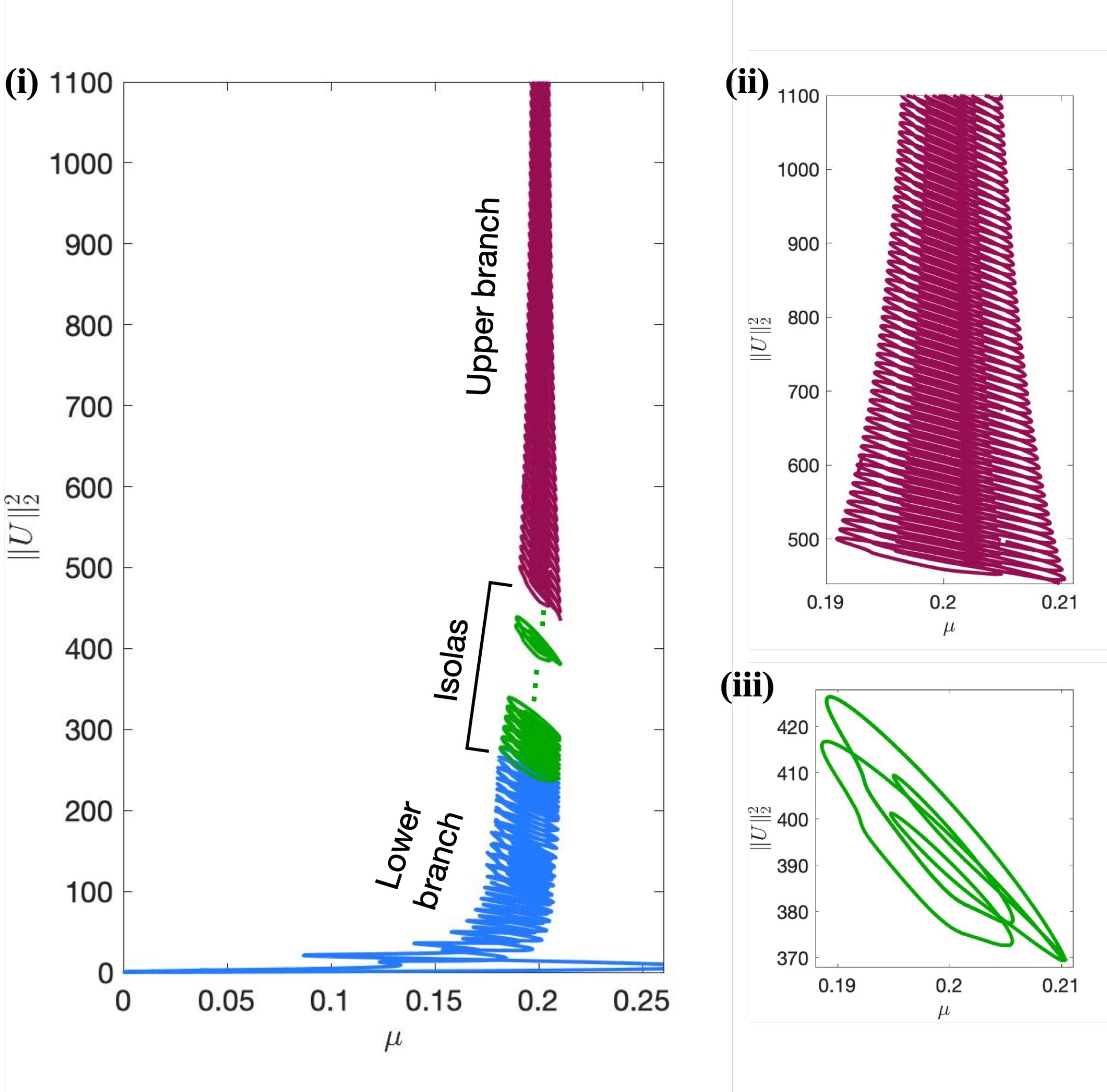}
    \caption{Snaking bifurcation branches of localized Spot A patterns in to the planar ($n = 2$) SHE \eqref{SHE_nDim}. (i) The full bifurcation diagram is fragmented into three distinct components: a lower snaking branch (blue), numerous isolas (green), and an upper snaking branch (purple). Only a representative few of the isolas are plotted as there are $>20$ which appear stacked upon each other. (ii) The upper snaking branch appears to narrow towards the Maxwell point at $\mu = 0.2$ in this case. (iii) A single isola from the bifurcation diagram. Spot B patterns exhibit a similar fragmented bifurcation diagram; see \cite{mccalla2010snaking}.}
    \label{fig:Snaking_SHE2D}
\end{figure}

Continuing the axisymmetric patterns of the previous subsection from $\mu \approx 0$ into larger values of $\mu$ reveals a complex bifurcation structure that bears little resemblance to the familiar 1D snakes and ladders profiles \cite{mccalla2010snaking}. The situation appears to be qualitatively the same for both dimensions $n = 2,3$ and Figure~\ref{fig:Snaking_SHE2D} presents a numerically obtained bifurcation diagram for visual reference. Initially the Spot A and Spot B branches follow the usual snaking structure wherein fold bifurcations flip back and forth over a compact region of parameter space with the folds appearing to align along two vertical asymptotes. Moreover, like the 1D snaking phenomenon, SPPs are added that extend the region of localization for the axisymmetric pattern, in turn causing the $L^2$-norm of the solution to increase, thus resulting in the vertical ascension in the bifurcation diagram. 

\begin{figure}[t]
    \centering
    \includegraphics[width=0.9\linewidth]{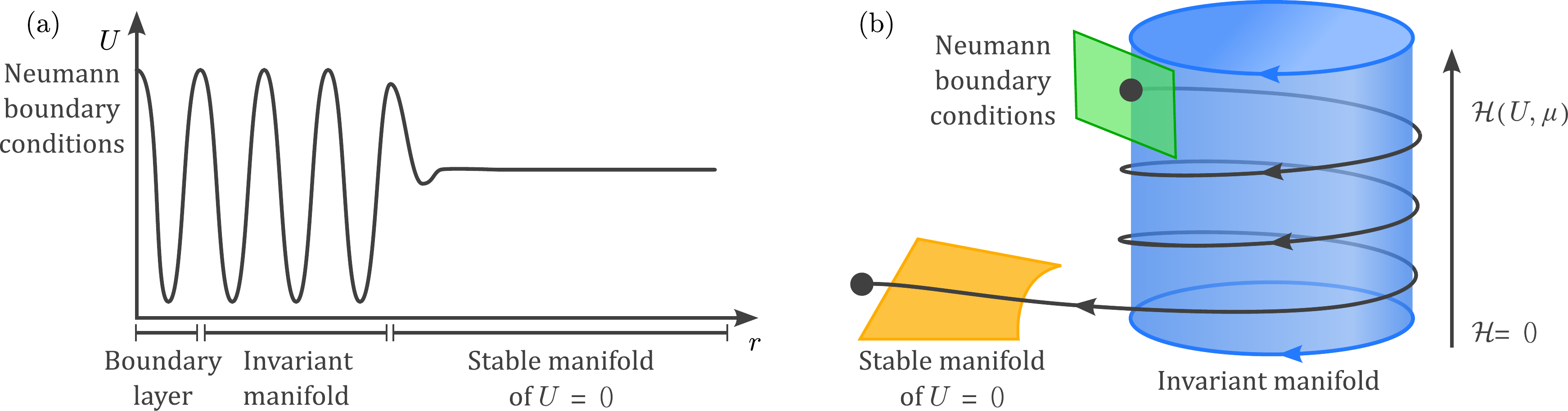}
    \caption{Left: The anatomy of a radial SPP solution used in the proofs of \cite{bramburger2019localized}. Right: In the phase-space of \eqref{SHE_spatial_Rad} localized radial SPPs wrap around an invariant cylinder in phase-space for a finite time to produce the SPP. }
    \label{fig:1+EpsProfile}
\end{figure}

This vertical growth along the bifurcation diagram is not long lived. Eventually both the Spot A and Spot B branches reach a maximal finite height in the bifurcation diagram where they turn into rings and begin decreasing in norm by subtracting SPPs. Following \cite{mccalla2010snaking}, Spot A connects to the Ring A branch and Spot B the Ring B branch, as defined in the previous subsection. Above both of these terminal snaking branches in the bifurcation diagram lie stacks of isolas - closed curves in the bifurcation diagram. Along each isola the solution profiles change in intricate ways and switch between both ring- and spot-like patterns. In both 2D and 3D there are only finitely many isolas with the 2D patterns extending vertically in the bifurcation diagram until they have approximately 38 SPPs, while in 3D the isolas terminate with fewer SPPs. Finally, above the isolas there is another snaking branch which consists of two intertwined arms that appear to ascend vertically in the bifurcation diagram without bound. This snaking branch is referred to as the {\em upper snaking branch}, while the branch containing Spot A or Spot B patterns that emerges from $(U,\mu) = (0,0)$ is termed the {\em lower snaking branch}. The upper branch differs significantly from the lower branch in that SPPs are added near the core of the pattern, i.e. near $r = 0$, as opposed to at the tail like in 1D. Based on numerical observations, McCalla and Sandstede further conjectured in \cite{mccalla2010snaking} that the upper branch collapses onto a single parameter value as the number of SPPs in the localized pattern goes to infinity.    
Bramburger et al. \cite{bramburger2019localized} attempted to better understand the bifurcation structure of localized axisymmetric patterns in the two- and three-dimensional SHE from an analytical perspective. As in the previous section, their work focused on the radial spatial dynamical system \eqref{SHE_spatial_Rad}. When $n = 1$ the nonautonomous terms disappear, reverting back to the 1-dimensional setting \eqref{SHE_spatial_1D} where all dynamics take place inside the level sets of the Hamiltonian function \eqref{SHE_Hamiltonian}.
Importantly, when $n = 1$ SPPs of the SHE \eqref{SHE_nDim} correspond to periodic orbits of \eqref{SHE_spatial_Rad} which is only a slice of a `stack' of periodic orbits forming a normally hyperbolic invariant manifold $\mathcal{P}$ that can be parameterized by the value of $\mathcal{H}$. The result is a cylinder of stacked periodic orbits in phase-space, as illustrated in Figure~\ref{fig:1+EpsProfile}. Due to the conserved quantity $\mathcal{H}$, localized SPPs of the Swift--Hohenberg equation in 1D can only be constructed through connections with the periodic orbit in $\mathcal{H}^{-1}(0)$, the zero level set of $\mathcal{H}$, to the rest state $U = 0$ for each fixed $(\mu,\nu)$. 

For any $n > 1$ the function $\mathcal{H}$ is no longer a conserved quantity of \eqref{SHE_spatial_Rad}. However, when $\mathcal{O}(|n-1|/r)$ is small we expect that system \eqref{SHE_spatial_Rad} presents itself as a perturbation from the $n = 1$ equation, and therefore $\mathcal{P}$ will persist as an integral manifold in the nonautonomous, non-conserved phase-space. It is important to note however that the dynamics on the perturbed invariant manifold will no longer be periodic, which potentially causes trajectories to leave the perturbed cylinder after a finite amount of time through its top or bottom.  

To better understand the system \eqref{SHE_spatial_Rad} (and in turn axisymmetric localized patterns), McCalla and Sandstede proposed introducing a ``dimensional perturbation'' in the form of $n = 1 + \varepsilon$ \cite{mccalla2010snaking}. Although meaningless in the context of the dimension of the underlying space, when $0 < \varepsilon \ll 1$ this dimensional perturbation casts \eqref{SHE_spatial_Rad} as a true perturbation from the well-understood spatial system \eqref{SHE_spatial_1D} in 1D. In particular, \cite[Theorem~2.7]{bramburger2019localized} shows that snaking branches persist for plateau lengths $L \leq \mathrm{exp}(b/\varepsilon)$, with constant $b > 0$. Such an upper bound is the result of determining the maximal number of times a trajectory can be guaranteed to wrap around the perturbed cylinder in complete generality. Surprisingly, numerical continuations of localized radial solutions to \eqref{SHE_spatial_Rad} at fixed values of $\varepsilon$ reveal that the upper bound $\mathrm{exp}(b/\varepsilon)$ closely tracks with the maximum of the lower snaking branches, even well beyond the asymptotic regime $0 < \varepsilon \ll 1$ and into $\varepsilon = 1,2$ (dimensions $n = 2,3$).

Bramburger et al. \cite{bramburger2019localized} also provide results on the nature of the snaking branches of the system \eqref{SHE_spatial_Rad} in dimensions $n = 2,3$. Using a combination of analytical and numerical results, it is shown that the snaking branches of the SHE must collapse onto the Maxwell point as one ascends the bifurcation diagram. This explains the upper snaking branch becoming increasingly localized about the Maxwell point and further proves the conjecture of \cite{mccalla2010snaking}.

There are a number of questions that remain outstanding related to the bifurcation curves of localized radial patterns. Primary among them is that there is still no explanation for the intermediate stack of isolas that splits the lower and upper snaking branches. Similarly, it is not clear why the upper branch forms a connected curve or if it is even the uppermost branch in the bifurcation diagram. That is, it is entirely possible that the observed upper branch terminates much like the lower branch at a finite height in the bifurcation diagram, possibly giving way to more isolas or another snaking branch. Such a finding would not necessarily be surprising as the work \cite{lloyd2009localized} originally presented evidence that the lower branch extends indefinitely and it wasn't until \cite{mccalla2010snaking} that the termination of the lower branch was observed far up in the bifurcation diagram. Finally, the reason for the difficulty in predicting the specific shape of the snaking diagram with $n = 2,3$ is that numerical investigations indicate that the structure is driven by the behavior of solutions away from the invariant manifold of periodic orbits. This means that the curves depend on the global dynamics of the dynamical system \eqref{SHE_spatial_Rad} and not just the local properties near the perturbed cylinder.

\subsection{Time Periodic Patterns}

Throughout this section we have only discussed time-independent localized solutions. This is primarily due to the fact that time-dependent patterns cannot be put into the spatial dynamics formulation, as was done for the SHE \eqref{SHE_spatial_Rad}. Nonetheless, as we saw in \S\ref{subsec:1DOscillons}, there is significant documentation in the mathematical and physical literature of localized time-periodic patterns, termed oscillons and/or breathers. In dimensions greater than 1, notable numerical documentation includes the axisymmetric and non-axisymmetric localized oscillons in \cite{Alnahdi2018} and the localized hexagon oscillons in \cite{coyle2023localised}. 

Oscillons as solutions to PDEs present a serious challenge mathematically as even those that are axisymmetric must be functions of both space and time to account for the time-periodicity. For this reason, analytical results on oscillations are sparse in the mathematical literature. One analytical result showed that the nonlinear Schr\"odinger equation proposed in \cite{miles1984parametrically} to model Newtonian fluids supports ring-type oscillons in two and three spatial dimensions \cite{barashenkov2002two}. Moreover, these ring oscillons are shown to be stable. Weakly nonlinear analysis has further provided insight into the existence of oscillons in Ginzburg--Landau models in one spatial dimension \cite{dawes2010localized,burke2008classification}.

One method of approaching the existence of axisymmetric oscillons is to return to the forced complex Ginzburg--Landau equation \eqref{e:FCGL}. In higher spatial dimensions this PDE takes the form
\begin{equation}\label{OscillonCGL}
    U_t = (1 + \mathrm{i}\alpha)\Delta U + (- \mu + \mathrm{i}\omega)U - (1 + \mathrm{i}\beta)|U|^2U + \gamma \bar{U},
\end{equation}
where $\alpha,\beta,\gamma,\mu,\omega \in \mathbb{R}$, $U = U(x,t) \in \mathbb{C}$, and $x \in \mathbb{R}^n$. Recall from \S\ref{subsec:1DOscillons} that although not explicitly time-periodic, \eqref{OscillonCGL} is argued to be an amplitude equation for periodically forced systems near a Hopf bifurcation. Assuming radial symmetry, i.e. $U(x) = U(r)$ only, we find that \eqref{OscillonCGL} is transformed to a complex ODE in the single spatial variable $r = |x| \geq 0$. McQuighan and Sandstede extended the 1D  existence of monotone tailed oscillons (as mentioned in \S\ref{subsec:1DOscillons}) of the first type to two spatial dimensions with the following theorem:

\begin{thm}[Radial oscillons, {\cite[Theorem~1]{mcquighan2014oscillons}}]\label{thm:OscillonRad}
    Fix $\alpha \omega < \mu < \beta \omega$ and let $\gamma = \sqrt{\mu^2 + \omega^2} - \varepsilon^2$. Then there is an $\varepsilon_0 > 0$ so that \eqref{OscillonCGL} with $x \in \mathbb{R}^2$ has a nontrivial localized radial solution of amplitude $\mathcal{O}(\varepsilon)$ for each $\varepsilon \in (0,\varepsilon_0)$.
\end{thm}

\begin{figure}[t]
    \centering
    \includegraphics[width=0.7\linewidth]{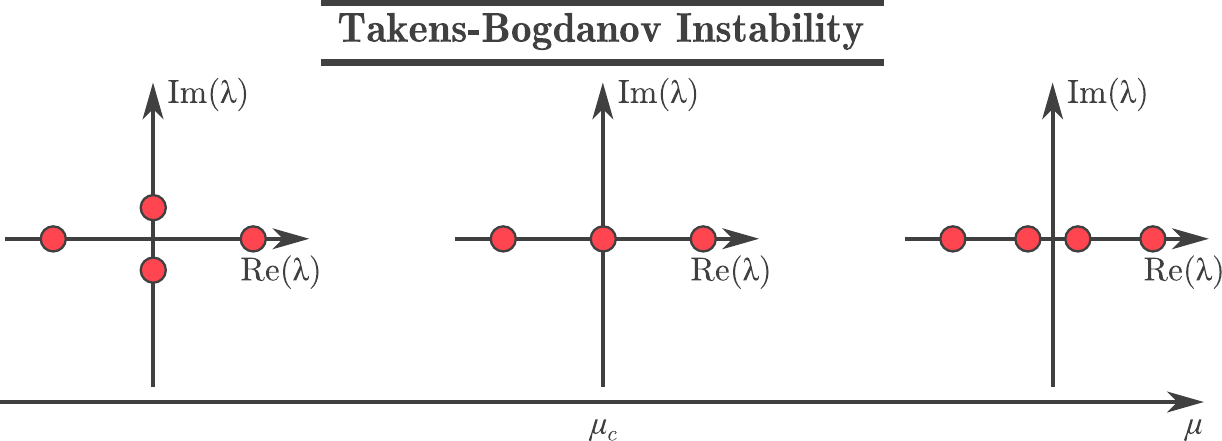}
    \caption{Spatial eigenvalues for the reversible Takens--Bogdanov instability.}
    \label{fig:TB-transition}
\end{figure}

This result again makes use of radial spatial dynamics and provides the existence as emergence from a bifurcation. In the setting of \eqref{OscillonCGL}, the curve $\gamma = \sqrt{\mu^2 + \omega^2}$ corresponds to a reversible Takens--Bogdanov bifurcation in the associated radial spatial dynamical systems; see Figure~\ref{fig:TB-transition}. The condition $\mu < \beta \omega$ provides that the bifurcation is subcritical, while we recall that similar subcriticality conditions were necessary for the existence of rings and the Spot B solution in Theorems~\ref{thm:RingRad} and \ref{thm:SpotBRad}, respectively. The difference is that in the SHE the centralizing bifurcation is of Hopf type which leads to the Turing patterns in space. In both cases, it is the subcriticality that leads to localized pattern formation. The condition $\mu > \alpha \omega$ provides that solutions in the far-field decay monotonically to zero as $r \to \infty$, leading to the monotonic tails of solutions coming from Theorem~\ref{thm:OscillonRad}.   

Further numerical investigations of \eqref{OscillonCGL} reveal a wealth of localized patterns. In the parameter region $\mu < \alpha \omega$ the far-field equations exhibit damped oscillations that decay to zero, which in turn can lead to the existence of localized solutions with oscillatory tails. Further explorations have revealed the presence of reciprocal oscillons exhibiting both monotone and oscillatory tails. The existence of these oscillon patterns remains to be proven, while a rigorous justification of \eqref{OscillonCGL} as a normal form for periodically forced RD systems near a supercritical Hopf bifurcation is also lacking. Proofs in either of these directions would represent a major advancement in the theory of oscillons.

\section{Fully Localized Planar Patterns}\label{ssec:FullyLocal}

\begin{figure}[htb]
    \centering
    \includegraphics[width=0.6\linewidth]{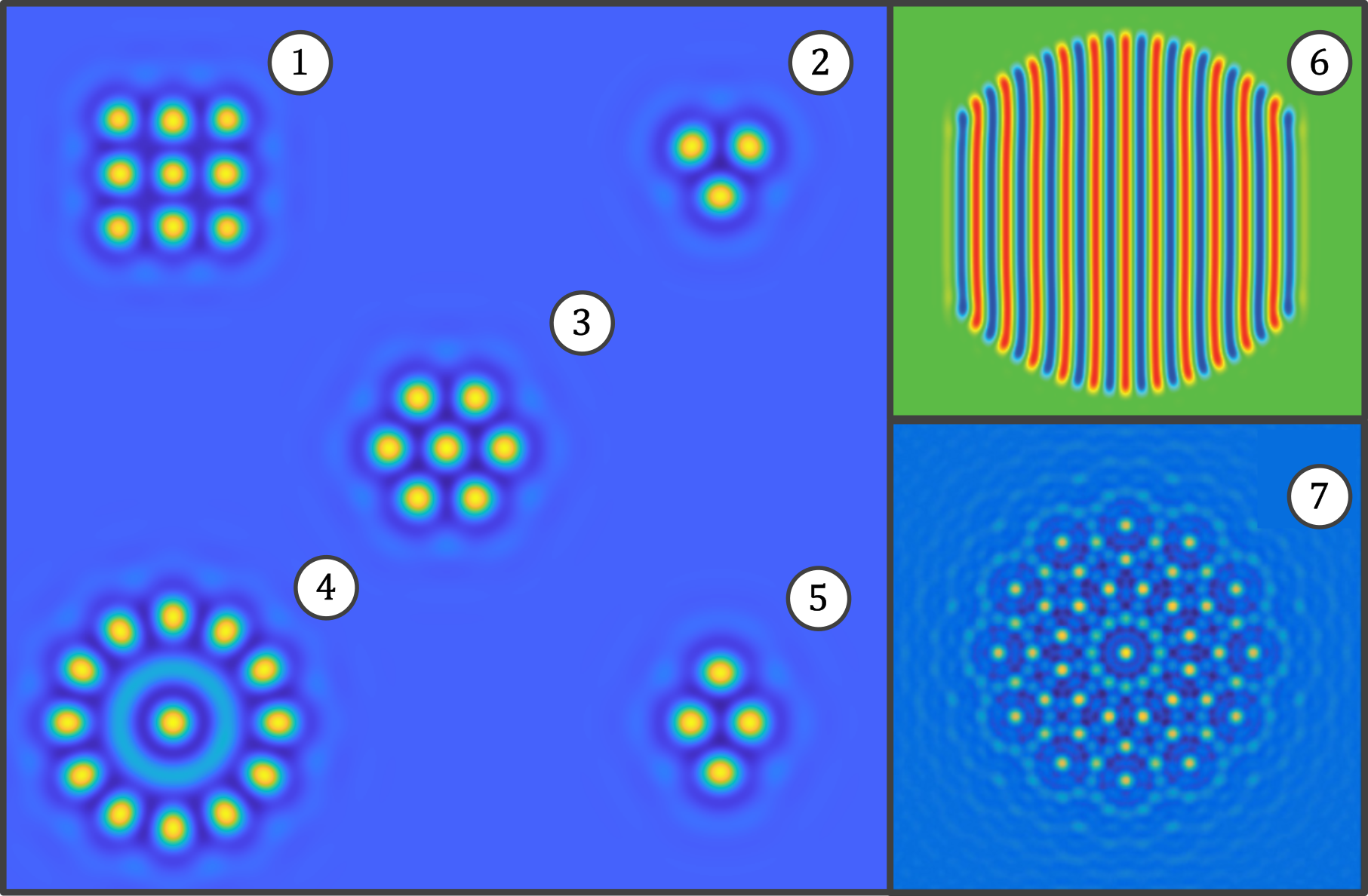}
    \caption{Fully localized planar patterns, including (1) squares, (2) triangles, (3) hexagons, (4) dodecagons, (5) rhomboids, (6) stripes (from \cite{avitabile2010snake}), and (7) quasipatterns (from \cite{Subramanian2018LocalizedPFC})}
    \label{fig:LocHexPatch}
\end{figure}

In this section, we turn our attention to fully localized planar patterns that are not radially-symmetric. Examples are shown in Figure~\ref{fig:LocHexPatch}, including the now famous localized hexagon patches found in the SHE \cite{lloyd2008localized}. In the SHE alone there is further documentation of localized patterns exhibiting rhombic and square symmetries \cite{zimanowski1997britte,sakaguchi1996stable}, while similar localized patterns exhibiting dihedral symmetries in other pattern-forming RD systems abound \cite{Vladimirov2002Clusters,mcsloy2002computationally,menesguen2006optical,Subramanian2018LocalizedPFC,Ophaus2021localizedPFC,buffoni2018variational,buffoni_groves_wahlen_2022,rankin2014continuation,umbanhowar1996localized,Aranson1998localizedGranular,lo_jacono_bergeon_knobloch_2013,sakaguchi1997stable,lloyd2015homoclinic,Hertkorn2021QuantumFerrofluids}. Notable examples come from vegetation models \cite{vonHardenberg2001,JAIBI2020vegetation} where these localized patterns represent symmetrically-arranged patches of dense vegetation surrounded by sparser ones. 

We have seen over the past two sections that some approaches to planar patterns rely on reducing the problem to one that is similar to the 1D setting. The case of fully localized patterns that are neither axisymmetric nor bounded in one direction  unfortunately does not exhibit a similar reduction to a 1D problem, and consequentially, has no spatial dynamics formulation. Due to the lack of a spatial dynamics interpretation, analytic approaches to localized patterns in 2D have been scattered and are significantly less developed than their 1D counterparts. Investigations typically require one to fixate on a limited parameter regime, such as near a Turing bifurcation, or deploy some creativity in studying proxy systems that seem to explain some of the observed phenomena.  

Here we will review the creative investigations of and approaches to fully localized (non-radial) patterns. We begin by reviewing the recent approach from \cite{hill2023approximate,hill2022dihedral} in \S\ref{subsec:Dihedral} to better understand the emergence of localized patterns with dihedral symmetries from the Turing bifurcation point $\mu = 0$ in the SHE. Then, we proceed to \S\ref{subsec:HexagonPatch} with a review of the extensive numerical investigation of localized hexagon patches. This subsection has a particular focus on the work \cite{lloyd2008localized} that inspired much of the work in the past decade on localized planar patterns. We conclude in \S\ref{subsec:SquareLattice} with a review of localized patterns to lattice dynamical systems of the form \eqref{2Dlattice}.

\subsection{Emergence of Localized Dihedral Patterns}\label{subsec:Dihedral}

The works \cite{hill2022dihedral,hill2023approximate} attempted to predict the types of fully localized patterns one should expect to emerge in RD systems from a Turing bifurcation. All of what follows holds for general two-component RD equations under standard Turing bifurcation hypotheses, but as in the previous sections we will focus exclusively on the SHE for illustration. We consider {\em localized dihedral patterns}, meaning fully localized patterns that are invariant under the action of a dihedral symmetry group, $\mathbb{D}_m$ for some $m \geq 2$. Writing the planar variables $(x,y)\in\mathbb{R}^2$ in polar form $(r,\theta)$, this means that a steady-state localized dihedral pattern to a planar evolution equation satisfies the rotational symmetry condition $U(r,\theta + 2\pi/m) = U(r,\theta)$ and the flip condition $U(r,-\theta) = U(r,\theta)$. Periodicity in the $\theta$-component along with the symmetry constraints imply that a dihedral pattern can be expanded as a cosine series
\begin{equation}\label{CosineExpansion}
    U(r,\theta) = u_0(r) + 2\sum_{n = 1}^\infty u_n(r)\cos(n m \theta). 
\end{equation}
The above Fourier decomposition essentially makes all phase dependence in $U(r,\theta)$ trivial and requires one only to identify the amplitudes $u_n(r)$, $n = 0,1,2,\dots$. The drawback, of course, is that substituting this expansion into the planar SHE \eqref{SwiftHohenberg} results in an infinite system of coupled nonautonomous ODEs with independent variable $r$.   

Since an infinite system of ODEs is generally difficult to manage, \cite{hill2023approximate,hill2022dihedral} introduced a relaxation of the problem that helps to inform what should be expected of localized dihedral patterns. Motivated by numerical treatments of pattern-forming systems, they truncate the expansion \eqref{CosineExpansion} at a finite value $N \geq 1$, which when put into the planar quadratic-cubic SHE \eqref{SwiftHohenberg} results in the coupled fourth-order ODEs
\begin{equation}\label{SHE_Galerkin}
    0 = -\bigg(1 + \partial_r^2 + \frac{1}{r}\partial_r - \frac{(mn)^2}{r^2}\bigg)^2u_n - \mu u_n + \nu\sum_{i + j =n}u_{|i|}u_{|j|} - \sum_{i + j + k = n} u_{|i|}u_{|j|}u_{|k|}, \qquad n = 0,1,\dots, N,  
\end{equation}
where the summations indices run over $i,j,k = -N,\dots, N$. For any finite $N$, system \eqref{SHE_Galerkin} provides a Galerkin expansion that can at best only approximate a solution to the true SHE. However, the results that follow below for $N \geq 1$ can be used as an initial guess to perform numerical continuation of localized dihedral patterns and in so doing identify previously unidentified phenomena in the realm of localized pattern formation. 

Using radial normal form theory, \cite{hill2021localised} provides the following theorem which describes the emergence of Spot A-type localized dihedral patterns in \eqref{SHE_Galerkin}. Such solutions emerge from the Turing point $\mu = 0$, much like the axisymmetric solutions discussed in Section~\ref{sec:AxiEmerg}. 

\begin{thm}[{\cite[Theorem~2.2]{hill2021localised}}]\label{thm:SpotCell} 
	Fix $m,N \in \mathbb{N}$ and assume the constants $\{a_{n}\}_{n=0}^{N}$ are nondegenerate solutions of the nonlinear matching condition 
		\begin{equation}\label{MatchEqSpot}
    a_{n} = 2\sum_{j=1}^{N-n} \cos\left(\frac{m\pi(n-j)}{3}\right) a_{j} a_{n+j} + \sum_{j=0}^{n} \cos\left(\frac{m\pi(n-2j)}{3}\right) a_{j}a_{n-j},
\end{equation}
    for each $n = 0,1,\dots, N$. Then, there exist constants $\mu_0,r_0,r_1 > 0$ such that the system \eqref{SHE_Galerkin} has a solution of the form
\begin{equation}\label{RadialProfile}
    u_{n}(r) = \frac{\sqrt{6}\, a_{n}}{\nu \sqrt{\pi}}(-1)^{m n}
    \begin{cases}
         \mu^{\frac{1}{2}}\sqrt{\frac{\pi}{2}}\,J_{mn}(r) + \mathcal{O}(\mu) & 0\leq r\leq r_{0},  \\
         \mu^{\frac{1}{2}} r^{-\frac{1}{2}}\,\cos( \psi_n(r)) + \mathcal{O}(\mu), & r_{0} \leq r \leq r_1\mu^{-\frac{1}{2}},\\
         \mu^{\frac{1}{2}} r^{-\frac{1}{2}}\,\textnormal{e}^{(r_1- \mu^{\frac{1}{2}}r)/2}\,\cos( \psi_n(r)) + \mathcal{O}(\mu), & r_1\mu^{-\frac{1}{2}} \leq r,
    \end{cases}
\end{equation} 
for each $\mu\in(0,\mu_{0})$, $n = 0,1,\dots,N$, where $\psi_n(r):= r - \frac{mn \pi}{2} - \frac{\pi}{4}$ and $J_{mn}$ is the $(mn)^\mathrm{th}$ order Bessel function of the first kind.
\end{thm} 

\begin{figure}[t]
    \centering
    \includegraphics[width=\linewidth]{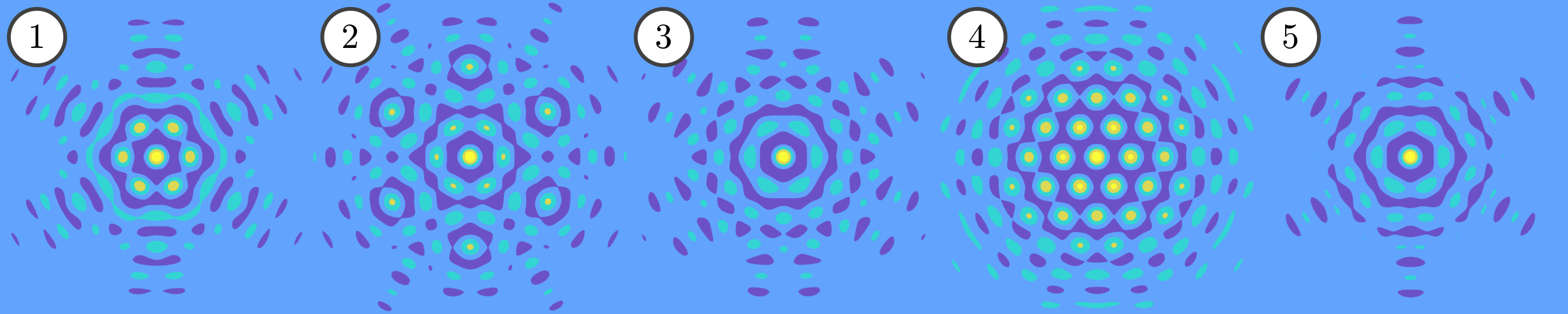}
    \caption{Solving \eqref{MatchEqSpot} with $(m,N) = (6,3)$ leads to the five distinct solutions in the form \eqref{RadialProfile} shown here for $\mu = 0.0006$ }
    \label{fig:DihedralSpotSolns}
\end{figure}

\begin{figure}[ht!]
    \centering
    \includegraphics[width=\linewidth]{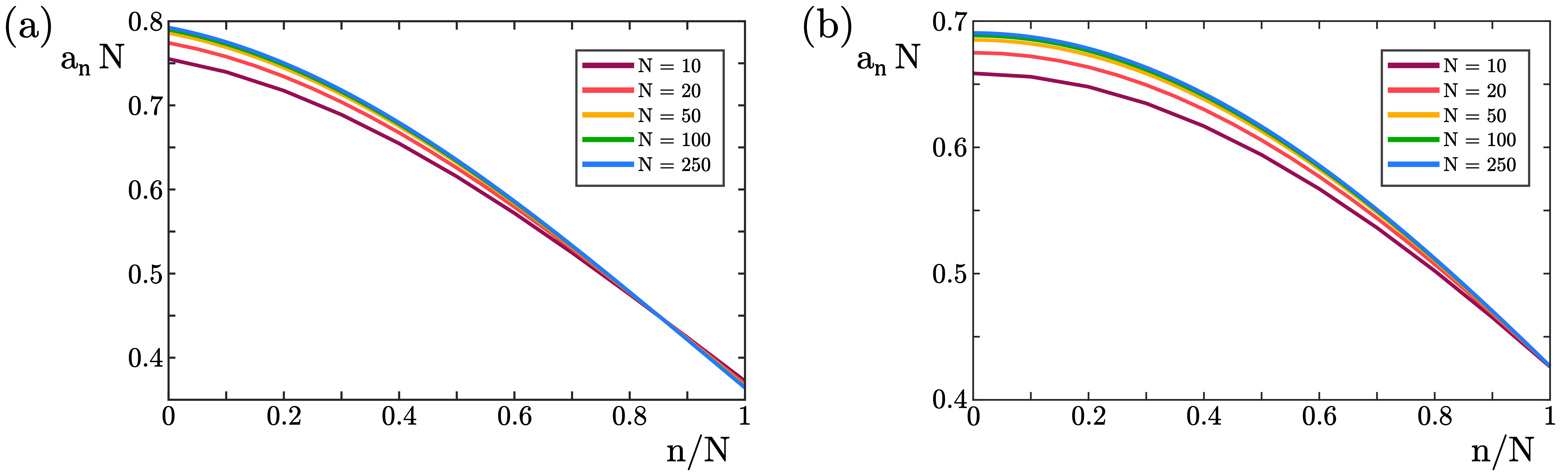}
    \caption{Scaled solutions of the matching condition (a)\eqref{MatchEqm6} and (b) \eqref{MatchEqRing} (with $m$ even) for large values of $N$. Figure reproduced from~\cite{hill2022dihedral,hill2023approximate}.}
    \label{fig:DihedralLargeN}
\end{figure}

Notice that everything comes down to satisfying the matching equation \eqref{MatchEqSpot}. For small $N$ one can do this by hand (or with symbolic computation) and one finds that the number of solutions for a given $m$ quickly becomes overwhelming as $N$ increases. For example, the hexagonal case $m = 6$ gives only 2 distinct solutions with profiles \eqref{RadialProfile} when $N = 2$, but 5 solutions when $N = 3$ (see Figure~\ref{fig:DihedralSpotSolns}), and many more for each $N \geq 4$. When $6 \mid m$ the matching equation \eqref{MatchEqSpot} becomes
\begin{equation}\label{MatchEqm6}
    a_{n} = 2\sum_{j=1}^{N-n} a_{j} a_{n+j} + \sum_{j=0}^{n} a_{j}a_{n-j},  
\end{equation}
losing all cosine terms, and numerically we observe that there exists exactly one solution satisfying $a_n > 0$ for all $n = 0,1,\dots,N$, for every value of $N$; as plotted in Figure~\ref{fig:DihedralLargeN}(a). For $m = 6$ this positive solution is the `standard' hexagon pattern, as documented in \cite{lloyd2008localized} and discussed in the next subsection. The existence of such a solution to \eqref{MatchEqm6} is proven to exist for all sufficiently large $N$. This is achieved by noticing that when $N$ is large, the matching equations \eqref{MatchEqm6} (after rescaling the $a_n$) resemble a Riemann sum discretization of the continuum matching equation
\begin{equation}\label{MatchEqSpotContinuum}
    \alpha(t) = \int_0^{1-t} \alpha(s)\alpha(s+t) \mathrm{d}s + \int_0^t \alpha(s)\alpha(t-s)\mathrm{d}s, \qquad t \in [0,1].
\end{equation}
A strictly positive solution $\alpha^*(t) > 0$ is proven to exist using computer assisted methods, following \cite{vandenberg2015Rigorous}, which in turn is proven to lead to the strictly positive solution of \eqref{MatchEqm6} roughly by $a_n \approx \alpha^*(n/(N+1))/(N+1)$; see Figure~\ref{fig:DihedralLargeN}.

Like localized dihedral spots, localized dihedral ring patterns are well-documented throughout the physical sciences. For example, such patterns arise as optical solitons \cite{menesguen2006optical,KIVSHAR1998Dark,Vladimirov2002Clusters} and as fairy circles \cite{Zelnik2015Gradual,Zhao2021Fairy,Getzin2015Fairy,Getzin2016FairyCircles}. Their emergence is also amenable to analysis in the Galerkin system \eqref{SHE_Galerkin}. The emergence of localized dihedral ring patterns from the Turing point $\mu = 0$ is detailed by the following theorem.   

\begin{thm}[{\cite[Theorem~2.2]{hill2022dihedral}}]\label{thm:RingCell} 
	Fix $m,N \in \mathbb{N}$ and assume the constants $\{a_{n}\}_{n=0}^{N}$ are solutions of the nonlinear matching condition 
	\begin{equation}\label{MatchEqRing}
        a_{n} = \sum_{i+j+k=n} (-1)^{\frac{m(|i| + |j| - |k| - n)}{2}}a_{|i|}a_{|j|}a_{|k|},
    \end{equation}
    for each $n = 0,1,\dots, N$ and $-N \leq i,j,k \leq N$. Then, there exist constants $q_0,\mu_0,r_0,r_1 > 0$ such that the system \eqref{SHE_Galerkin} has a solution of the form 
\begin{equation}\label{RadialProfileRing}
    u_{n}(r) = 2 a_{n}
    \begin{cases}
         \mu^{\frac{3}{4}}\sqrt{\frac{\pi}{2}} q_0r J_{mn+1}(r) + \mathcal{O}(\mu), & 0\leq r\leq r_{0},  \\
         \mu^{\frac{3}{4}} q_0r^{\frac{1}{2}} \sin(\psi_{n}(r)) + \mathcal{O}(\mu), & r_{0} \leq r \leq r_1\mu^{-\frac{1}{2}},\\
         \mu^{\frac{1}{2}}q(\mu^{\frac{1}{2}}r)\sin(\psi_{n}(r)) + \mathcal{O}(\mu), & r_1\mu^{-\frac{1}{2}} \leq r,
    \end{cases}
\end{equation} 
for each $\mu\in(0,\mu_{0})$, $n\in[0,N]$, where $\psi_n(r):= r - \frac{mn\pi}{2} - \frac{\pi}{4}$, $J_{mn+1}$ is the $(mn+1)^\mathrm{th}$ order Bessel function of the first kind, and $q(s)$ is the nontrivial homoclinic solution \eqref{GinzbLand:asymp} of the radial Ginzburg--Landau equation \eqref{Ginzburg-Landau_Rad}.
\end{thm}

\begin{figure}[t]
    \centering
    \includegraphics[width=\linewidth]{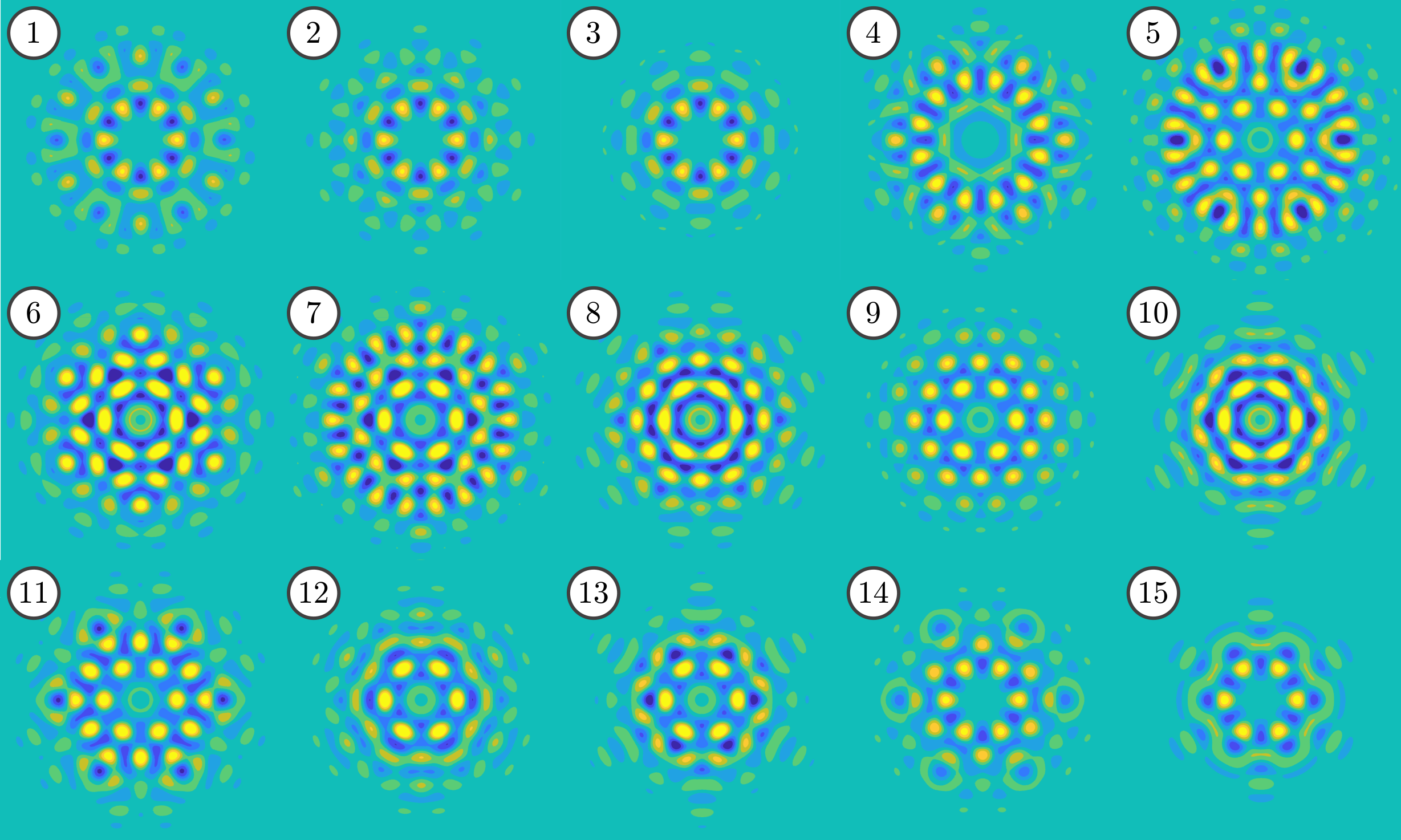}
    \caption{Solving \eqref{MatchEqRing} with $(m,N) = (6,3)$ leads to the fifteen distinct solutions in the form \eqref{RadialProfileRing} shown here for $\mu = 0.08$}
    \label{fig:DihedralRingSolns}
\end{figure}

There are marked differences between the spot profiles of Theorem~\ref{thm:SpotCell} and the ring profiles of Theorem~\ref{thm:RingCell}, similar to the spots and rings encountered in Section~\ref{sec:radial}. Primary among them is that the spots have a local maximum at the origin of the plane, $(x,y) = (0,0)$, while all rings satisfy $u(0) = 0$, which is a defining characteristic of ring solutions to the SHE; see Figure~\ref{fig:DihedralRingSolns}. Another difference is the scaling which mimics the scalings of axisymmetric Spot A solutions and rings in Section~\ref{sec:AxiEmerg}. That is, as $\mu \to 0^+$ ring solutions scale as $\mu^{3/4}$, while the spots in Theorem~\ref{thm:SpotCell} take on the usual Turing bifurcation scaling of $\mu^{1/2}$. Both of these differences carry over from the analysis on axisymmetric solutions and so the novel differences appear in the matching equations \eqref{MatchEqSpot} and \eqref{MatchEqRing}. In particular, notice that the ring solutions do not require a non-degeneracy assumption, while the spots do. In the ring case the non-degeneracy condition can be proven to hold for any solution of the matching equation \eqref{MatchEqRing}. This result simply leverages the non-degeneracy results of \cite{vandenberg2015Rigorous} on non-degenerate axisymmetric solutions to a Ginzburg--Landau PDE as was similarly done for the axisymmetric rings in \cite{mccalla2013spots}. This greatly eases the analysis of ring patterns as one only requires identifying solutions to the matching equation \eqref{MatchEqRing}, while in the case of the spots in Theorem~\ref{thm:SpotCell} we have the extra step of confirming non-degeneracy to arrive at a localized profile.    

The most interesting distinction between the dihedral spot and ring patterns comes from how they emerge from the Turing instability at $\mu = 0$. Notice that the cubic order of the ring matching equation \eqref{MatchEqRing} endows the $\mathbb{Z}_2$ symmetry that if $\{a_n\}_{n = 0}^N$ is a solution, then so is $\{-a_n\}_{n=0}^N$. This means that for any $(m,N)$ the emergence of one localized dihedral pattern $u$ is accompanied by another pattern related by negating it. The result is that rings emerge from $\mu = 0$ in the Galerkin system \eqref{SHE_Galerkin} through pitchfork bifurcations. This is not the case of the spots as they emerge along a single branch without a symmetric counterpart due to the lack of symmetry in their matching equation \eqref{MatchEqSpot}. 

When $m$ is even we may obtain a similar continuum formulation to \eqref{MatchEqSpotContinuum} for the ring matching equation \eqref{MatchEqRing}, as shown in Figure~\ref{fig:DihedralLargeN}(b). The result is that one may use such a continuum formulation to prove the existence of solutions to the matching equation (when $m$ is even) for all sufficiently large $N$. Despite results of this kind, the reader is reminded that one always has a finite truncation of the Fourier-cosine series \eqref{CosineExpansion}, meaning that we can only approximate solutions to the SHE. Extending these solutions to the full SHE is a nontrivial task without a clear course of action. One limitation comes from the fact that the proofs of Theorems~\ref{thm:SpotCell} and \ref{thm:RingCell} require $\mu_0 \to 0^+$ as $N \to \infty$, meaning that the region of validity for a solution to exist shrinks as the Galerkin expansion approximation is thought to be improving. Thus, one may require employing techniques such as Nash--Moser iteration that can provide a fast convergence algorithm as $N \to \infty$ to beat the quickly shrinking region of validity $\mu \in (0,\mu_0)$. A guide towards this may come from work that proves the existence of quasipatterned states to the SHE using Nash--Moser iteration to overcome similar issues in moving from a truncated Fourier expansion to a full one \cite{iooss2019existence}.

\subsection{Localized Patches Beyond Onset}\label{subsec:HexagonPatch}

The work reviewed in the previous subsection gives evidence for a plethora of different types of fully localized patches that are possible in the SHE and related models. While we were previously concerned with the emergence of localized dihedral patterns, here we turn our attention to their bifurcation structure. 

\begin{figure}[htb]
    \centering
    \includegraphics[width=0.9\linewidth]{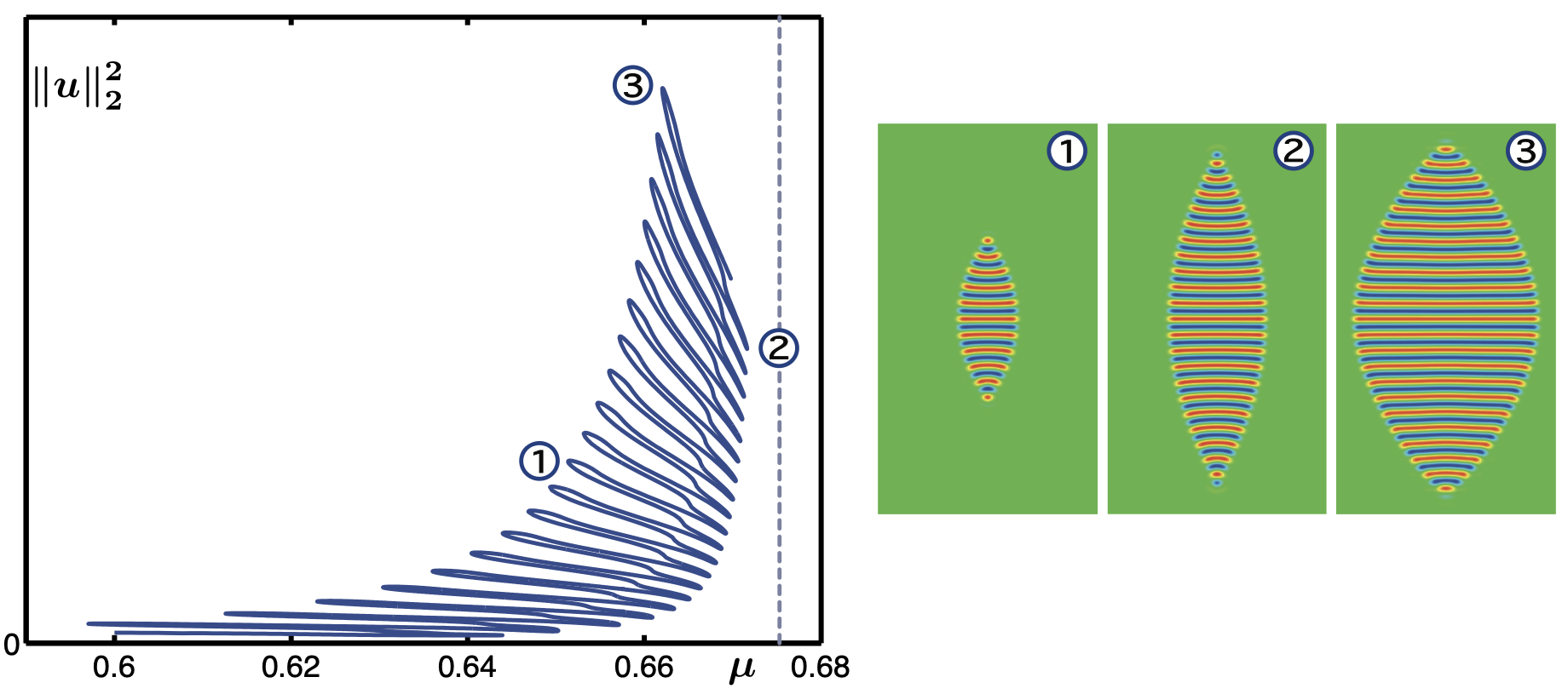}
    \caption{The bifurcation diagram of fully localized stripes for $\nu=2$ in \eqref{e:35SHE}. The Maxwell point for 1D stripes occurs at the vertical dashed line. Image originally appears in \cite{avitabile2010snake}. }
    \label{fig:worm_nu2}
\end{figure}

\begin{figure}[htb]
    \centering
    \includegraphics[width=0.9\linewidth]{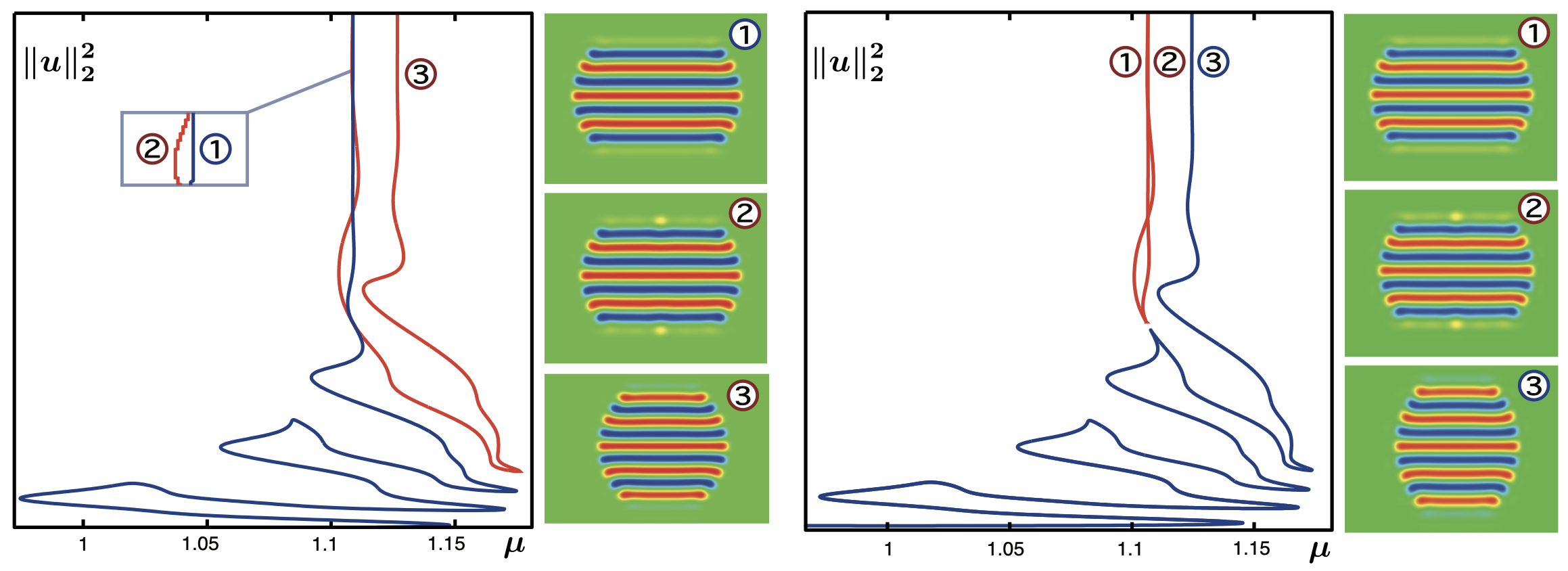}
    \caption{Fully localized stripes for the cubic-quintic SHE \eqref{e:35SHE} with $\nu=2.655$ (left) and $\nu=2.52$ (right). As $\nu$ decreases branches of localized stripes `pinch' off and form isolated branches. Image originally appears in  \cite{avitabile2010snake}.}
    \label{fig:worm_2}
\end{figure}

Localized patches of stripes (sometimes also known as worms) have been investigated in the SHE with a cubic-quintic nonlinearity~\cite{avitabile2010snake}
\begin{equation}\label{e:35SHE}
    0 = -(1+\Delta)^2u - \mu u + \nu u^3 - u^5.
\end{equation}
Here they find that the patches appear to undergo snaking behavior, as shown in Figure~\ref{fig:worm_nu2}. However, the branch eventually terminates once the patch gets sufficiently large with various isolated branches of larger patches found nearby, as shown in Figure~\ref{fig:worm_2}. Numerical investigations strongly suggest that the various branches of the localized stripes with $n$ interior SPPs should approach vertical asymptotes given by 1D pulses with vanishing Hamiltonian that consist of $n$ SPPs, but the reason for this remains a mystery. 

\begin{figure}[htb]
    \centering
    \includegraphics[width=0.75\linewidth]{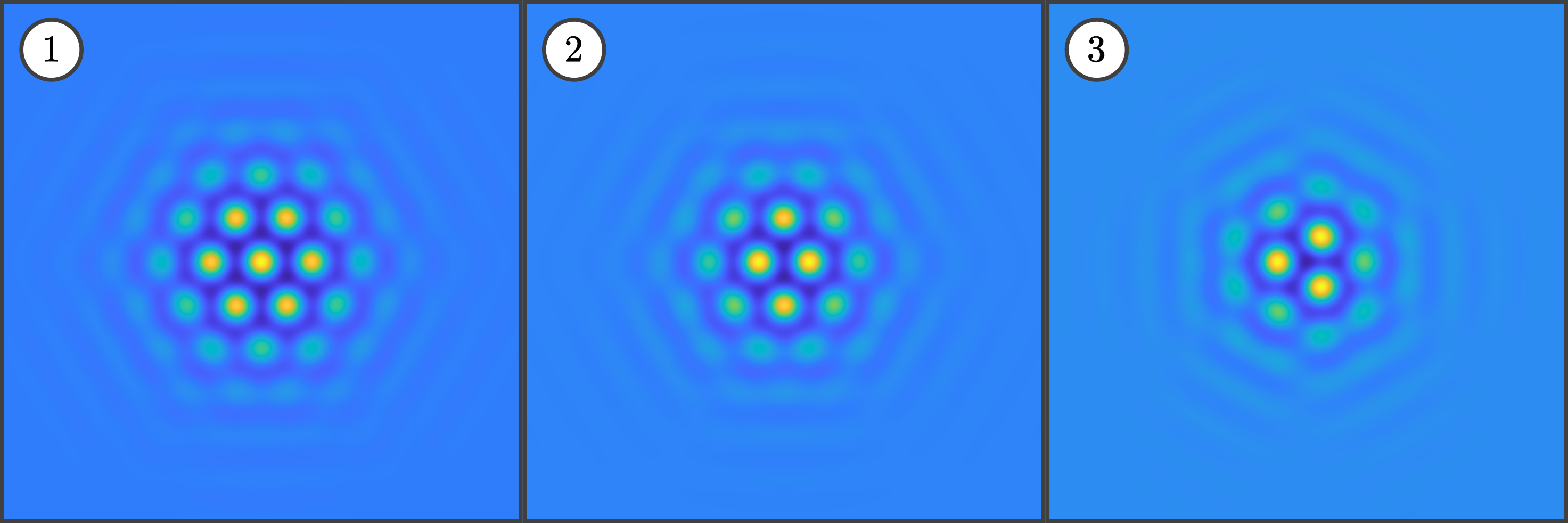}
    \caption{Illustrative examples of localized hexagon patches with (1) $\mathbb{D}_{6}$ symmetry (hexagons), (2) $\mathbb{D}_{2}$ symmetry (rhomboids), and (3) $\mathbb{D}_{3}$ symmetry (triangles). 
    }
    \label{fig:various_hex_patches}
\end{figure}

Most of the analysis on localized patches has been on hexagon patterns. As is shown in Figure~\ref{fig:various_hex_patches}, various different localized patches of cellular hexagon pattern to the SHE have been identified far from onset. The main difference in all of these patterns comes from whether the center of the localization occurs at the center of a hexagon cell (leading to localized structures with $\mathbb{D}_6$ symmetry termed \textit{standard hexagons}) or off-center (leading to localized structures with either $\mathbb{D}_{2}$ or $\mathbb{D}_3$ symmetry, breaking the $\mathbb{D}_6$ symmetry of the hexagonal superstructure, termed \textit{rhomboids} and \textit{triangles}, respectively). In the SHE \eqref{SwiftHohenberg} with $\nu = 1.6$ whether or not a pattern is centered or not leads to two different types of bifurcation diagrams. While both on-center and off-center patches appear to bifurcate from the trivial state $U = 0$ at $\mu=0$, the on-center localized patch branch quickly collides with the radial Spot A branch while the off-center patch branches do not interact with the radial spot and eventually undergo snaking behavior. At $\mu\approx0.15$, there is another symmetry-breaking ($\mathbb{D}_6$) bifurcation of the Spot A branch leading to a large amplitude on-center hexagon patch bifurcating and subsequent snaking behavior. The large amplitude branch of on-center localized hexagon patches is shown in Figure~\ref{fig:hex_snake} along with the Spot A branch it bifurcates from. Rigorous existence of localized square, hexagon and octagon patches using validated numerics has recently been established by Cadiot et al.~\cite{Cadiot_Lessard_Nave_2024} for specific parameter values while also establishing criteria for the convergence of spatially periodic solutions to a localized solution. 

\begin{figure}[htb]
    \centering
    \includegraphics[width=0.9\linewidth]{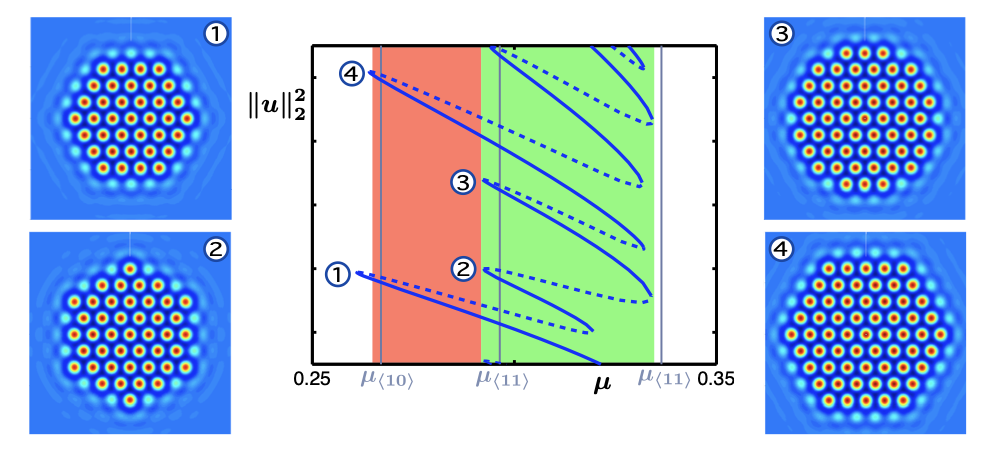}
    \caption{A zoom of the hexagon patch bifurcation diagram in the SHE with $\nu=1.6$ showing how a new layer of cells is added to a hexagon patch. Image originally appears in \cite{lloyd2008localized}.}
    \label{fig:hexagon_patch_bif}
\end{figure}

\begin{figure}[htb]
    \centering
    \includegraphics[width=0.9\linewidth]{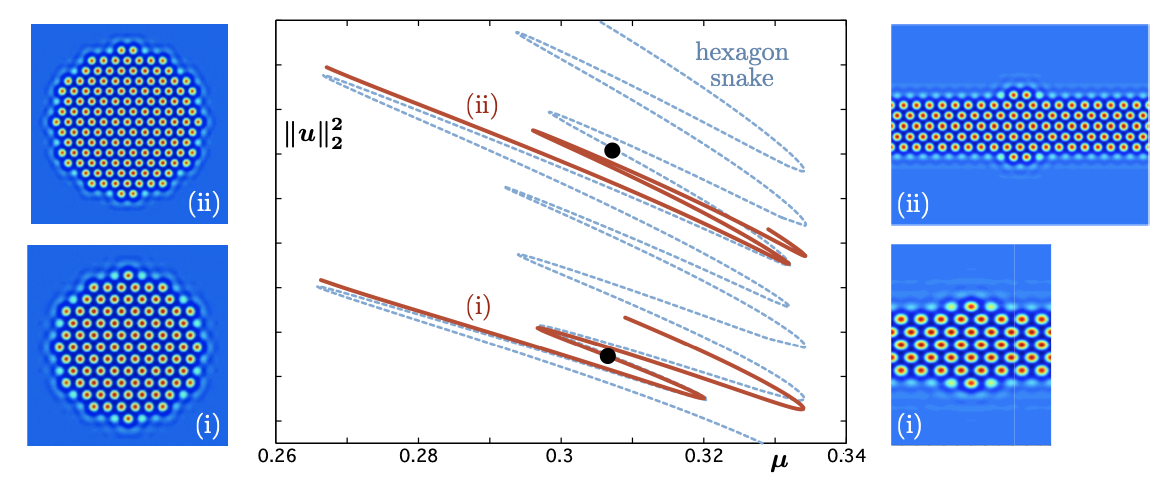}
    \caption{Comparison of the hexagon patch and front snaking bifurcation diagrams showing the emergence of cells along the $\langle10\rangle$-front interface. Image originally appears in \cite{lloyd2008localized}.}
    \label{fig:hexagon_front_patch_comp}
\end{figure}

What is most interesting about continuing the hexagon patches is that their snaking bifurcation diagrams are significantly more complicated than the planar fronts or 1D patterns encountered previously. They also bear little resemblance to the axisymmetric curves we saw in Section~\ref{subsec:RadialSnaking} in that localized hexagons appears to lie along a single curve that causes the pattern to expand into the quiescent state without bound. As can be seen in Figure~\ref{fig:hexagon_patch_bif}, the bifurcation curve is quite complicated with numerous folds on either side that do not appear to align at common values of $\mu$ as one ascends the diagram. What is apparent in the diagram is that large leftward excursions represent the localized hexagon patch ``completing'' itself wherein the region of localization makes a near-perfect hexagon shape. Considering the localized hexagon patch as the makeup of various planar hexagon fronts (as in \S\ref{subsec:HexFront}), the large leftward excursions are found to be well explained by the growth of individual cells along the edge of the hexagon planar fronts, as shown in Figure~\ref{fig:hexagon_front_spot_snaking} and Figure~\ref{fig:hexagon_front_patch_comp}. In between these leftward excursions the patterns grow by adding activated cells along the perimeter of the pattern in increasingly complicated ways. 

\begin{figure}[htb]
    \centering
    \includegraphics[width=0.7\linewidth]{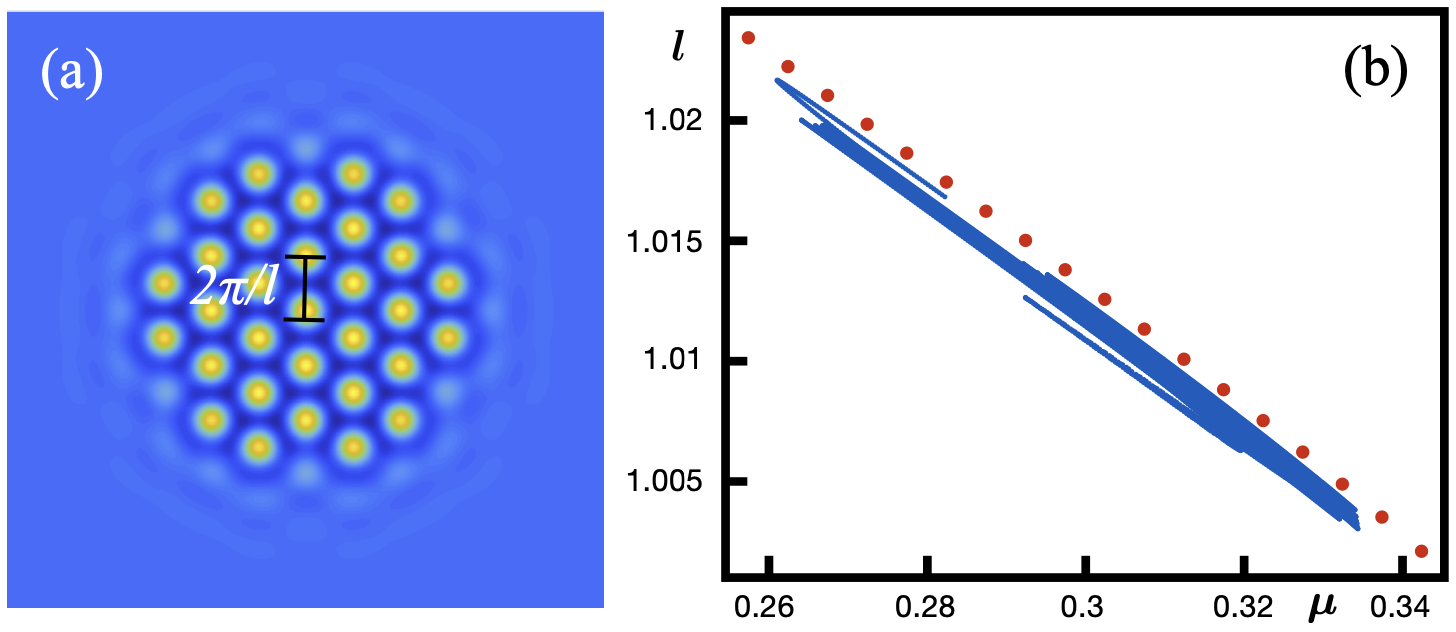}
    \caption{(a) A localized hexagon patch of the SHE \eqref{SwiftHohenberg} with $(\mu,\nu)=(0.3,1.6)$. (b) The selected wavenumber of the hexagon cell in the snaking region. Image originally appears in \cite{lloyd2021}.}
    \label{fig:hexagon_patch_wavenumber_selection}
\end{figure}

\begin{figure}[htb]
    \centering
    \includegraphics[width=0.9\linewidth]{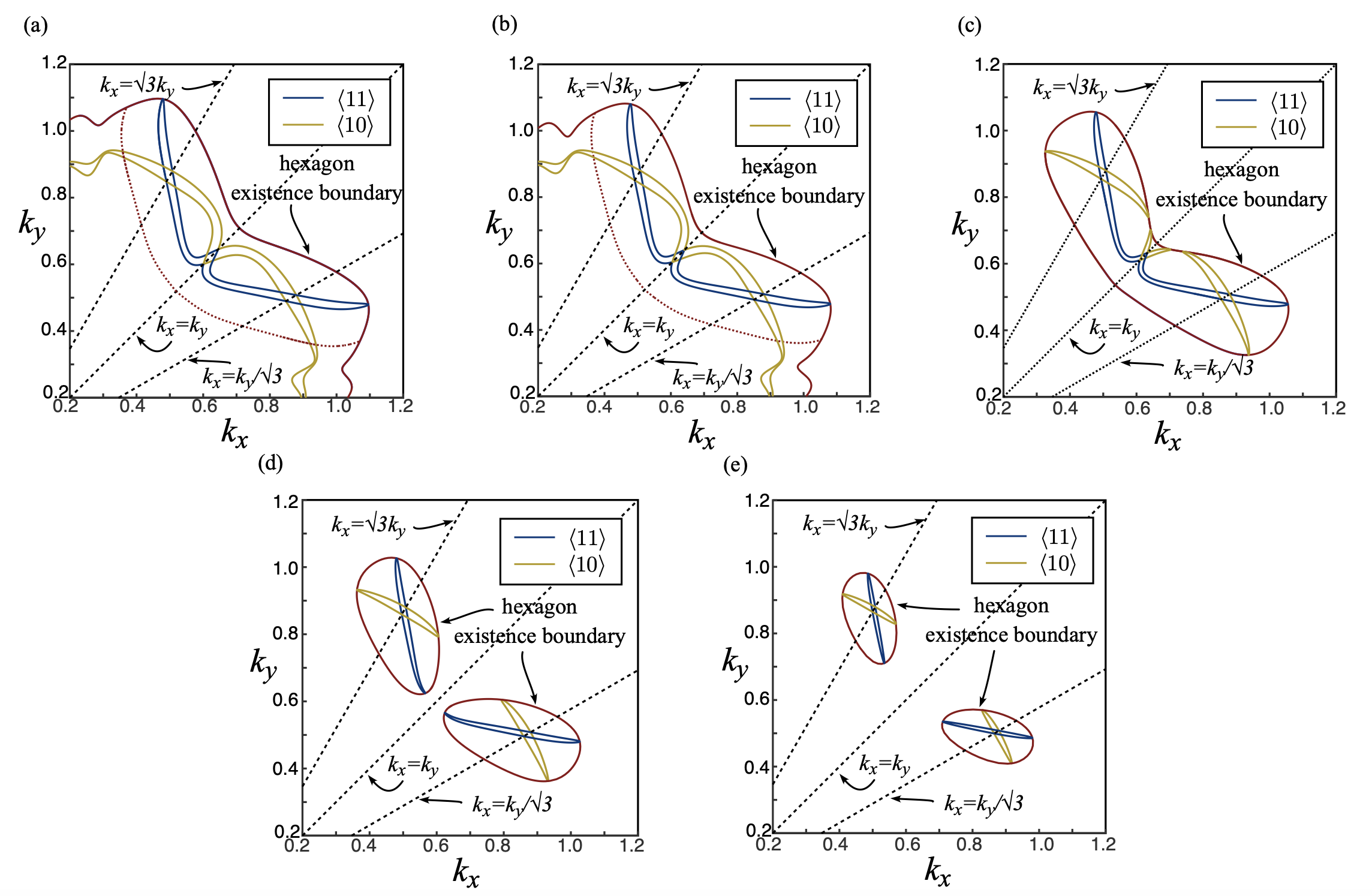}
    \caption{Compatibility diagrams for stretched dihedral hexagons in the snaking region for $\nu=1.6$ SHE \eqref{SwiftHohenberg} for (a) $\mu  = 0.267$, (b) $\mu  = 0.28$, (c) $\mu  = 0.3$, (d)
$\mu  = 0.32$, and (e) $\mu  = 0.345$. Image originally appears in \cite{lloyd2021}.}
    \label{fig:hexagon_patch_wavenumber_selection_2}
\end{figure}

Lloyd~\cite{lloyd2021} extended this concept of considering a hexagon patch made up of various planar fronts to propose a pattern selection criterion that predicts the selected wavenumber of the pattern cells in the patch. While the hexagon patch can be considered as being made up of several fronts, we consider just two fronts oriented perpendicular to each other, connected together by the same pattern cell. This is illustrated in Figure~\ref{fig:hex_front}(c) where the horizontal front is denoted by $\langle10\rangle$ and the vertical front is denoted by $\langle11\rangle$ and the matching pattern cell is highlighted as a white box. From the planar front theory in \S\ref{subsec:HexFront}, we know that for a given vertical wavenumber $k_y$, the horizontal front must select a corresponding horizontal wavenumber $k_x$. Similarly, the vertical front must select a corresponding $k_y$ for a given $k_x$. Thus, for the pattern cell to exist, a necessary condition is that the selected $k_x$ and $k_y$ values lead to a compatible pattern.  Tracing out the selected wavenumbers $(k_x,k_y)$ of the $\langle10\rangle$- and $\langle11\rangle$-fronts as we vary $k_y$ or $k_x$, respectively, leads to the compatibility diagrams shown in Figure~\ref{fig:hexagon_patch_wavenumber_selection_2}. These diagrams show various slices through the snaking region of the hexagon patches. Where the blue and gold curves intersect one can expect to find compatible fronts to occur. It is found that intersections of the curves occur along the lines $k_y=\sqrt{3}k_x$ (perfect hexagons), $k_x=k_y$ (perfect squares), and just off the perfect hexagon lines. For the intersections along the perfect hexagon lines, these lead to excellent predictions of the selected wavenumber of the hexagon patch. The patch wavenumber selection argument can be extended to systems that do not have a spatial Hamiltonian structure, as well as temporally invading patches using the far-field core numerics  \cite{lloyd2021}. The methods can also be extended to other localized patterns, such as localized squares.

Sakaguchi and Brand~\cite{sakaguchi1997stable} were the first to find stationary localized patches involving squares in a modified SHE, given by 
\[
    U_t = -d(q_0^2 + \Delta)^2 U + a U + b U^3 - c U^5 + e\nabla\cdot[(\nabla U)^3].
\]
We suspect these square patches also undergo snaking bifurcations similar to the hexagon patches and the lattice problem shown in Figure~\ref{fig:LatticeSnake}. In fact, continuations of many of the various dihedral patterns predicted by the work of the previous subsection in \cite{hill2023approximate} show similarly complicated bifurcation structures. Despite these similarities, little rigorous investigation has gone into understanding the bifurcation structure of localized patches beyond the hexagons patterns we have reviewed here.  

\begin{figure}[htb]
    \centering
    \includegraphics[width=0.9\linewidth]{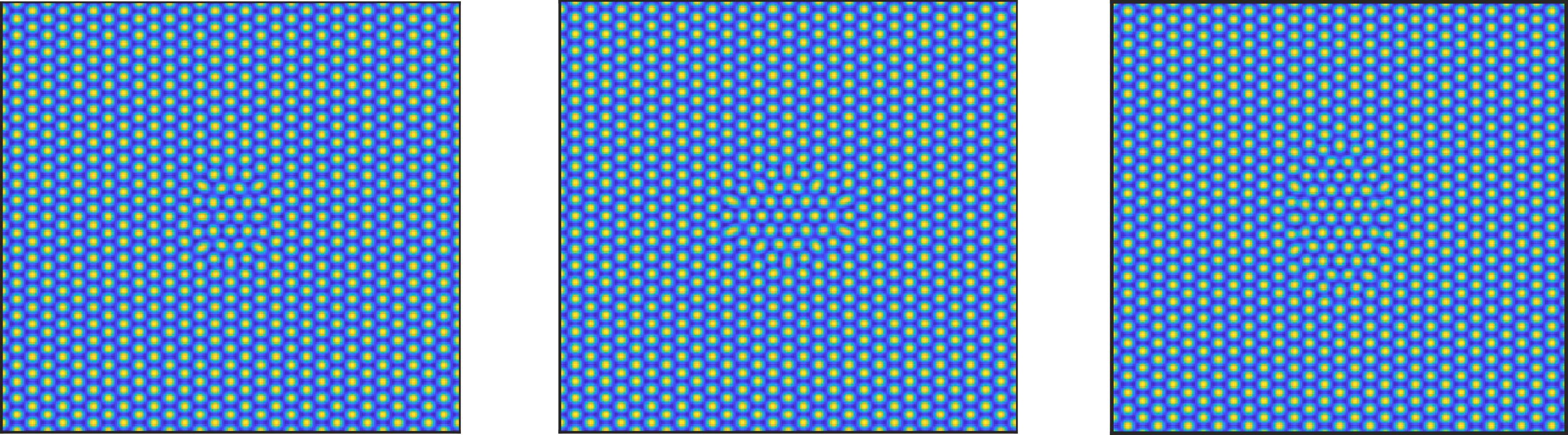}
    \caption{Various different penta-hepta defect hexagon-hexagon patches where the interior hexagon pattern patch is rotated by an angle of $\pi/6$ to the exterior hexagon pattern. Image originally appears in \cite{subramanian2021}.
    }
    \label{fig:hexagon_hexagon_patch}
\end{figure}

Localized patches involving quasi-patterns embedded in a quiescent state have also been documented in \cite{Subramanian2018LocalizedPFC} for a phase field crystal model. These structures undergo a succession of folds leading to the quasi-pattern taking up more of the domain. Near onset, these structures should be captured by the weakly nonlinear analysis described in Section~\ref{subsec:Dihedral}, but away from this limited parameter regime very little is known analytically about these structures. The same authors have further documented localized hexagon patches embedded in a hexagon state \cite{subramanian2021}, as shown in Figure~\ref{fig:hexagon_hexagon_patch}. Continuations of these structures shows that they typically lie along isolas in parameter space. However, even numerically, their existence is not well understood and many questions about them remain.

\subsection{Planar Lattices}\label{subsec:SquareLattice}

\begin{figure}[htb]
    \centering
    \includegraphics[width=\linewidth]{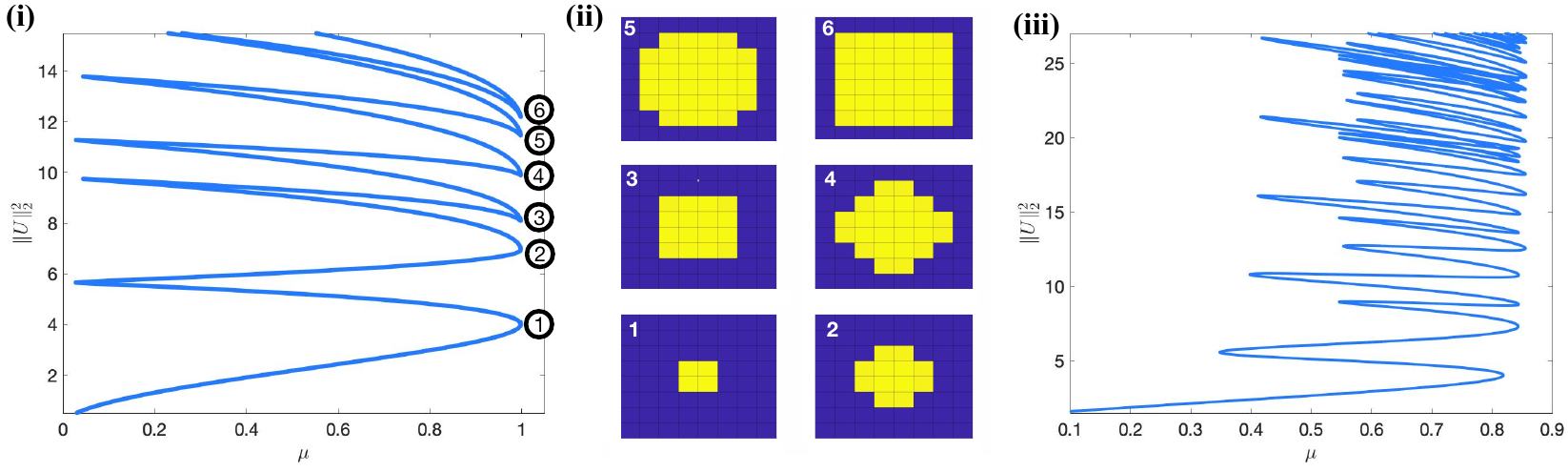} 
    \caption{Sample profiles and snaking bifurcation curves of planar localized $\mathbb{D}_4$-symmetric off-site patterns of the lattice system \eqref{2Dlattice}. Panel (i) takes $d = 0.001$ and panel (ii) provides solution profiles along the first six rightward saddle-node bifurcations on the curve. Panel (iii) shows the same bifurcation diagram, but now with $d = 0.1$. Bifurcation curves for on-site patterns exhibit the same behavior when increasing $d$; see \cite{bramburger2020localized}.} 
    \label{fig:LatticeSnake}
\end{figure}

While much attention is rightfully paid to localized patterns in the SHE and other RD systems, fully localized steady-state solutions have also been well-documented in spatially discrete lattice systems. This is particularly true in models from nonlinear optics \cite{bortolozzo2004bistability,chong2009multistable,firth2007homoclinic,tlidi1994localized,Vladimirov2002Clusters,yulin2011snake}. Relevant to our discussion in this section, numerical continuations of localized steady-state square patterns on two-dimensional lattices have been shown to exhibit snaking bifurcation behavior that bears a striking resemblance to the hexagons in the previous subsection \cite{taylor2010snaking,kusdiantara2019snakes,chong2009multistable}. For demonstration, compare the existence curves in panels (i) and (iii) of Figure~\ref{fig:LatticeSnake} for the lattice system  
\begin{equation}\label{2Dlattice35}
    \dot{u}_{n,m} = d[u_{n+1,m} + u_{n-1,m} + u_{n,m+1} + u_{n,m-1} - 4u_{n,m}] - \mu u_{n,m} + 2u_{n,m}^3 - u_{n,m}^5, \quad (n,m)\in\mathbb{Z}^2,
\end{equation}
to those in Figure~\ref{fig:hex_snake}. We remind the reader that, as in \eqref{Lattice1D} and \eqref{2Dlattice}, $d > 0$ represents the strength of interaction between the elements $u_{n,m}$ arranged over the integer lattice $\mathbb{Z}^2$, while the coupling comes in the form of a typical 5-point discretization of the usual Laplace operator in two spatial dimensions. 

The coupling strength plays a critical role in determining the shape of the bifurcation curves as panels (i) and (iii) in Figure~\ref{fig:LatticeSnake} show a very regular snaking curve for $d = 0.001$ that resembles a well-organized bifurcation curve typically seen in 1D system. Furthermore, much like 1D systems and the hexagons of the previous subsection, Figure~\ref{fig:LatticeSnake} demonstrates the existence of two types of patterns: on-site and off-site. On-site patterns are square ($\mathbb{D}_4$-symmetric) equilibria of \eqref{2Dlattice} which have their center of symmetry on a lattice point, while off-site have it in the middle of four lattice points, similar to the localized hexagons in the SHE.  

To further understand the structure of fully localized $\mathbb{D}_4$ patterns in \eqref{2Dlattice35}, Bramburger and Sandstede \cite{bramburger2020localized} provided a detailed investigation that uses both rigorous analysis and computational methods. Although this work is applicable to systems well beyond that of \eqref{2Dlattice35}, we review the results only in the context of \eqref{2Dlattice35} for clarity. Analytical results can be obtained by noticing that setting $d = 0$ in \eqref{2Dlattice35}, the so-called {\em anti-continuum limit}, completely decouples elements arranged over the lattice, resulting in the steady-state equation 
\begin{equation}
    0 = - \mu u_{n,m} + 2u_{n,m}^3 - u_{n,m}^5, \quad (n,m)\in\mathbb{Z}^2.
\end{equation}
For each $\mu \in [0,1]$ the non-negative roots of the above quintic polynomial are $0,\sqrt{1 \pm \sqrt{1 - \mu}}$, with collisions of these roots taking place in a saddle-node bifurcation at $\mu = 1$ and a pitchfork bifurcation at $\mu = 0$ from the trivial branch. Importantly, taking $u_{n,m} \in \{0,\sqrt{1 \pm \sqrt{1 - \mu}}\}$ for each $(n,m) \in \mathbb{Z}^2$ allows one to create arrangements of patterns on the lattice. Away from the bifurcation points $\mu = 0,1$ one may evoke the implicit function theorem to guarantee the persistence of these arrangements as steady-state solutions to the lattice system \eqref{2Dlattice35} with $0 < d \ll 1$. Thus, the problem of understanding the organization of the bifurcation curves of localized squares in \eqref{2Dlattice35} with small $d$ essentially reduces to identifying how these continued branches of arrangements connect near the bifurcation points at $\mu = 0,1$.

Using a combination of Lyapunov--Schmidt reduction and blow-up methods, \cite{bramburger2020localized} proved that for sufficiently small $d > 0$ the bifurcation curves follow the neat organization demonstrated in Figure~\ref{fig:LatticeSnake}(i) and (iii), akin to the snaking branches in the 1D SHE. Moreover, the growth of the pattern as one ascends the bifurcation curve is predictable, adding activated lattice points (those continued in $d> 0$ from $\sqrt{1 + \sqrt{1 - \mu}}$) in a ring around a square to increase its size; see panel (ii) of Figure~\ref{fig:LatticeSnake}. This analysis goes further and provides spectral and nonlinear stability of the patterns along the curve as well. Outside of the instability results in \cite{makrides2019existence}, little is known about stability in the continuous spatial setting, and so lattice systems such as \eqref{2Dlattice35} present themselves as mathematical models that are amenable to analysis, at least when $d > 0$ is small.     

\begin{figure}[t]
    \centering
    \includegraphics[width=\linewidth]{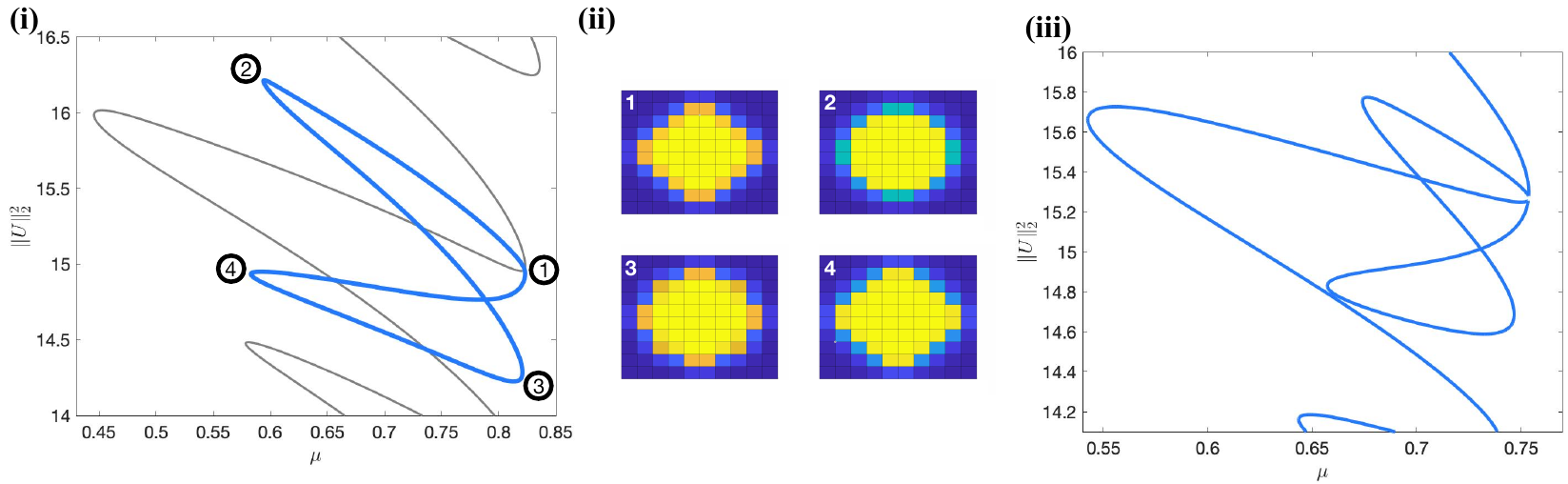}
    \caption{Switchbacks take place in \eqref{2Dlattice35} when an isola collides with the main snaking branch and attaches itself. (i) At $d = 0.12$ there is an isola (blue) that is about to collide with the primary solution branch of off-site solutions (grey). (ii) Sample profiles along the isola. (iii) The rearranged primary snaking branch at $d = 0.2$ due to the switchback taking place at a point $ d \in (0.12,0.2)$.} 
    \label{fig:Switchback}
\end{figure}

The question still remains: what happens when $d$ gets larger that leads to the perceived complexity of the snaking curves? One may be tempted to think that the curve in panel (i) of Figure~\ref{fig:LatticeSnake} is the same that in panel (iii) with only a diffeomorphic deformation mapping one to the other as one increases $d$. It was originally observed by Taylor and Dawes in \cite{taylor2010snaking} that this is not the case. They observed that the bifurcation diagram begins to turn back on itself after an isola collides with the primary snaking branch of localized square patterns. This phenomenon is referred to as a ``switchback'' and happens when a fold along the isola and a fold along the primary solution branch collide with each other, resulting in a co-dimension 2 cusp bifurcation. The result of a switchback is illustrated in Figure~\ref{fig:Switchback}, while numerical results in \cite{bramburger2020localized} indicate that one should expect infinitely many switchbacks to occur upon increasing the coupling parameter $d$. Denoting the sequence $\{d_1,d_2,d_3,\dots\}$ to be the set of values at which one of the infinitely many switchbacks occurs, numerical evidence indicates that $d_n \to d_\infty \approx 0.068$ at an exponential rate. This means that increasing $d$ across $d_\infty$ would result in a near complete re-arrangement of the bifurcation diagram from order to complexity in the bifurcation curves seemingly all at once.     

Finally, we are also able to examine the asymmetric patterns that emerge through pitchfork bifurcations near each of the folds along the primary branches. One may evoke equivariant bifurcation theory \cite{stewart2000singularities,hoyle2006pattern}, and in particular the equivariant branching lemma, to determine exactly how many states bifurcate near each fold and what their one-dimensional eigenspaces will be when $d > 0$ is small. All bifurcating branches are initially unstable, but can become stable after a fold bifurcation. What is noteworthy here is that the branches appear to be unpredictable, as opposed to those in 1D \cite{beck2009snakes}, thus presenting another complexity in the study of localized planar pattern formation.

\section{Three Dimensional Patterns}\label{sec:3d}

Over the previous three sections, we have introduced the limited understanding of localized patterns in 2D and the different approaches required to study them. Increasing the spatial dimension to three now moves one even further into mostly uncharted territory. Despite the challenge of analyzing and simulating mathematical equations in three spatial dimensions, localized patterns have been identified in numerous settings. This includes water waves \cite{buffoni_groves_sun_wahlen_2013,buffoni2018variational,parau_vanden-broeck_cooker_2005,akers_milewski_2010}, fluid convection \cite{Lo_Jacono_2017,solomatov_2012,mercader_batiste_alonso_knobloch_2011,tumelty_beaume_rucklidge_2023}, turbulence \cite{pershin_beaume_tobias_2019,schneider_gibson_burke_2010,avila_mellibovsky_roland_hof_2013}, and crystal formation in soft matter \cite{Thiele2013,Subramanian2018LocalizedPFC}. What further makes the analysis of localization in three spatial dimensions difficult is that it can manifest itself in one, two, or three directions. We refer to each respective case as (1+2)D patterns, (2+1)D patterns, and fully localized 3D patterns. 

As in previous sections, numerical continuation provides an invaluable tool to study the bifurcation structure of localized 3D patterns. In some of the following examples, solutions to three-dimensional equations can be computed directly, whereas some other examples require restrictions to particular symmetry groups via pseudo-spectral methods. In the case of the latter, bifurcation diagrams are then also restricted to such symmetry groups and so do not reveal the entire bifurcation structure of the problem. A new tool for simulating localized 3D patterns is the MATLAB package \textit{pde2path}~\cite{uecker_wetzel_rademacher_2014}, which is an elliptic PDE solver with numerical continuation and bifurcation detection. Uecker and Wetzel \cite{uecker2020snaking} considered the prototypical Brusselator model in three spatial dimensions and, using \textit{pde2path}, studied the bifurcation structure of snaking branches of planar fronts connected to body centered cubic patterns (BCCs) - 3D patterns that are localized in a single direction. This was a significant computational undertaking that highlights the potential utility of \textit{pde2path} for pushing forward the theory of localization in 3D with computational insight.

Similar to localized fronts in 2D discussed in \S\ref{sec:fronts}, most analytical results for (1+2)D patterns and (2+1)D patterns in 3D consist of reducing the problem to a lower dimensional system where the tools and techniques discussed in the previous sections can be applied. Usually this requires the non-localized directions to be bounded or periodic, although we will note some examples where this is not the case. Many of the above examples in 3D are modeled by quasilinear PDEs, and so often require more delicate analysis than semilinear systems like the SHE and RD systems.  

Notable examples of 3D patterns come from the study of water waves, where localized patterns form on a 2D free surface between two 3D domains, and so it is natural to perform some form of dimensional reduction. There have been a number of analytical results for localized 3D patterns in water waves, for which we direct readers to relevant review articles \cite{Haziot2022EbbandFlow,Groves2004WaterWaves}. We will briefly mention some of the key methods employed in this area and summarize recent results that represent the extent of current analytical tools. However, we remind the reader that results in 3D are even less unified than those in 2D, making this section a venture into the frontiers of localized patterns.

\subsection{Localized (1+2)D Patterns}\label{ss:3D-fronts}

We begin by considering patterns that exhibit localization in a single unbounded direction, typically with either bounded domains or periodic profiles in the two remaining transverse directions. The predominant example of such patterns are two-dimensional solitary waves, which have been well-studied both analytically and numerically since the first observations of John Scott Russell in 1834 (see the historical review \cite{allen1998early}). In more recent years, localized states have been observed in fluid convection, where they are known as \emph{convectons}, as well as in magnetohydrodynamics and turbulent fluids. Techniques in these fields fall into roughly three approaches: (1) numerical continuation where significant advances have been made to make the computations feasible (such as time-stepping and equation-free methods), (2) semi-analytic weakly nonlinear analysis~\cite{tumelty_beaume_rucklidge_2023} where one has to numerically compute the null space of the linearization of the bifurcating state, and (3) a minimal Fourier decomposition and asymptotic analysis of the reduced model~\cite{dawes_2007}. While significant progress has been made in numerically studying these structures, any analytic results remain an open problem.

\begin{figure}[htb]
    \centering
    \includegraphics[width=0.75\linewidth]{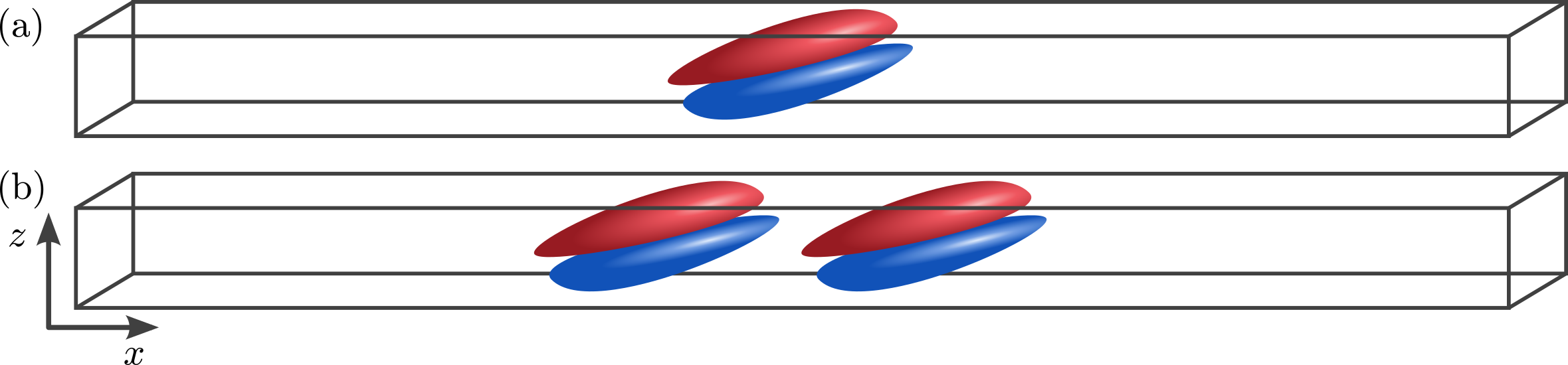}
    \caption{An illustration of the 3D doubly diffusive convectons found in \cite{beaume_bergeon_knobloch_2018}. The box is rotated so that gravity acts to the left. Plotted are isosurfaces of the velocity in the $x$-direction where positive values are shown in red and negative values in blue.}
    \label{fig:doublydiff_conv}
\end{figure}

The study of convectons goes back to the work of Blanchflower \cite{blanchflower1999magnetohydrodynamic} and has been predominantly undertaken in two models for fluid convection. The first of which is \emph{binary fluid} convection, where two horizontal layers of miscible fluids are heated from below~\cite{mercader_batiste_alonso_knobloch_2011,watanabe_iima_nishiura_2016,watanabe_iima_nishiura_2012, watanabe_toyabe_iima_nishiura_2011}, and the second is \emph{doubly diffusive} convection, where they studied a two-component fluid subject to horizontal gradients of temperature and concentration~\cite{Beaume_2013_doubly_diffs,beaume_rucklidge_tumelty_2022,tumelty_beaume_rucklidge_2023,Beaume_nonsnaking_2013} and with vertical gradients~\cite{beaume_bergeon_knobloch_2011}. Various localized solutions can be found via direct numerical simulations of the models, and numerical continuation can be used to further understand their bifurcation diagrams~\cite{Mercader_2010,mercader_batiste_alonso_knobloch_2011,mercader_batiste_alonso_knobloch_2013,bergeon_knobloch_2008,beaume_bergeon_knobloch_2011,beaume_bergeon_knobloch_2018}. In particular, localized solutions can form bound states called \emph{multiconvectons}~\cite{mercader_batiste_alonso_knobloch_2011}, which are similar to the multipulse solutions studied by Knobloch et al. \cite{knobloch2011isolas}. In closed box doubly diffusive convection, two branches of symmetric convectons emerge from the conduction state 
and undergo homoclinic snaking, where numerous secondary bifurcations occur along each branch~\cite{Beaume_2013_doubly_diffs,beaume_bergeon_knobloch_2018}. Each of the bifurcating solutions possess less symmetry than the original convecton, and undergo secondary snaking similar to that observed in the SHE \cite{lloyd2008localized}.  In Figure~\ref{fig:doublydiff_conv}, we show an example of some of the numerically computed doubly diffusive convectons from~\cite{beaume_bergeon_knobloch_2018} from the lowest kinetic energy branch. 

Beyond the standard convection problems, localized patterns have also been studied in Rayleigh--B\'enard convection~\cite{solomatov_2012,solomatov_jain_2021}, rotating fluids~\cite{beaume_bergeon_kao_knobloch_2013,beaume_kao_knobloch_bergeon_2013}, and magnetohydrodynamics,~\cite{dawes_2007,blanchflower1999magnetohydrodynamic,blanchflower_weiss_2002,lo_jacono_bergeon_knobloch_2011}. In most of these examples, there exist conserved quantities that result in slanted snaking (recall \S\ref{ss:slant}) rather than the standard homoclinic snaking. In particular, for convection in rotating fluids, Beaume et al.~\cite{beaume_bergeon_kao_knobloch_2013} determined that the presence of a conserved quantity results in the nonlocal amplitude equation 
\begin{equation*}
    A_{XX} + \mathrm{i}\left(\gamma A_{X} + a_1 |A|^2 A_{X} + a_2 A^2 \,\overline{A}_{X}\right) + \mu A + b|A|^2 A - |A|^4 A = 0
\end{equation*}
where $a_1,a_2$ are fixed constants, while the `nonlocal' coefficients $\mu,\gamma, b$ depend on the spatial average of $A$. Furthermore, localized solutions no longer require SPPs to bifurcate subcritically in order to emerge and persist outside the bistability region. Localized $(1+2)$D patterns have also been observed in turbulence of plane Couette flow~\cite{schneider_gibson_burke_2010,schneider_marinc_eckhardt_2010,pershin_beaume_tobias_2019}; however, these solutions have a strong connection to chaos, and so they are extremely difficult to study using deterministic methods~\cite{avila_mellibovsky_roland_hof_2013}.

One area of significant investigation of (1+2)D patterns is into localized solutions in the water wave problem. Some notable examples include waves induced by a combination of gravity and surface tension (\emph{gravity-capillary waves}), waves induced by stratified flows within a fluid (\emph{internal waves}), and waves propagating through a fluid with an elastic cover, such as an ice sheet (\emph{hydroelastic waves}). In these models such solutions are referred to as \emph{two-dimensional solitary waves}. The scope of research in this area is much too broad and merits its own review altogether, so we do not cover it in any significant detail here. Instead, we will present a brief summary of the relevant tools for studying two-dimensional solitary waves, and refer the reader to previous reviews \cite{Miles1980Solitary,Ilichev2000Solitary,Groves2004WaterWaves,Dias2003WaterWaves} and the many references therein for more details. The two main approaches for numerically computing solitary waves are via boundary integral or series truncation methods, as summarized in \cite{vanden2007solitary}. By utilizing ideas from complex analysis, one can compute water wave solutions by solving a contour integral (in the case of the former) or by taking a finite Laurent expansion and solving the resulting nonlinear algebraic system (in the case of the latter). In order to numerically capture solitary waves, the domain is often taken sufficiently large so that the solution is not affected by the boundary. Alternatively, one can compute periodic wave solutions and then take increasing values of the wavelength, so that the periodic solution tends to a solitary wave profile.

The analytic study of localized patterns in the water wave problem again relies heavily on the methods of spatial dynamics, as introduced by Kirchg\"assner \cite{Kirchgassner1982SpatialDyn}. The governing equations are first formulated as a (usually ill-posed) evolution equation
\begin{equation}\label{eqn:ww-evo}
    u_{\xi} = L u + N(u),
\end{equation}
where the traveling spatial direction $\xi=x-ct$, with wave speed $c\in\mathbb{R}$, is taken to be an unbounded time-like variable. The linear operator $L$ contains derivatives in the vertical direction and admits an infinite-dimensional phase-space, much like we saw with the approach to stripe and hexagon front patterns in \S\ref{subsec:Spatial2D}. Hence, one must perform some form of reduction in order to analyze the vector field of \eqref{eqn:ww-evo}. Numerous rigorous reduction techniques have been developed for the study of water waves, including the use of Dirichlet--Neumann operators, Lyapunov--Schmidt reductions and center--manifold reductions. These approaches have become standard within the water waves community, and so we direct readers to previous reviews \cite{Groves2004WaterWaves,Dias2003WaterWaves} for more details. Due in part to these developments, the study of localized (1+2)D patterns is the most advanced of the 3D localized patterns and almost all the applications and models are in the fluid mechanics context.

\subsection{Localized (2+1)D patterns}\label{ss:3d-planar}

Next, we consider patterns that exhibit localization in two spatial dimensions, referred to as (2+1)D patterns. These localized patterns appear as fully localized 2D patterns extended in a third spatial dimension. Examples again include water waves and binary fluid convection, as well as patterns on the surface of a ferrofluid in the Rosensweig instability experiment. As detailed in \S\ref{ssec:FullyLocal}, fully localized patterns in two spatial dimensions are still not well-understood mathematically, and this similarly extends to (2+1)D patterns as well. Hence, the primary focus of this subsection is on the numerical computation of localized planar patterns, as well as several analytic studies of solutions that are either axisymmetric or that decay algebraically in one direction.

\begin{figure}[t]
    \centering
    \includegraphics[width=0.8\linewidth]{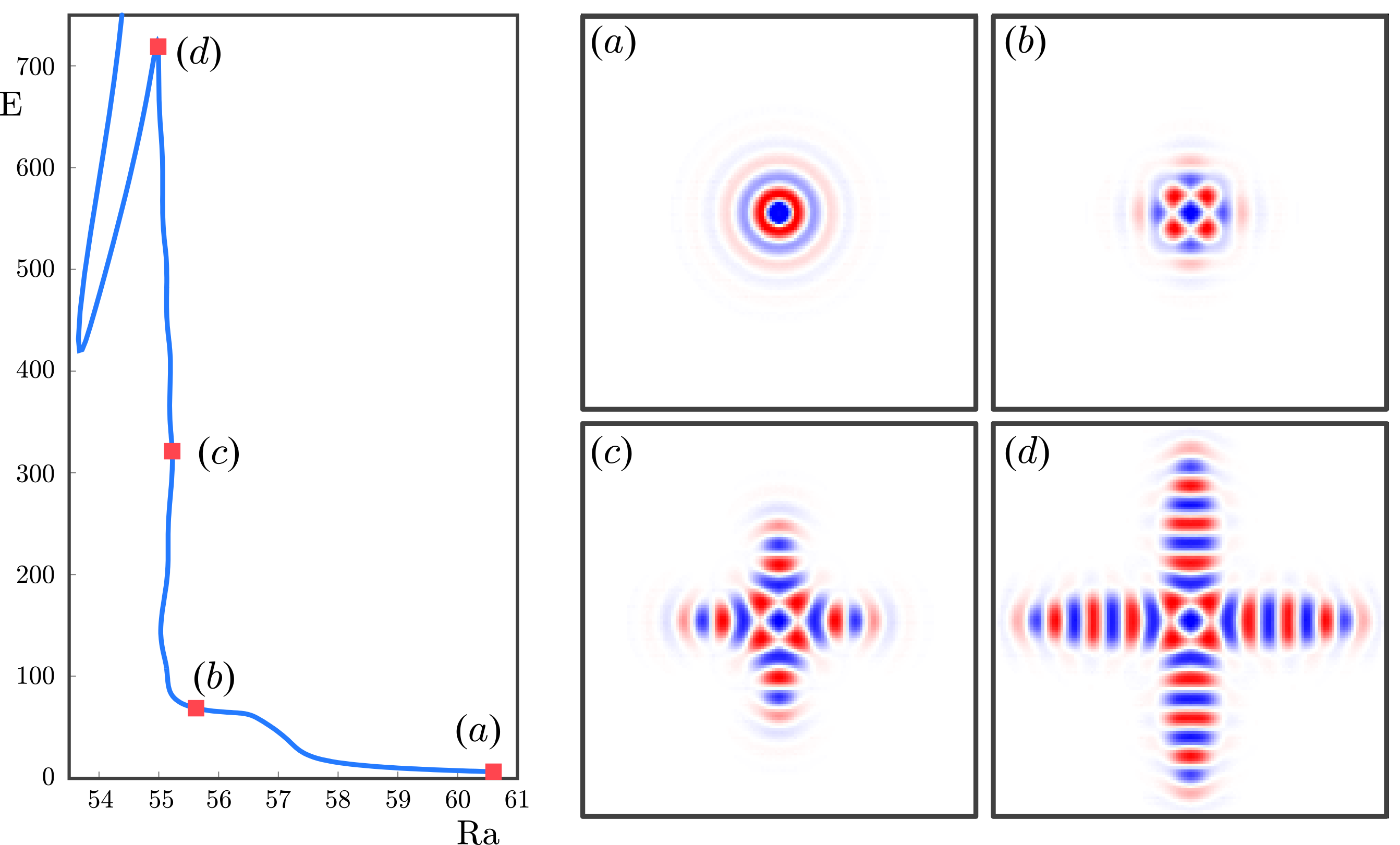}
    \caption{A bifurcation diagram of localized planar binary fluid convectons in a porous medium, where $\mathrm{E}$ and $\mathrm{Ra}$ are the kinetic energy and Rayleigh number, respectively. (a)-(d) 2D slices of 3D localized patterns for binary-fluid convection, plotted from the mid-plane velocity.}
    \label{fig:2Dconvectons_growth}
\end{figure}

Localized planar patterns have been observed in various hydrodynamic settings, including binary fluid convection~\cite{Lo_Jacono_2017}, magnetohydrodynamics~\cite{blanchflower_weiss_2002,houghton_bushby_2011}, and turbulence of plane Couette flow~\cite{brand_gibson_2014}. Between these examples, the only thorough investigation of the bifurcation structure of localized planar patterns comes from the work of Lo Jacono et al.~\cite{lo_jacono_bergeon_knobloch_2013,Lo_Jacono_2017} on binary fluid convection. Here, the authors use spectral element methods to numerically identify localized rhombic, i.e. $\mathbb{D}_{2}$-symmetric, and square, i.e. $\mathbb{D}_{4}$-symmetric, patterns that are then continued in a system parameter to trace out their bifurcation diagrams. Localized planar patterns with both $\mathbb{D}_{2}$ and $\mathbb{D}_{4}$ symmetries are found to emerge and undergo snaking, where the pattern grows in width. However, in contrast to the localized dihedral patterns observed in \cite{hill2023approximate,lloyd2008localized}, the patterns do not grow in a regular fashion by maintaining a convex core of periodic pattern. Rather, as one can see in Figure~\ref{fig:2Dconvectons_growth}, grow takes place through the extension of SPPs in a symmetric arm-like configuration. The `arms' of each localized planar pattern exhibit similar behavior to the worm solutions studied by Avitabile et al. \cite{avitabile2010snake} for the cubic-quintic SHE. Lo Jacono et al. further investigated localized planar patterns in binary fluid convection in various sizes of domains~\cite{Lo_Jacono_2017}. They observed that, for relatively small domains, localized rhombic and square patterns grow until they fill the domain, generating spatially extended convection patterns. In contrast, if the domain is large enough, each pattern maintains a `convex core - extended arms' structure as observed previously. As such, the pattern never fills the domain and does not generate any kind of spatially extended convection pattern, as shown in Figure~\ref{fig:2Dconvectons_growth}.

Beyond fluid convection, many of the key results for (2+1)D patterns again come from the water waves community, which perhaps represent the most natural example of such localized planar patterns in the form of \emph{three-dimensional solitary waves}. In 2005, Parau et al.~\cite{parau_vanden-broeck_cooker_2005} numerically computed three-dimensional gravity-capillary solitary waves by extending boundary integral methods for two-dimensional waves. The resulting solutions exhibit decaying oscillations in the direction of propagation and monotone decay in the transverse direction. Similar results have been determined for internal~\cite{parau_vanden-broeck_cooker_2007} and hydroelastic waves~\cite{parau_vaden-broeck_2011,trichtchenko_parau_vaden-broeck_milewski_2018}, and further numerical studies have focused on the nonlinear interaction between multiple solitary waves~\cite{wang_milewski_2012,akers_milewski_2010}.

\begin{figure}[t]
    \centering
    \includegraphics[width=\linewidth]{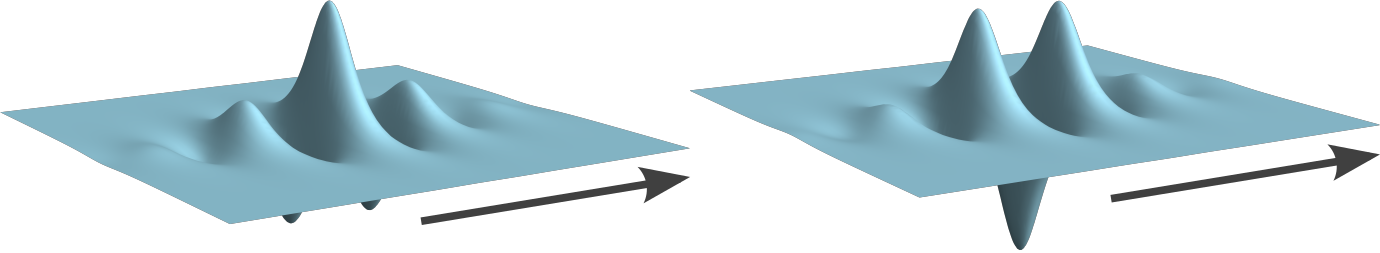}
    \caption{An illustration of fully localized solitary waves traveling in the direction shown by the arrow.}
    \label{fig:3Dww}
\end{figure}

The majority of rigorous results for localized planar patterns utilize variational techniques, such as variational Lyapunov--Schmidt reductions~\cite{groves_sun_2008,buffoni_groves_sun_wahlen_2013,buffoni2018variational,Ehrnstrom2018fullyloc}. Localized solutions are then found as minimizers of an energy functional, such as for the Davey--Stewartson equation, and can usually be found by using mountain-pass or concentration-compactness methods. More recently, Buffoni et al. \cite{buffoni_groves_wahlen_2022} presented a non-variational Lyapunov--Schmidt reduction that maps the infinite-depth water wave problem to a perturbation of the two-dimensional (stationary) nonlinear Schr\"odinger equation
\begin{equation}\label{eq:2DNLS}
    -\frac{1}{2}\zeta_{xx} - \zeta_{yy} + \zeta - \frac{11}{16}|\zeta|^2\zeta = 0, 
\end{equation}
for a complex field $\zeta = \zeta(x,y) \in \mathbb{C}$. Localized planar patterns are then approximated by a unique ground state of \eqref{eq:2DNLS}. Figure~\ref{fig:3Dww} provides an illustration of such minimizers. For a more detailed review of the functional analytic approaches for localized planar patterns in water waves see \cite[Section~8.3]{Haziot2022EbbandFlow}. 

\begin{figure}[t]
    \centering
    \includegraphics[width=0.9\linewidth]{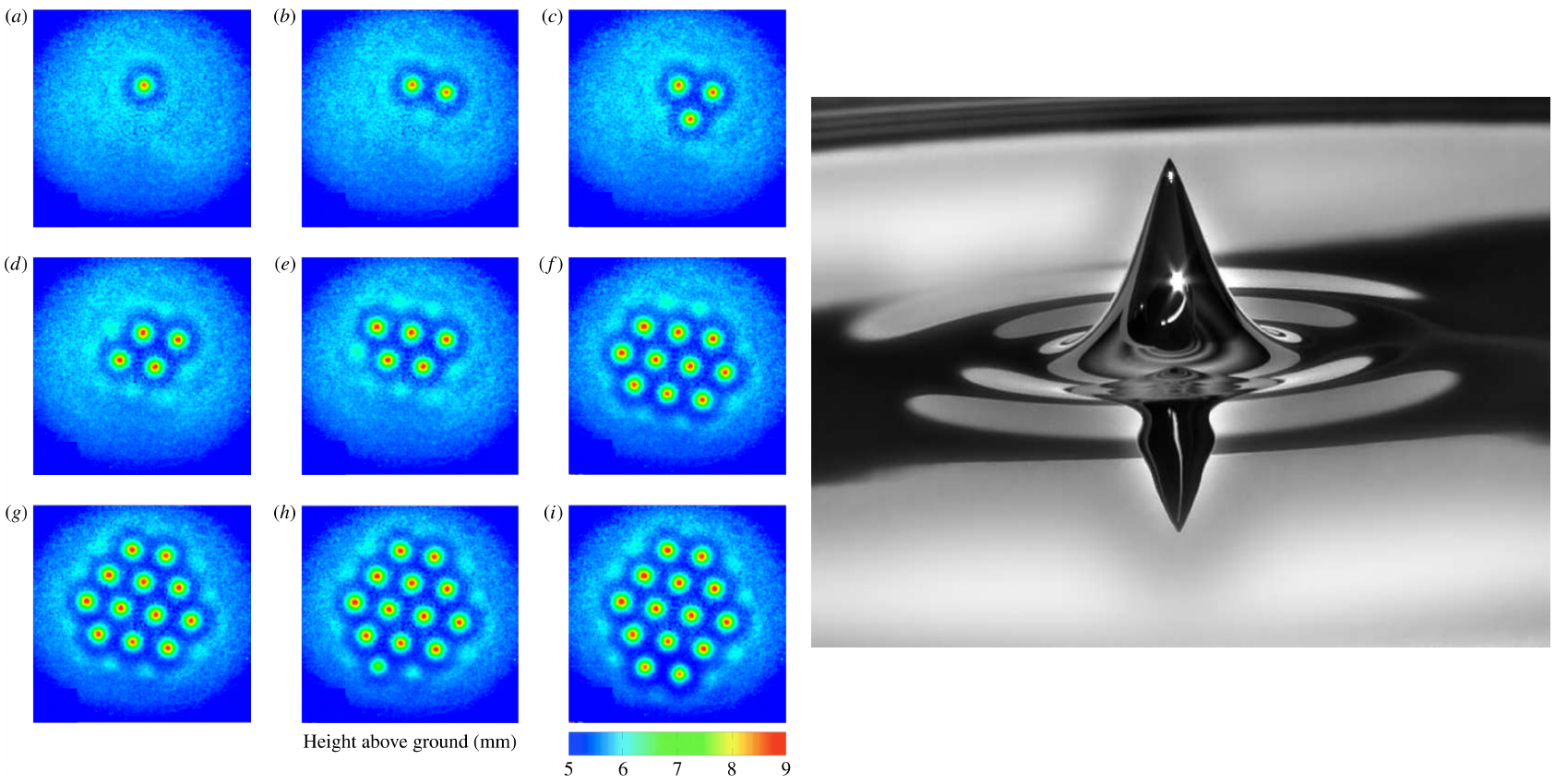}
    \caption{Left: Localized ferrofluid patterns for increasing applied magnetic field strength, originally appearing in \cite{lloyd2015homoclinic}. Right: A photo of a localized spike on the surface of a ferrofluid, originally appearing in \cite{Richter2011Mountains}.}
    \label{fig:ff_growth}
\end{figure}

A notable example of (2+1)D patterns comes from the so-called Rosensweig instability experiment. Here, a circular dish of magnetic fluid, known as a ferrofluid, is placed between two Helmholtz coils that induce a constant magnetic field vertically up through the fluid. At a critical applied field strength, a surface instability occurs and spikes begin to form on the surface of the previously quiescent fluid. In 2005, the experiments of Richter \& Barashenkov~\cite{richter_barashenkov_2005} revealed that localized spikes can form in the bistable region and remain stable, as can be seen in Figure~\ref{fig:ff_growth}.

\begin{figure}[t]
    \centering
    \includegraphics[width=\linewidth]{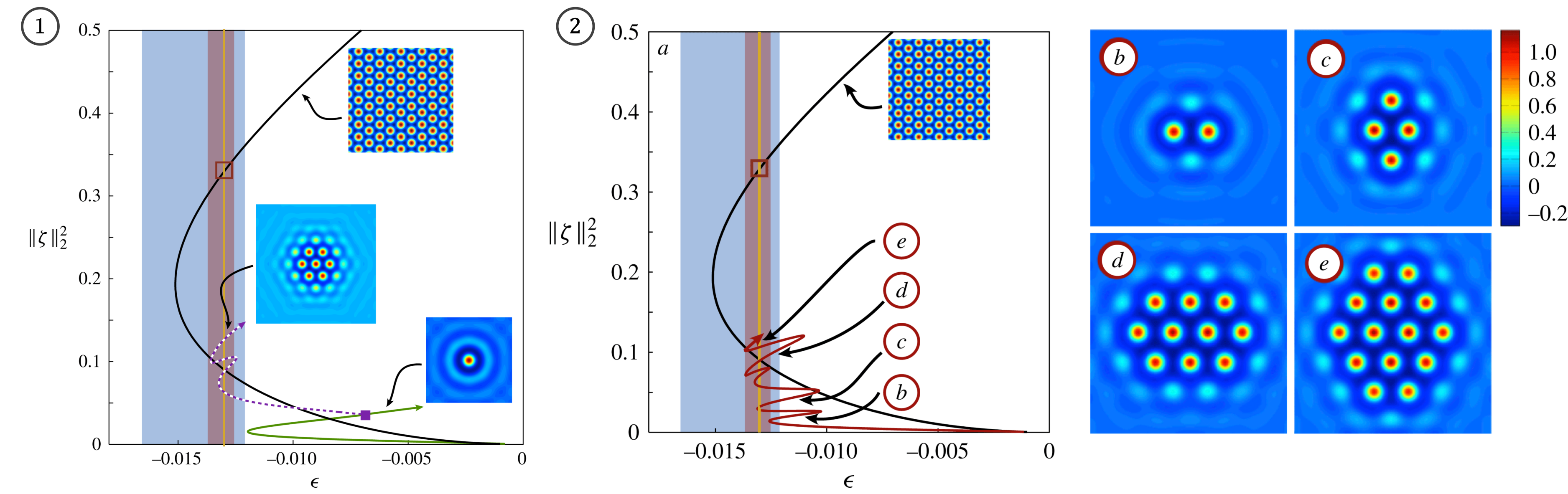}
    \caption{Bifurcation of localized (1) hexagonal and (2) rhombic ferrofluid patterns. Image originally appears in \cite{lloyd2015homoclinic}. 
    }
    \label{fig:ff_snake}
\end{figure}

In 2015, Lloyd et al. \cite{lloyd2015homoclinic} aimed to further study the emergence of localized planar ferrofluid patterns using a combination of experimental and numerical tools. The non-dimensional ferrohydrostatic equations take the form
\begin{equation}\label{eqn:ferrofluidic}
    \begin{aligned}
        \Delta\phi^{-} ={}& 0, &\qquad  & z\in (-D,\eta), & \qquad \left(\mu\nabla \phi^{-} - \nabla \phi^+\right)\cdot\mathbf{\hat{n}} ={}& 0, &\qquad  &z=\eta,\\
         \Delta\phi^{+} ={}& 0, &\qquad  & z\in (\eta,D), & \qquad \left(\nabla \phi^{-} - \nabla \phi^+\right)\times\mathbf{\hat{n}} ={}& \mathbf{0}, &\qquad  &z=\eta,\\
        \nabla\phi^{\pm}\cdot\mathbf{\hat{n}} ={}& 0, &\qquad  & z=\pm D,&\qquad 
        \frac{\mu}{2}|\nabla\phi^{-}|^2 - \frac{1}{2}|\nabla\phi^{+}|^2 + F(\phi^{-},\phi^{+},\eta) ={}& 0, &\qquad  &z=\eta,\\
    \end{aligned}
\end{equation}
for respective magnetic potentials $\phi^-,\phi^+$ in the ferrofluid and the air, and free surface $\eta$. Here $\mu$ denotes the magnetic permeability of the ferrofluid and the function $F(\phi^{-},\phi^{+},\eta)$ models the effects of the magnetic field, gravity, and surface tension on the interface. Beyond the axisymmetric spikes found previously, the authors observed the emergence of localized patches of dihedral patterns, including the hexagons pictured in Figure~\ref{fig:app_collage}(1). Starting with a single axisymmetric spike, they were able to produce dihedral patches of spikes in a hexagonal arrangement, where the number of spikes increased proportionally with the strength of the applied magnetic field, as shown in Figure~\ref{fig:ff_growth}. In order to numerically simulate these patterns, the ferrohydrostatic problem was formulated as an energy minimization problem that can then be approximated via pseudo-spectral methods. Through this approach, Lloyd et al. were able to numerically study localized planar ferrofluid patterns with axisymmetry as well as rhombic ($\mathbb{D}_{2}$) and hexagonal ($\mathbb{D}_{6}$) symmetry. Figure~\ref{fig:ff_snake}(1) shows that, similar to the SHE, localized hexagons in the ferrofluid problem emerge from a symmetry-breaking bifurcation on the solution branch for the axisymmetric spot, whereas localized $\mathbb{D}_{2}$ patterns emerge from the trivial state in Figure~\ref{fig:ff_snake}(2). Both the rhombic and hexagonal patterns undergo snaking, with both solution branches orbiting the Maxwell point for the domain covering hexagonal pattern. The growth dynamics of the patch, whereby spikes are added to the patch one-by-one as the applied field strength increases, suggest that the snaking branches may be slanted (recall \S\ref{ss:slant}) because of certain modeling assumptions such as choice of magnetization law or boundary effects.

Recent attempts to prove the existence of localized (2+1)D ferrofluid patterns have been restricted to considering cylindrical patterns, such as those observed in experiments by Richter and Barashenkov in 2005~\cite{richter_barashenkov_2005}. These cylindrical patterns are axisymmetric and localized in two dimensions, while being extended and bounded in the third direction. In order to study the emergence of localized axisymmetric patterns in \eqref{eqn:ferrofluidic}, Hill et al. \cite{hill2021localised} presented a formal reduction to the center subspace of the linearized problem about the quiescent state in the far-field region, i.e. $r \gg 1$, using the theory of strong stable foliations previously applied to localized oscillons for the complex Ginzburg--Landau equation \eqref{OscillonCGL} in \cite{mcquighan2014oscillons}. The radial ferrohydrostatic problem, restricted to the center subspace, can then be written in an equivalent form to the radial SHE \eqref{SHE_spatial_Rad}, where the weakly nonlinear theory of \cite{lloyd2009localized,mccalla2013spots} is then applied to find localized spots and rings, similar to those reviewed in \S\ref{sec:AxiEmerg}. However, a rigorous existence result for localized axisymmetric solutions to the ferrofluid equations remains an open problem.

The understanding of these patterns is largely driven and limited by our understanding of the planar localized patterns in the SHE described in \S\ref{ssec:FullyLocal}. Hence, any advances in understanding localized planar patterns in the SHE will lead to new insights for localized (2+1)D patterns, and vice-versa.

\subsection{Fully Localized 3D Patterns}

The final class of localized patterns in 3D are fully localized patterns, where localization is exhibited in all three directions. These are less well-studied than the previous front and planar patterns, and real-world examples are currently limited to nonlinear optics \cite{Brambilla2004LightBullets,Tlidi2021LightBullets} and soft matter crystals \cite{Subramanian2018LocalizedPFC}. As discussed at the end of Section~\ref{sec:AxiEmerg}, axisymmetric spot solutions were proven to exist in the three-dimensional SHE in \cite{mccalla2013spots} using the same tools from radial spatial dynamics as in the two-dimensional case. However, there is currently no existence proof for rings in three dimensions, or for fully localized patterns with non-trivial angular dependence. The weakly nonlinear Galerkin approach of 
Hill et al. \cite{hill2022dihedral,hill2023approximate} for planar dihedral patterns might be extended to three-dimensional problems, especially in the case of patterns with discrete rotational symmetries. Such patterns have already caught the interest of the soft matter community, particularly in regard to quasicrystals. 

\begin{figure}[t]
    \centering
 \includegraphics[width=0.8\linewidth]{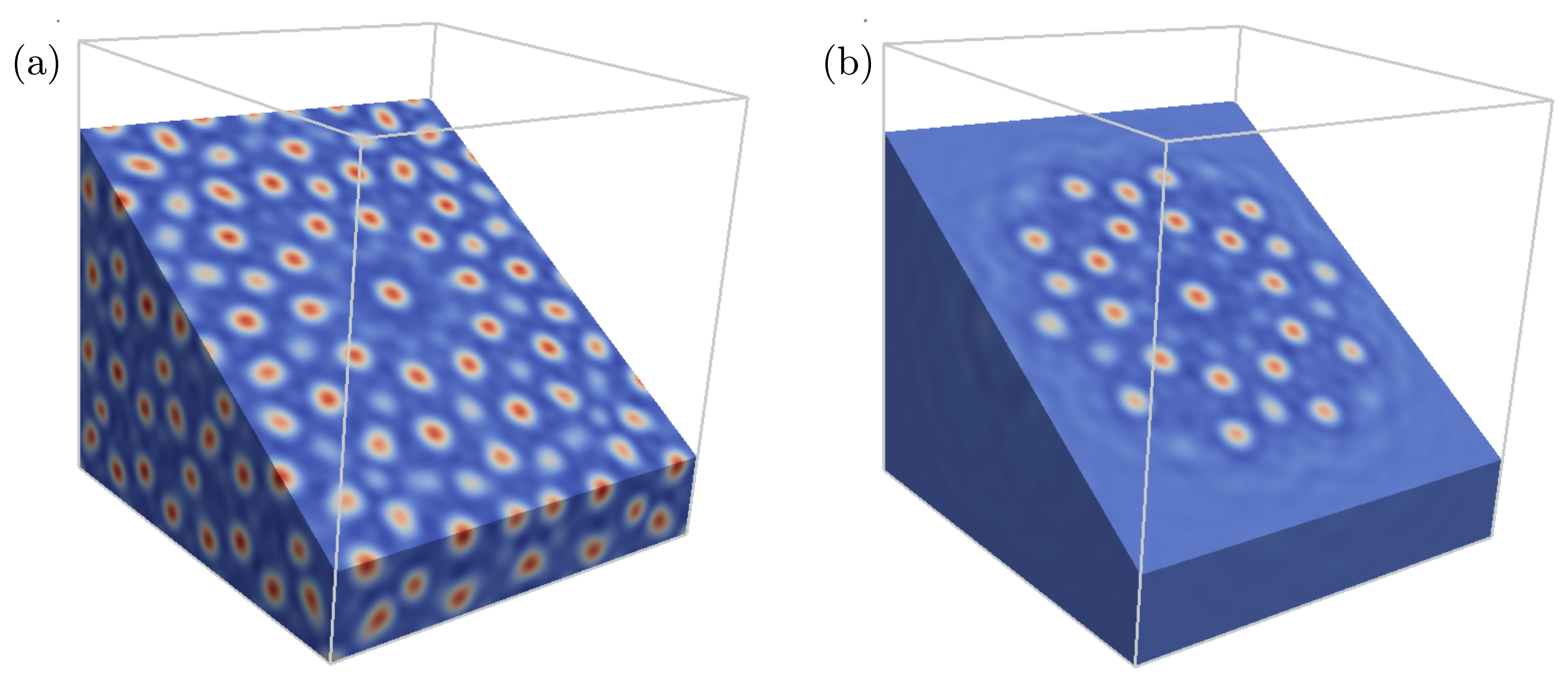}
    \caption{Domain covering and localized quasicrystals in the PFC model. Image originally appears in \cite{Subramanian2018LocalizedPFC}.
    }
    \label{fig:Quasicrystals}
\end{figure}

Localized quasicrystalline structures were investigated by Subramanian et al. in \cite{Subramanian2018LocalizedPFC} for the Phase-Field-Crystal (PFC) model. These states minimize the energy potential
\begin{equation*}
    \Omega(U) = \int\left[-\mu U -\frac{1}{2}U\mathcal{L}U  - \frac{Q}{3} U^3 + \frac{1}{4}U^4\right]\mathrm{d}^3\mathbf{x},
\end{equation*}
where the linear operator $\mathcal{L}$ induces two distinct wavenumbers $k=1$, $k=q<1$ (in comparison to the SHE that induces only $k = 1$). Using numerical methods to minimize $\Omega(U)$, the authors obtain not only thermodynamically-stable domain covering quasicrystals, but also metastable fully localized quasicrystalline structures. These structures retain the rotational symmetry of their domain covering counterparts, and can be thought of as a domain covering quasicrystal overlaid with a fully localized envelope. An example of a domain covering and localized quasicrystals are provided in Figure~\ref{fig:Quasicrystals}.

It is clear that fully localized 3D patterns have not received much attention, mostly due to the computational demands to compute them as well as the problem of visualization and classification of the different types of patterns. For this area to advance, it is clear that more numerical computations are required to investigate the solutions to in turn lead to further mathematical undertakings.

\section{Conclusion and Open Problems}\label{sec:Conclusion}

The study of localized patterns has seen significant development over the past 30 years. It has motivated new mathematical methods and techniques that can be used for a variety of applications that go far beyond localized patterns. The use of numerical continuation and the increase in speed of modern computers has allowed researchers to uncover new phenomena and provide evidence for interesting mathematical questions. It is clear the understanding of one-dimensional localized patterns is strongest and most complete, while the community's grasp weakens as the dimension the patterns arise on increases.

In many respects multidimensional localized patterns have a lot in common with their 1D localized counterparts, and so the simpler 1D setting provides an optimal testing ground to understand more complex and realistic patterns. Some common features include that localized stripes and axisymmetric patterns have a well-developed mathematical theory, while localized 2D patches appear to have snaking behavior that is governed by localized front patterns. Despite the numerical evidence for the existence of fully localized 2D patches, one of the biggest open mathematical problems is that there currently exists no mathematical theory for the emergence of such patches from a pattern forming instability. The recent work of Hill et al.~\cite{hill2022dihedral} has provided some advancement in this area that could be further developed mathematically or used as a point of initiation for a computer-assisted proof.

{\bf Future directions and open problems}

While it is clear from the review that developing mathematical theory for 3D localized patterns is a major and fruitful area for future research, there exists other avenues that might be more tractable in the immediate future. Many have been highlighted as we progressed through the individual sections, but here we enumerate some outstanding problems that would significantly advance the field.

\begin{enumerate}

\item In one spatial dimension, periodic-to-periodic (PtoP) and quasi-periodic to quasi-periodic heteroclinic connections remain largely unstudied. Kaheman et al.~\cite{Kaheman_Bramburger_Kutz_Brunton_2023} used similar techniques to \cite{beck2009snakes,bramburger2019localized} to prove the existence of PtoP connections in Hamiltonian systems using local Poincar\'e mappings, which could be applied to the spatial system \eqref{SHE_spatial_1D}. This work may be useful to analytically justifying the recent preliminary numerical findings for the 1D SHE and variations \cite{Bandara2023,Bentley2021}, while deeper techniques would be required to better understand 2D and 3D stripe grain boundaries~\cite{Lloyd2017,Haragus2007,Haragus2022,Buffoni2023,Haragus2012}. The main challenge from a computational perspective is that the boundary conditions become part of the problem to be solved. This is because the heteroclinic orbits typically select a wavenumber of the far-field spatially periodic orbits, which inhibits the numerical investigations that have largely driven the field to date. 

\item Localized patterns in the presence of spatial inhomogeneities is an active area of research in the singular perturbation limit~\cite{Doelman2016,Wong2021} in one spatial dimension. Most common is the study of either a delta or step function inhomogeneity but more general types of small amplitude inhomogeneities have also been considered~\cite{Bastiaansen2020,Doelman2018}. Near the Turing instability limit various numerical studies have been carried out for domain covering  and localized patterns~\cite{Eamonn2023,Kao2014} but there is little mathematical analysis beyond linear instability analysis. 

\item Many of the proofs for the results reviewed in Section~\ref{sec:radial} rely upon the fact that axisymmetric RD systems have a spatial dynamics representation, with the radial variable acting as a `time-like' quantity. This spatial dynamics formulation does not construct a single equation for all of $r$ and leads to a very involved matching process of the core and far-field manifolds. It would be advantageous to future studies to develop a reduction that does not rely on such an involved matching process and have a more direct proof of the axisymmetric localized patterns. Furthermore, when looking to axisymmetric (2+1)D patterns in 3D with a bounded third direction, developing a general radial center-manifold reduction for quasi-linear PDEs and nonlocal equations is of immediate importance.  

\item The approximate theory of localized dihedral patches of Hill et al.~\cite{hill2022dihedral} is a first step to understanding the emergence of localized patches from a Turing instability. Extending the analysis to prove the existence of the patches in PDEs remains an open problem. It is clear from the numerics that the bifurcation diagram of the patches is intrinsically linked to the bifurcation behavior of the hexagon fronts, but currently there is no proof of this.

\item The majority of work on localized patterns in lattice equations described in~\S\ref{subsec:SquareLattice} relies heavily on the form of coupling between lattice sites. What is not fully understood is how the connection topology of the lattice informs both the patterns that emerge and the bifurcations of these patterns that one observes. Some work has been made in this direction. Tian et al.~\cite{tian2021snaking} attempted to better understand the effect of different couplings on ring lattices by looking at how longer-range couplings affect the bifurcation diagram of localized solutions, while \cite{bramburger2023pattern} has initiated a study into predicting patterns that emerge from Turing bifurcations on random networks. This latter study could potentially form the basis for an understanding of localized pattern formation and the resulting snaking bifurcations on random networks that were observed in \cite{mccullen2016pattern}. Beyond just spatially-discrete RD systems, localization has further been observed in networks of coupled oscillators \cite{papangelo2017snaking,papangelo2018multiple}. Little is understood regarding these localized oscillations, while recent numerical work by Nicolaou and Bramburger \cite{nicolaou2024complex} provides numerical evidence for entirely new routes to localization beyond Turing instabilities. A better understanding of these results both analytically and numerically remains outstanding.

\item Developing a general stability theory in particular for 3D localized patterns in \S\ref{sec:3d} is very much an open problem. This could potentially be tackled through a combination of variational and spatial dynamical methods. Spatial inhomogeneities can have a significant impact on pattern selection, and there are some recent advances in this direction~\cite{Jaramillo2015,Jaramillo2019}. 

\item Studying localized patterns in the context of disordered, random, and noisy systems is a very open area for future research with very limited work carried out so far~\cite{Lord2012,Inbal2010,Descalzi2015,Cartes2012,Cartes2019}. Most works have been limited to either numerical simulation or finding a mean-field equation and then applying the methods described in this review. Developing a more systematic theory would greatly increase the range of applications of localized patterns. 
\end{enumerate}

While localized pattern formation is a maturing field, the fundamental problems of why emergent, localization with structure occurs in continuum and discrete models and their behaviors such as pattern selection remain very much open. The importance of developing new mathematical and computational tools to study these structures is a very exciting area of active research. 

\section*{Acknowledgments}

This work was carried out with support from the Alexander von Humboldt Foundation (DJH) and an NSERC Discovery Grant (JJB). We gratefully acknowledge Mark Groves (Figure~\ref{fig:3Dww}), Edgar Knobloch (Figure~\ref{fig:2Dconvectons_growth}) and Priya Subramanian (Figures~\ref{fig:hexagon_hexagon_patch} and \ref{fig:Quasicrystals}) for providing data to reproduce some of the figures presented in this work. We also thank C\'edric Beaume, Alan Champneys, Jonathan Dawes, Arjen Doelman, Gregory Kozyreff, Edgar Knobloch, Alastair Rucklidge,  Priya Subramanian, Edgardo Villar-Sep\'ulveda, and Michael Ward for their helpful comments on the first draft of the review.

\bibliographystyle{abbrv}
\bibliography{Bibliography.bib}
\end{document}